\documentstyle[times,psfig,emulateapj]{article}

\def\vvmax{$\langle V_e/V_a \rangle$}
\def\veva{$\langle V_e/V_a \rangle$}
\def\avgz{$\langle z \rangle$}
\def\avgf{$\langle f \rangle$}
\def\fxfr{$f_x/f_r$}
\def\nupeak{$\nu_{peak}$}
\def\Lbol{$L_{bol}$}

\lefthead{Rector et al.}
\righthead{The Properties of EMSS XBLs}

\begin{document}

\title{The Properties of the X-Ray-Selected EMSS Sample of BL Lac Objects}
\author{Travis A. Rector\altaffilmark{1,2,3,4} and John T.
Stocke\altaffilmark{1,2,3}}
\affil{Center for Astrophysics and Space Astronomy, University of Colorado,
Boulder, Colorado 80309-0389}

\author{Eric S. Perlman\altaffilmark{2,3}}
\affil{Johns Hopkins University, Baltimore, MD 21218}

\author{Simon L. Morris}
\affil{Dominion Astrophysical Observatory, 5071 West Saanich Road, Victoria, BC V8X
4M6, Canada}

\and

\author{Isabella M. Gioia\altaffilmark{5}} 
\affil{Istituto di Radioastronomia del CNR, via Gobetti 101, I-40129, Bologna Italy}

\altaffiltext{1}{Visiting Astronomer, Kitt Peak National Observatory, National
Optical Astronomy Observatories, which is operated by the Association of
Universities for Research in Astronomy, Inc. (AURA) under cooperative agreement
with the National Science Foundation.}
\altaffiltext{2}{Visiting Astronomer, Multiple Mirror Telescope Observatory.  MMTO
is owned and operated by the Smithsonian Astrophysical Observatory and the
University of Arizona.}
\altaffiltext{3}{Visiting Astronomer, National Radio Astronomy Observatory.  NRAO
is a facility of the National Science Foundation operated under cooperative
agreement by Associated Universities, Inc.}
\altaffiltext{4}{Current address:  National Optical Astronomy Observatories, 950 N. Cherry
Avenue, Tucson, AZ  85719}
\altaffiltext{5}{Current Address: Institute for Astronomy, University of Hawaii,
2680 Woodlawn Dr., Honolulu,
HI 96822}



\begin{abstract}

We present updated and complete radio, optical and X-ray data for BL Lacs in the {\it
Einstein} Medium Sensitivity Survey (EMSS).  The complete ``M91" sample first
presented in Morris et al. is updated to include 26 BL Lacs in total
and we define a new, virtually complete
sample consisting of 41 EMSS BL Lacs (the ``D40" sample). 
New high signal-to-noise,
arcsecond-resolution VLA observations are also presented for eleven EMSS BL Lacs, completing
VLA observations of the M91 sample.
The addition of four new objects, as well as updated X-ray flux and redshift information, has
increased the \vvmax\ value for the M91 sample to $0.399\pm0.057$; and \vvmax\ $=
0.427\pm0.045$ for the newly defined D40 sample.  
In conjunction with other studies of X-ray-selected BL Lac (XBL) samples,
these results solidify negative evolution for XBLs, especially for more extreme
high-energy-peaked BL Lacs, for which we find \vvmax\ $=0.271\pm0.77$.  
The observed \vvmax, spectral and radio properties of XBLs are 
completely consistent with being the beamed population of low-luminosity, FR--1
radio galaxies.
However, our VLA observations do confirm that XBLs are too core-dominated to be
consistent with a beamed population of FR--1s seen at intermediate angles, as
suggested by the unified model, if XBLs have moderate outflow velocities ($\gamma \sim
5$).  

\end{abstract}




%

\section{Introduction}

The {\it Einstein} Medium Sensitivity Survey (EMSS; Gioia et al. 1990; Stocke et al.
1991, hereafter S91; Maccacaro et al. 1994) contains 835 faint ($f_x > 7$ x $10^{-14}$
ergs cm$^{-2}$ s$^{-1}$ in the 0.3--3.5 keV soft X-ray band) X-ray sources
discovered serendipitously with {\it Einstein} Imaging Proportional Counter (IPC)
images obtained of various targets at high Galactic latitude ($b > 20$\arcdeg). At
the time of this writing, the EMSS is still the largest statistically-complete
sample of faint X-ray sources that can be used to define complete samples of X-ray
selected AGN, clusters of galaxies, Galactic stars and BL Lac objects. The primary
virtue of the EMSS compared to recent survey work done with ROSAT (e.g., Bade et al. 1998)
is that the EMSS sample has been thoroughly scrutinized over the years, with
periodic updates of identifications and properties (e.g., Maccacaro et al. 1994; Wolter et
al. 1994; Perlman et al. 1996a, hereafter P96; Rector, Stocke \& Perlman 1999, hereafter
RSP99). And more detailed observations of this sample exist than for any other BL Lac sample,
excepting perhaps the 1Jy radio-selected BL Lac sample (Stickel et al. 1991; Rector \&
Stocke 2000, hereafter RS00). The primary purpose of this paper is to provide 
updated information on the EMSS sample of BL Lacertae objects.


Morris et al. 1991, hereafter M91, first defined a plausibly
complete subsample of 22 EMSS BL Lacs with $f_x > 5$ x $10^{-13}$ ergs cm$^{-2}$
s$^{-1}$ in the 0.3--3.5 keV IPC band and declination $\delta \geq -20$\arcdeg. 
M91 also obtained high signal-to-noise spectra of these objects, from which
redshifts were derived for most of them. The M91
BL Lac sample was drawn from a subset of bright EMSS sources which were thought to be
completely and correctly identified.  However, investigations of
this sample continued, first by obtaining ROSAT PSPC soft X-ray spectra of the M91
sample to measure X-ray spectral indices to determine more accurate X-ray flux densities
from IPC count rates and to determine $K$-corrections (P96). P96 also delineated the
criteria by which we identified sources as low-luminosity BL Lac candidates.  Then by
using the ROSAT High Resolution Imager (HRI) we identified additional BL Lacs which were
originally misidentified as clusters of galaxies (RSP99). These new X-ray images and
X-ray spectral indices, in conjunction with new optical spectroscopy and VLA mapping,
have increased the M91 BL Lac sample to 26 objects and the full EMSS sample to 44 objects.


The properties of the EMSS BL Lac objects have been presented in part in several
publications. The original description of the X-ray selected BL Lac (XBL) class was given in
Stocke et al. (1985) using EMSS and HEAO-1 A2 BL Lacs, in which the original distinctions
between XBLs and radio-selected BL Lacs (RBLs) were described. The original discussions of the
source counts of XBLs and the ``negative" cosmological evolution of this class can be found in
Maccacaro et al. (1983).  Subsequent papers give updates on the sample membership (S91; P96;
RSP99), source counts (e.g., Wolter et al. 1991) and the luminosity function,
cosmic evolution and \vvmax\ test (M91; P96; RSP99). High signal-to-noise optical
spectra were presented by M91 and RSP99, high dynamic range VLA radio maps of many
of the M91 sample by Perlman \& Stocke (1993; hereafter PS93) and RSP99, ROSAT
PSPC X-ray spectra for the M91 sample by P96, deep optical imaging for host
galaxy morphology and luminosity by Wurtz, Stocke \& Yee (1996), clustering
environment by Wurtz, Stocke, Ellingson \& Yee (1997), infrared photometry by Gear
(1993) and optical polarimetry by Jannuzi, Smith \& Elston (1993, 1994).

The purpose of this publication is to present the EMSS BL Lac sample in its entirety, with
additions to the M91 sample as well as the introduction of the new, probably complete ``D40"
sample (\S 2).  We update and describe the observational properties (\S 3),
with attention to the criteria adopted to define the sample (\S 4).  Individual objects are
discussed in
\S 5, and we discuss the completeness of the sample in \S 6.  With the updated M91 and new
D40 samples we derive the X-ray luminosity function (XLF), \vvmax\ and implied cosmological
evolution for XBLs (\S 7). In \S 8 we discuss the implications of this work for unification
models.  Finally, we summarize the most important
results as well as suggest future work (\S 9).  Throughout this paper we assume a
standard cosmology with $H_0 = 50$ km s$^{-1}$ Mpc$^{-1}$ and $q_0 = 0.0$.  We
also assume the notation $f(\nu) \propto \nu^{+\alpha}$.

\section{The EMSS Sample of BL Lacertae Objects}

The revised sample of 44 EMSS BL Lac objects is listed in Table~\ref{tbl-4.1} along
with a summary of their basic properties. Twenty-six of these BL Lacs now satisfy
the flux and declination criteria for the M91 statistically-complete sample (see
also P96 \& RSP99), increasing the original sample size by four.
Browne \& March\~a (1993, hereafter BM93) recognized that a few of the EMSS
sources originally identified as clusters of galaxies instead might be BL Lacs.
RSP99 tested the BM93 hypothesis by obtaining ROSAT HRI images of ten
of the most suspect sources and found that four of them were unresolved  
at the position of a radio galaxy in the cluster. Subsequent optical
spectroscopy showed evidence for a depressed CaII break, which either met or came
close to meeting the criteria suggested by S91 for inclusion in the BL Lac class
(see \S 4.1). The other properties of these sources (e.g., radio luminosity and
structure, overall spectral energy distributions) were also a good match to those
of other XBLs, strongly suggesting that these are low luminosity BL Lacs as
predicted by BM93. All four sources which appeared point-like
to the HRI (MS 1019.0+5139, MS 1050.7+4946, MS 1154.1+4255 \& MS 1209.0+3917) had
count rates as measured by the ROSAT HRI, which were comparable to or greater than
that expected from the IPC discovery count rates using the observed soft X-ray
spectral indices. Two of these objects, MS 1019.0+5139 \& MS 1050.7+4946, match the
M91 selection criteria and are now included in the M91 sample.  A fifth source (MS
0011.7+0837) was partially resolved by the HRI (RSP99) and the HRI count rate was
considerably lower than could account for the full IPC discovery flux. In RSP99 we
postulated that both a BL Lac point source and diffuse cluster emission were
present in this case but assigned the bulk of the IPC discovery flux to the cluster
so that this potential BL Lac is not included in the complete EMSS sample. The
optical image of MS 0011.7+0837 (Gioia \& Luppino 1994) shows a cluster surrounding this
source, consistent with this interpretation.

A similar HRI observation was made of MS 1333.3+1725 by Hattori \& Morikawa (private
comm).  We have analyzed this observation in the manner of RSP99 and have confirmed the
presence of an X-ray point source coincident with a radio source and an elliptical galaxy 
which has a depressed CaII break.  This source is now included in the EMSS sample as 
well. Because RSP99 targeted all of the suspect clusters in the M91
sample and most of the suspect clusters at lower flux densities and declinations
(including MS 1333.3+1725, which was originally on our target list), we are
confident that the M91 sample is virtually complete as of this writing and the
entire EMSS BL Lac sample is largely complete for reasons discussed in detail in
RSP99.

In addition to the two low-luminosity BL Lacs discussed above, two 
previously identified BL Lac objects, MS 2336.5+0517 \& MS 2347.4+1924, are now included in
the M91 complete sample due to a measured or estimated soft X-ray spectral index which is
steeper than
$\alpha_x = -1$ (as was assumed by M91 for all sources). P96 found that the soft
X-ray spectral indices of EMSS XBLs are quite steep ($\langle \alpha_x \rangle =
-1.38$), not only altering the conversion of IPC count rate to flux units but also
providing a useful discriminant between X-ray emission dominated by diffuse thermal
bremsstrahlung (which has a flat energy distribution with $\alpha_x \sim 0$ in the
IPC band) from emission dominated by a nonthermal BL Lac nucleus.  Using these new
spectral data to compute flux densities from IPC count rates raises the observed flux of
these two sources just above the M91 flux
limit.  The reader should be aware that the use of ROSAT PSPC spectral indices
obtained years after the {\it Einstein} IPC discovery observations does not take
into account potential X-ray spectral variability, by which sources usually harden
as they brighten (Giommi et al. 1990). P96 discusses this and the other effects
that source variability can have on the EMSS complete BL Lac source catalogue.

In this paper we present a new, nearly completely identified subsample
(hereafter called the ``D40" sample) of 41 EMSS BL Lacs bounded by $f_x \geq 2$ x
$10^{-13}$ ergs cm$^{-2}$ s$^{-1}$ in the IPC band and by the declination of our
completed VLA observations ($\delta \geq -40$\arcdeg).  This sample includes all
the objects in the M91 sample plus 15 more; only three known EMSS BL Lacs are
not members of the D40 sample because they are too far south.

Table~\ref{tbl-4.1} includes the following basic data by column: (1) object name; 
(2) the complete sample (M91 or D40) to which the object belongs (if either).  Note that all
members of the M91 sample are also in the D40 sample; 
(3) redshift as
determined by cross-correlation methods (see below); (4) X-ray power-law energy spectral
index as measured by the ROSAT PSPC in the (0.2--2.4 keV)
band or estimated using the Padovani \& Giommi (1996)
$\alpha_x - \alpha_{ox}$ correlation (entries of this latter sort are denoted as
uncertain); (5) the {\it Einstein} IPC discovery flux in $10^{-13}$ ergs cm$^{-2}$
s$^{-1}$ in the (0.3--3.5 keV) band assuming the X-ray spectral indices from
column (4) and Galactic neutral hydrogen columns from Gioia et al. (1990). The
flux densities quoted are all unabsorbed as was done for all extragalactic sources in
Gioia et al. (1990); (6) the monochromatic IPC X-ray flux in $\mu$Jy at 2 keV
assuming the spectral indices in column (4); (7) the $V$-band magnitude (see S91 for
discussion); (8) the total radio flux at 5 GHz from S91. The entries for MS 0622.5--5256 \&
MS 1333.3+1725 are previously
unpublished values obtained at the Australia Telescope Compact Array (ATCA) and
the VLA respectively; (9) and
(10) the two-point optical-to-X-ray ($\alpha_{ox}$) and radio-to-optical
($\alpha_{ro}$) spectral indices where $\alpha_{ox}$ = --log($S_{\rm 2 keV} / S_{\rm
2500A}$)/2.605 and $\alpha_{ro}$ = log($S_{\rm 5 GHz} / S_{\rm
2500A}$)/5.38 (S91; see Section 2.3 for details).  X-ray positions are given in Gioia et al.
(1990); radio and optical positions are given in S91.

Table~\ref{tbl-4.2} gives the monochromatic X-ray, optical and radio luminosities for
EMSS BL Lacs, by column: (1) object name; (2) the X-ray luminosity at 2 keV in the
BL Lac rest frame including $K$-corrections based upon column (3) of
Table~\ref{tbl-4.1}; (3) the optical luminosity at 5500\AA\ in the BL Lac restframe
assuming $\alpha_o = -1$; (4) the radio luminosity at 5 GHz in the BL Lac restframe
assuming $\alpha_r = -0.5$.  All luminosities are given in W Hz$^{-1}$.  Luminosities
marked with colons are based upon uncertain redshift information.  A single colon indicates
a probable redshift, and a double colon indicates a highly speculative redshift.

\section{Observations and Reduction}

\subsection{Optical Spectroscopy}

One of the virtues of X-ray selected BL Lac objects (XBLs) is that their
optical spectra often show the presence of weak stellar absorption lines,
consistent with the optical imaging which shows an underlying early-type galaxy
morphology (e.g. Urry et al. 2000: Scarpa et al. 2000; Wurtz, Stocke \& Yee 1996; Abraham et
al. 1991). Due to this characteristic most of the EMSS BL Lac objects have measured redshifts
as shown in Table~\ref{tbl-4.1}. In order to determine consistent redshifts we have employed
the cross-correlation techniques as described in Ellingson \& Yee (1994),
which have been used on moderate to low signal-to-noise ratio (SNR) spectra of
normal cluster galaxies to some advantage (e.g., Yee, Ellingson \& Carlberg 1996).
This cross-correlation method obtains accurate redshift values ($\pm100$ km
s$^{-1}$) and quantifiable reliability assessments based upon the height of the
peak in the cross-correlation (the ``$R$ value"). Based upon numerous
experiments using simulated and actual data, Ellingson \& Yee (1994) concluded that
a redshift determination with $R > 4$ was ``firm" ($\geq 99$\% confidence)
while those with $3 < R < 4$ were ``possible" (90--95\% confidence). 
Because our BL Lac spectra have much higher
SNR but lower contrast features, it was not clear how robust this cross-correlation
technique would be given the somewhat different use. However, all but one of the M91 
redshift
determinations, which were made by identifying individual spectral features, were
recovered by this process to within $\pm0.001$ in $z$. For the one remaining object
(MS 0205.7+3509), the cross-correlation technique on the original spectrum failed
to recover the redshift quoted in M91 of $z=0.318$ but a subsequent
cross-correlation using a residual spectrum obtained after subtracting a power-law
best-fit in the blue ($< 5000$\AA; see Figure~\ref{fig-4.1}) did recover this
redshift. CaII absorption weakly appears in both of our spectra of this BL Lac.  Not
all spectra were available in a form amenable to the cross-correlation process;
redshifts quoted for these remaining objects were obtained by averaging the measured
location of individual lines.  None of these spectra appear to be at all ambiguous
as to their redshift values. However, a few of these spectra are quite modest to poor in
SNR because the only spectra available for these sources (MS
0331.3--3629; MS 0350.0--3712, MS 1205.7--2921 and MS 1219.9+7542) are the original EMSS
classification spectra obtained at the Steward Observatory Bok 90" and Las Campanas
Observatory (LCO) DuPont 100" telescopes. While these spectra meet or nearly
meet the minimum SNR required to include or reject an AGN from the BL Lac class, better
quality spectra should be obtained to provide more definitive data for these
four objects.

In several other cases the BL Lac redshift placed the CaII H\&K lines near the
``join" in the original M91 spectrum obtained using the Double Spectrograph (Oke
\& Gunn 1982) at the
Palomar 200" telescope. Because the SNR of these spectra are significantly poorer
and systematics considerably worse near the join than at other wavelengths, we
obtained new spectra of these few BL Lacs at the Multiple Mirror Telescope
Observatory (MMTO), which verified the redshifts in all cases. New spectra were
also obtained at the MMTO for BL Lac candidates at lower X-ray flux densities and
declinations than the M91 sample objects. All BL Lac candidates suggested by S91
are confirmed as BL Lacs by these spectra and, in many cases, redshifts
were obtained.

We have also used the spectra in Figure~\ref{fig-4.1} to determine
the CaII H\&K ``break strength" and the equivalent width or limits on equivalent
width for any emission lines that might be found within the wavelength range of our
observations. In the manner of Dressler \& Shectman (1987) we have used the
rest-frame wavelength regions, region $A= 3750-3950$\AA\ and region $B=
4050-4250$\AA\, to define the flux depression across the H\&K ``break". We define the CaII
``break strength" as $B_{4000} = (f_B - f_A)/f_B$. This is related to the $D(4000)$ break
defined by Dressler \& Shectman (1987) by
$B_{4000} = 1 - 1/D(4000)$.  By this scheme a normal elliptical galaxy or radio galaxy in
the current epoch has $B_{4000} = 0.5 \pm0.1$. S91 suggested a limit on $B_{4000}$ values for BL Lac 
objects of $B_{4000} \leq 0.25$ but RSP99 showed that some EMSS BL Lacs exceeded that
limit slightly; and March\~a et al. (1996) estimate that galaxies with $B_{4000} \leq
0.4$ have a high probability ($\sim$95\%) of possessing a weak, non-thermal continuum
in addition to the starlight in their spectra. Emission line equivalent widths and
3$\sigma$ limits are measured in the normal fashion then corrected to rest-frame
values in order not to bias the BL Lac class
against higher redshift objects. Non-detection limits are obtained assuming a
velocity width of 1000 km s$^{-1}$, which is roughly twice the resolution of the spectra.
We note that equivalent widths are measured against the total continuum, which includes the
nonthermal continuum of the AGN as well as the starlight from the host galaxy.


Table~\ref{tbl-4.3} presents the basic information obtained from these optical spectra
including (by column number): (1) EMSS alphanumeric name; (2) redshift determined by
cross-correlation.  A single colon indicates an uncertain redshift, whereas a double colon
indicates a highly speculative redshift (see
\S 5 for discussion of individual sources); (3) ``R-value" of the cross-correlation, which
measures the confidence level of the redshift determination (see above); (4) the CaII H\&K
``break strength" as defined above with
$1\sigma$ limits; (5) the minimum emission line $W_{\lambda}$ detectable in the spectrum at
$\sim 5000$\AA\ (in
\AA).   The best
quality spectrum is used if more than one is available. Notice that a few of
these $W_{\lambda}$ limits slightly exceed the BL Lac criterion suggested in S91; (6-9)
Rest-frame emission-line 3$\sigma$ luminosity limits for MgII $\lambda$2798, [O II]
$\lambda$3727, [O III] $\lambda$5007 and H$\alpha$ respectively in log (ergs cm$^{-2}$
s$^{-1}$ \AA$^{-1}$).  Actual detections are listed in bold; and tentative line
identifications are marked with colons. A blank entry indicates that
the wavelengths of these lines are not present in the spectra in hand.
Figure~\ref{fig-4.1} presents the spectra for the entire EMSS sample, which were
obtained at the MMTO, at the Canada France Hawaii 3.6m Telescope (CFHT) at the Las
Campanas (LCO) Du Pont 100" Telescope or at the KPNO 2.1m as indicated on the
figures themselves.  Figure~\ref{fig-4.1} does not include spectra for MS 0622.5--5256 and MS
1219.9+7542, which are described in Stocke et al. (1985) but are not available, and MS
2316.3--4222, which is presented in Crawford \& Fabian (1994).

\subsection{Radio Continuum Imaging}

Six centimeter (5 GHz) snapshot observations have been obtained for all
extragalactic and unidentified sources in the EMSS at $\delta \geq -40$\arcdeg\
(S91). These VLA snapshots have typical flux limits of 1 mJy.  Radio observations
with similar flux limits at 6cm have also been made for selected sources with
$\delta  < -40$\arcdeg\ with the ATCA and reported in Maccacaro et al. (1994). The
basic 6cm data obtained for all the sources are compiled in Table~\ref{tbl-4.1}. 
Twenty centimeter continuum radio maps for the brightest radio sources in the M91 BL Lac
sample were also obtained at the VLA and published in PS93. 
However, a concern has been that Doppler boosting could have biased this subset
towards the more highly core-dominated radio sources amongst the M91 sample, making the PS93
observations not entirely representative of EMSS XBLs. 

To provide an unbiased look at the radio structure of XBLs, we have obtained
new, high SNR ($\sigma \approx 0.05$ mJy) VLA A- and B-array 20cm radio maps of 
ten more EMSS XBLs.  A B-array ``snapshot" of MS 1207.9+3945 was also
obtained, thereby completing 20cm VLA observations of all the XBLs in the M91 sample.
The core flux was measured by fitting the core with a single Gaussian with the
synthesized beam's parameters.  The extended flux was determined by measuring
the total flux within a box enclosing the entire source and then subtracting the
core flux.  
The maps of resolved sources are presented in Figure~\ref{fig-4.3}.  RSP99 have already
presented the VLA map for MS 1050.7+4946, which was newly identified
as a low-luminosity BL Lac.  The radio data for these sources are given in
Table~\ref{tbl-4.5} (by column): (1) EMSS source name; (2) core source position (in J2000
coordinates); (3) the core flux (in mJy). The beam size in arcsecs is given below in parentheses; (4)
the extended flux (in mJy).  Limits for unresolved sources are estimated as described below;
(5) the uncorrected core radio power (in W Hz$^{-1}$); (6) the uncorrected (top) and
corrected (bottom) extended radio power (in W Hz$^{-1}$; see below for discussion of the
applied corrections); (7); the ratio ($f$) of uncorrected core flux to uncorrected
(top) and corrected (bottom) extended flux.  The radio
luminosities and the core to extended flux ratios in columns 5, 6 \& 7 are $K$-corrected by
assuming
$\alpha = -0.3$ for the core and $\alpha = -0.8$ for extended structure.  Similar data for the
other M91 XBLs can be found in PS93 in Table 2.  At this point all EMSS BL Lacs which qualify
to be members of the M91 sample have been mapped by us with the VLA. The flux and LAS
corrections are taken from PS93 and described below. 

Surprisingly, despite the quality of these new maps, extended flux was detected for only
four sources.  Where extended emission was not detected, conservative upper limits on
extended radio power levels were obtained by assuming that each source has uniformly bright
extended emission at the 1$\sigma$ detection level over a 3000 kpc$^2$ area surrounding the
core.  Because the extended emission seen in resolved XBLs have radio power levels typical
of FR--1s, we have corrected the extended and core powers and the core dominance values for
the effects of surface brightness dimming and physical resolution of FR 1 type sources as
described by PS93.  These corrections are strongly dependent on the source structure, which
for these objects is not well known; we assume a wide-angle-tail structure, which is the
more conservative correction of the two morphologies modeled in PS93.  The resulting values
are shown in Table~\ref{tbl-4.5}; the sources mapped in RSP99 are also included in
Table~\ref{tbl-4.5} since RSP99 did not make these corrections.  As can be seen in the maps
and from the extended power levels in Table~\ref{tbl-4.5}, as with those M91 XBLs already
mapped, XBLs in the M91 sample have extended power levels and morphologies similar to FR--1
radio galaxies. A comparison of the radio structure and power levels for EMSS XBLs and FR--1s
is presented in
\S 8; the radio structure of EMSS XBLs are compared to 1Jy RBLs in RS00.

\subsection{X-Ray Imaging}

\subsubsection{ROSAT PSPC Observations}

Soft X-ray spectral indices are required to accurately determine both the detection
flux densities from IPC count rates and $K$-corrections. The measured
entries for soft X-ray spectral indices in Table~\ref{tbl-4.1} come either from the
pointed observations of P96 and Blair, Georgantopolous \& Stewart (1997) or from observations
found in the White, Giommi \& Angelini (1994) Catalogue (WGACAT) of serendipitous sources
detected with the ROSAT PSPC. Where individual BL Lacs were observed on multiple occasions in
the WGACAT, we have used mean values. 
For those few EMSS BL Lacs without
soft X-ray spectral data, we have estimated spectral indices using the $\alpha_x -
\alpha_{ox}$ correlation found by Padovani \& Giommi (1996). While this correlation
has a substantial spread ($\Delta\alpha_x = 0.5$) at any single value of
$\alpha_{ox}$, these estimates are typically a factor of two more accurate than
using the $\langle \alpha_x \rangle$ of P96 for any individual object. The
uncertain $\alpha_x$ values obtained using the Padovani \& Giommi (1996)
correlation method are marked in Table~\ref{tbl-4.1} with colons.    

\subsubsection{ROSAT HRI Observations}

In addition to the observations presented in RSP99, a search of the ROSAT WGACAT
found serendipitous HRI observations for five more BL Lacs in the EMSS sample: MS
0331.3--3629, MS 0607.9+7108, MS 0737.0+7436, MS 1219.9+7542 and MS 1332.6--2935.  In
addition, a targeted observation of MS 1312.1--4221 was also found; and as such this is the
only source which was sufficiently ``on-axis" for the point-source analysis of RSP99 to
be applied.  The X-ray source is found to be unresolved and at the radio and optical
position.  MS 0331.3--3629, MS 0607.9+7108, MS 0737.0+7436 and MS 1219.9+7542 also appear
to be unresolved at the radio and optical position. 
MS 1332.6--2935 is too far off-axis to determine its
morphology without doubt, although the HRI observation does confirm the X-ray
source location to be coincident with the optical and radio positions.

\subsection{Broadband Spectral Properties}

As radio and X-ray surveys have been efficient in discovering them, BL Lacs historically have
been divided into the XBL and RBL classes based upon the discovery method.  Recently this
terminology has been superseded by a classification scheme based upon the broadband spectral
energy distributions (SEDs), wherein the luminosities of ``high-energy-peaked" BL Lacs (HBLs)
peak in the UV/X-ray and ``low-energy-peaked" BL Lacs (LBLs) peak in the IR/optical.  
The division between the two classes is based upon the ratio of the X-ray (1 keV) to
radio (5 GHz) flux densities \fxfr\ (e.g., Padovani \& Giommi 1995), with a somewhat arbitrary
dividing line between the HBL and LBL classes at log(\fxfr) $\sim -5.5$ (e.g., Wurtz et al.
1996).  Most XBLs are HBLs and most RBLs are LBLs, although there are exceptions in both
cases.  

In this paper we use the term XBL to describe EMSS BL Lacs because it is an X-ray selected
sample; radio emission is not a priori considered in the selection process.   
We note that a sharp division between HBL and LBL is no longer appropriate, as recent surveys
which use both radio and X-ray detection as selection criteria (e.g., the RGB;
Laurent-Muehleisen et al. 1999 and the DXRBS; Perlman et al. 1998) have found prodigious
numbers of ``intermediate" BL Lacs with log(\fxfr) $\sim -5.5$, indicating that there is a
continuum of spectral energy distributions among BL Lacs and that the historical HBL/LBL
division is primarily due to selection effects.

\section{BL Lac Classification}

Several papers (e.g., March\~a et al. 1996) have raised the question of what are
appropriate criteria for the BL Lac classification.  And several surveys, e.g., the RGB
(Laurent-Muehleisen et al. 1999), the DXRBS (Perlman et al. 1998) and the REX
(Caccianiga et al. 1999) contain many BL Lac-like objects which do not meet
the original BL Lac criteria.  Here we discuss the
classification scheme first proposed by S91 and suggest modifications based upon the EMSS
sample.

\subsection{Spectroscopic Criteria}

The spectroscopic criteria originally proposed by S91 and slightly modified by Stickel
et al. (1991) are: (\#1) a CaII H\&K ``break strength" $B_{4000} \leq 0.25$ and (\#2) a
rest-frame equivalent width of any possible emission lines present of $W_{\lambda} \leq
5$\AA.  RSP99 showed that the strict application of criterion \#1 is only
partially successful in that it would eliminate eight objects 
whose CaII break strengths slightly exceed the limit (see
Table~\ref{tbl-4.3}). Nevertheless the BL Lac classification
seems most appropriate for these objects given their luminous, point-like X-ray
emission and low-power, core-dominated radio emission. Except for MS 2306.1--2236, each of
these sources has been observed with the ROSAT HRI and found to be unresolved, confirming a
generic AGN classification. And MS 2306.1--2236 was found by the PSPC to be variable and to
possess a steep X-ray spectral index,  
making the X-ray emission inconsistent
with being a diffuse cluster source. Since these X-ray sources are much too
luminous to be normal low-power radio galaxies (see e.g., Fabbiano et al. 1987)
and since they lack the strong emission lines seen in other AGN, the BL Lac
classification seems the most appropriate.  Although this collection
of evidence seems conclusive enough, the detection of optical continuum
polarization would be definitive
and observations should be made to confirm these inferences.  Polarimetric observations 
blueward of the H\&K break are ideal because they will suffer the least
amount of starlight contamination from the host galaxy. While one might suspect that we have
misidentified other BL Lacs with even larger
$B_{4000}$ values, ROSAT HRI images of other EMSS clusters containing radio galaxies have
confirmed their extended nature (RSP99).  These new identifications stretch
criterion \#1 towards values found in normal ellipticals and FR--1 radio galaxies
(Owen, Ledlow \& Keel 1996), which also lack strong emission lines. For example, Dressler
\& Shectman (1985) found $< 5$\% of nearby cluster galaxies have $B_{4000} < 0.4$ and only
$\sim 1$\% of the full sample have both low break strengths and emission lines weak enough
to qualify as a BL Lac by criterion \#2. 
The continuum of
$B_{4000}$ values seen in the EMSS survey between BL Lac and radio galaxy (Figure 6 of RSP99)
indicates that the CaII break strength is a poor criterion when considered alone.  However
it is an important method for
discovering or confirming low luminosity BL Lac objects.  For example, in RSP99 we confirmed
the suggestion of Owen et al. (1996) that 3C264 and IC 310 are low luminosity
BL Lac objects, a suggestion based upon the
reduced CaII break strength of these nearby cluster radio galaxies.
A continuum of $B_{4000}$ values has also been seen in other samples of radio-loud AGN 
(e.g., Laurent-Muehleisen et al. 1998 and Caccianiga et al. 1999).

There are no BL Lac objects in Table~\ref{tbl-4.1} whose emission line rest equivalent
widths exceed the 5\AA\ limit of criterion \#2, although we note that a few objects have
poor spectra which do not completely rule out the presence of $\sim$5\AA\ lines (also see
the discussion of MS 1256.3+0151 in \S 5).  All weak-lined objects that we have
classified as QSOs in the EMSS have rest frame $W_{\lambda} \geq 30$\AA\ and several
have noticeably broad emission (see spectra of these weak-lined AGN in Figure
11 of S91). Also, the few weak-line QSOs with $W_{\lambda} < 50$\AA\ in
the full S91 sample are optically unpolarized (Jannuzi et al. 1993; M91) and only
one of these weak-lined AGN is radio-loud (MS 0815.7+5233). Since the EMSS BL Lacs
(and XBLs in general) were found to be only moderately optically variable
(typically $< 1$ mag.; Stocke et al. 1985; Jannuzi et al. 1993; S91), significant
changes in emission line equivalent widths would not be expected for XBLs depending
upon the epoch of observation. However, several radio-selected 1Jy BL Lacs have
emission lines which exceed the proposed equivalent width guideline and blur the
distinction between the BL Lac and ``blazar" categories (e.g., B2 1308+326;
Stickel et al. 1991; see also RS00 for other examples). 
Criterion \#2 therefore remains a sufficient discriminant in that no
EMSS object with rest frame 5\AA\ $\leq W_{\lambda} \leq 50$\AA\ has been excluded from
the BL Lac sample based upon criterion \#2 alone.

 
In summary, based upon current data the BL Lac spectroscopic criteria suggested by
S91 should be modified slightly in order to include all EMSS BL Lacs.
Several of the new EMSS BL Lacs require the CaII break strength criterion
to be relaxed to at least $B_{4000} < 0.33$.  We note that four EMSS XBLs have break strengths
which even exceed this new limit, but in all cases the spectra are too poor in quality to
accurately measure the break.  No modification of the
$W_{\lambda}$ criterion is required.  The less restrictive criteria proposed by March\~a et
al. (1996) includes all of the BL Lacs discovered within the EMSS.  However we note that their
criteria should be used only in conjunction with other observed properties, including, e.g., 
point-like X-ray emission, optical polarization, coincident
radio emission or a unique spectral energy distribution because, as described above,
some weak-lined AGN in the EMSS which meet the March\~a et al. (1996) criteria 
are optically unpolarized and radio quiet, arguing against a BL Lac identification.

\subsection{Photometric Criteria}

Besides optical spectroscopic criteria for the classification of all EMSS X-ray
sources, a secondary set of photometric criteria were also suggested by S91 to
verify the spectroscopic classifications. For the BL Lac class, these included:

(1) Radio Emission.  While originally not included as a BL Lac requirement in order
to leave open the possibility of radio-quiet BL Lac objects, none were found by S91
(see also Stocke et al. 1990).  To our knowledge, no BL Lac
has been found to be radio-quiet (e.g., Bade et al. 1998) and radio
emission has been detected for all EMSS BL Lacs (see Table~\ref{tbl-4.1}).
Fan et al. (1999) report the discovery of a ``radio quiet" quasar at $z=4.62$ 
which lacks emission lines, suggesting the first example of a radio-quiet BL Lac;
although the 6cm flux limit does not rule out the possibility that it is simply
a relatively radio-weak BL Lac.  

(2) Overall Energy Distribution. Given the radio detection for all EMSS XBLs,
two-point optical-to-X-ray ($\alpha_{ox}$) and radio-to-optical ($\alpha_{ro}$)
spectral indices, can be defined in the usual way (S91) using 2 keV, 2500\AA\ and 5
GHz rest-frame flux densities (see Table~\ref{tbl-4.1}). The resulting plot of these
two-point spectral indices (Figure~\ref{fig-4.2}) can be used to characterize the
overall energy distribution of BL Lacs and other AGN classes. As noted by several
authors previously (Stocke et al. 1985; Ledden \& O'Dell 1985; S91) the XBLs
congregate in a specific region of this diagram, a region largely distinct from the
other AGN classes (see S91), allowing an additional XBL discriminant. Such a
discriminant was tested using the X-ray sources in the {\it Einstein} ``Slew
Survey" (Elvis et al. 1992) and found to be $> 90$\% efficient at identifying 
BL Lac objects (Perlman et al. 1996b) using the region of Figure~\ref{fig-4.2} 
defined by the EMSS sample.    

The full EMSS sample is now plotted in Figure~\ref{fig-4.2}.  Unlike
previous plots of this type, we have corrected the
optical flux values by removing the contribution from the host galaxy's starlight.
This correction uses the optical slit spectroscopy in hand to determine the
fraction of V-band light due to the non-thermal source by assuming that the host
galaxy has an intrinsic $B_{4000} = 0.50$ and that the CaII break is depressed
by the nonthermal continuum from the AGN.  This fraction was applied
to the observed V-band magnitude of the object, correcting for the amount of
light that did not fall into the slit.  
We extrapolate from V-band to 2500\AA\ by assuming $\alpha_o = -1$ (S91).  We expect that
these optical flux measurements are accurate to better than the level of optical variability
seen in these sources ($\leq 1$ mag; S91).  The diagonal line in Figure~\ref{fig-4.2}
shows the direction and amount that a point would move given one magnitude of optical
variability.  
Figure~\ref{fig-4.2} illustrates that the EMSS BL Lacs are remarkably
homogeneous in their overall spectral energy distributions.

\section{Notes on Individual Sources}

MS 0205.7+3509: The redshift of this unusual XBL remains uncertain although we have
obtained high SNR spectra of it on two occasions in addition to the spectrum
presented in M91. In M91 a redshift of $z=0.318$ was proposed based upon the weak
detection of CaII H\&K and G-band. These CaII lines are close to the spectrograph
join so that this redshift was uncertain and was not confirmed by
cross-correlation methods. However, the MMTO spectrum appears to show CaII H\&K at
the same location and so we record $z=0.318$ as a tentative redshift in
Table~\ref{tbl-4.1}. Stocke, Wurtz \& Perlman (1995) suggested that this BL Lac might
be a gravitationally-lensed object based upon the presence of excess soft X-ray
absorption in its PSPC spectrum (P96) and the presence of an offset optical galaxy
(Wurtz, Stocke \& Yee 1996). 
Falomo et al. (1997) alternatively suggested that the offset between the BL Lac and
its surrounding  nebulosity was due to the presence of a companion galaxy; Falomo
et al. also claimed a detection of the BL Lac host galaxy. Presently it is not
known whether the tentative redshift is due to the host galaxy for this BL Lac or
the offset nebulosity. 

\begin{figure}
\plottwo{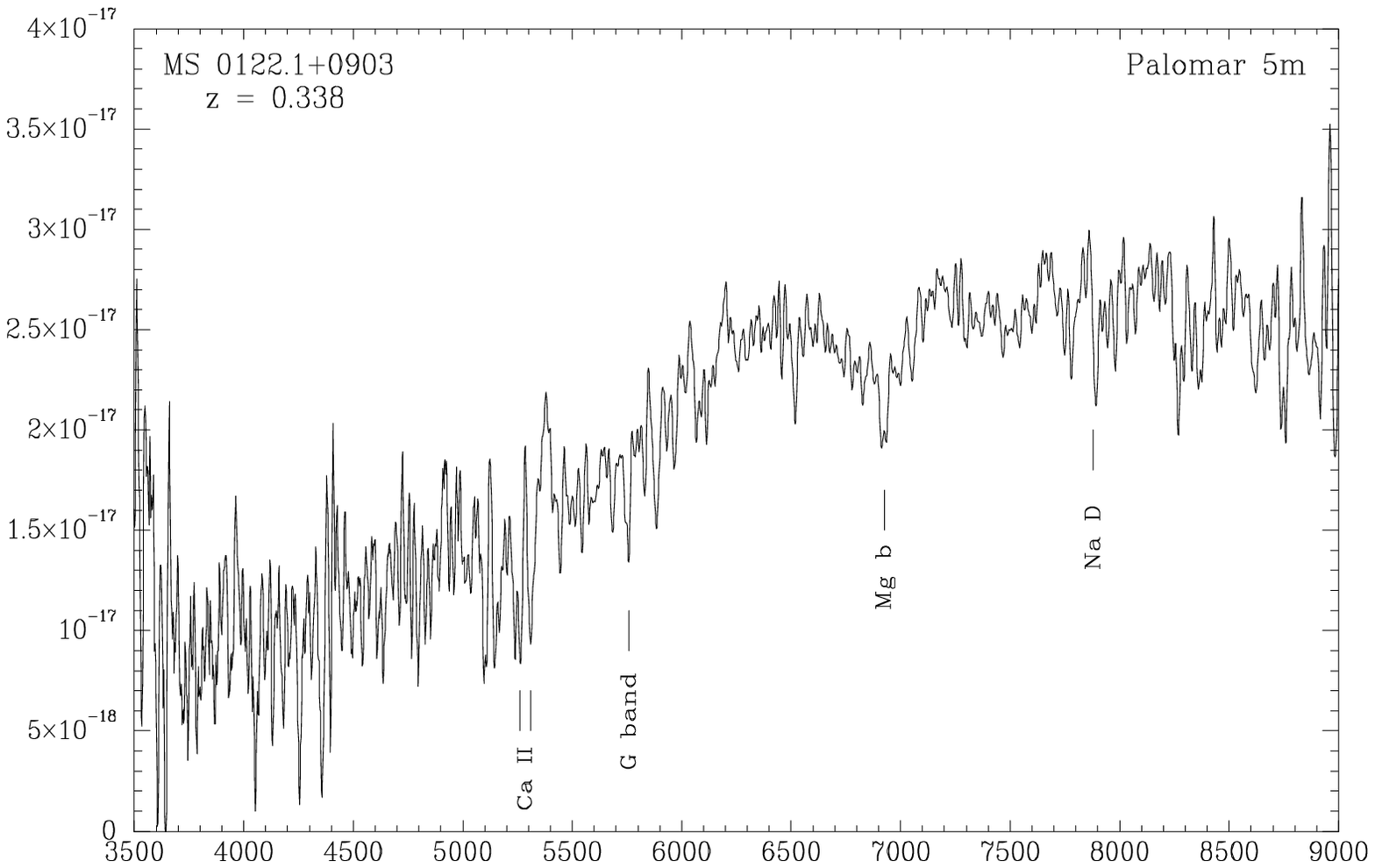}{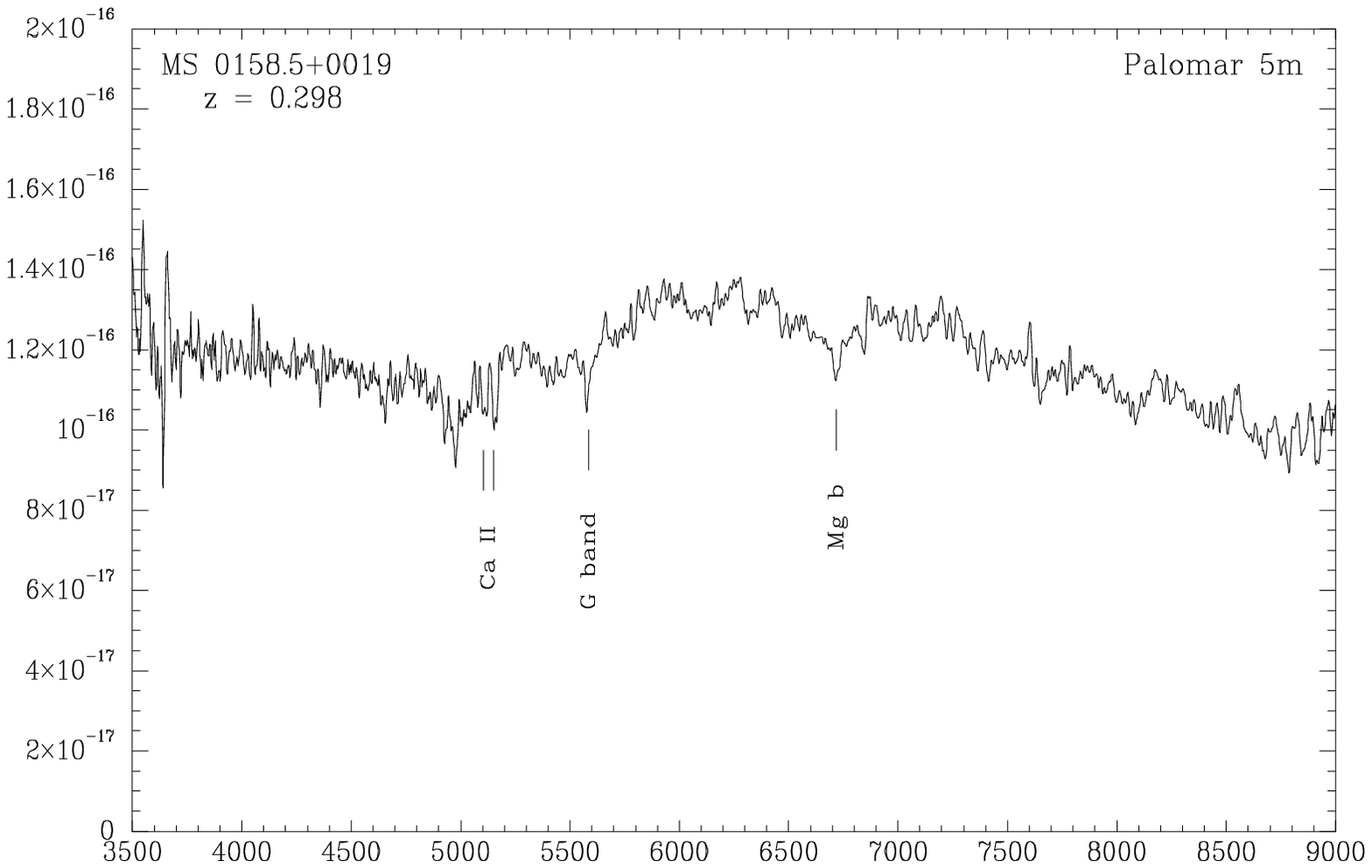}
\caption{Optical spectra of EMSS BL Lacs.  Spectra for all EMSS BL Lacs  are presented
here, except for MS 0622.5--5256 and MS 1219.9+7542, which were described in Stocke et al.
(1985) but are not available, and MS 2316.3--4222, which is presented in Crawford \&
Fabian (1994).  The flux scale is usually $F_{\lambda}$ in ergs s$^{-1}$ cm$^{-2}$ but
calibration was not available for several spectra (MS 0317.0+1834, MS
1229.2+6430, MS 1235.4+6315 \& MS 1552.1+2020, whose vertical scales
are in relative $F_{\lambda}$ and MS 1333.3+1725 \& MS 2347.4+1924, whose
vertical scales are in detector counts); the x-axis is wavelength
in \AA.  Possible features are labeled with a question mark. The symbol $\earth$
identifies features due to imperfect removal of telluric lines.  Fluxes, where shown, 
are only approximate due to the small slit width ($2\arcsec$). 
\label{fig-4.1}} 
\end{figure}

MS 0331.3--3629: The low-SNR spectrum of this BL Lac in Figure~\ref{fig-4.1} is from
the original EMSS identification program data. Despite the modest spectral
quality, a blue non-thermal component is clearly present as is a CaII break which
slightly exceeds the original limit of S91 for BL Lac objects. However, the
classification of this source as a BL Lac seems clear in that ROSAT HRI
observations find a point-source at the radio galaxy location and several PSPC
observations show that the source has a count rate which varies by a factor of two
and a steep X-ray spectral index (see Table~\ref{tbl-4.1}).

MS 0350.0--3712: The low-SNR spectrum in Figure~\ref{fig-4.1} shows a substantial blue
non-thermal component and a small CaII break in accordance with the BL Lac
definitions. Serendipitous PSPC and HRI observations verify that this source is
pointlike at the radio and optical positions, and that the X-ray source has a steep,
possibly variable spectral index and possibly variable count rate.

MS 0419.3+1943: Due to the faintness of this source, the blue side of the M91
spectrum is quite noisy so that the redshift was verified using only the red
spectrum in the cross-correlation procedure.

MS 0622.5--5256: Despite having only a very modest SNR spectrum of this object
in-hand (Stocke et al. 1985), the BL Lac nature of this source is verified by a
serendipitous PSPC observation which confirms that the X-ray, radio and optical
sources are coincident and provides a one-time measurement of a steep X-ray
spectral index.

MS 0922.9+7459: Due to the faintness of this source, the blue side of the M91
spectrum is quite noisy such that the redshift was verified using only the red
spectrum in the cross-correlation procedure.

MS 0950.9+4929: This BL Lac has been observed by us spectroscopically on three
separate occasions without detection of any emission or absorption features.
Further, the optical (Wurtz, Stocke \& Yee 1996) and radio (PS93)
images of this source are unresolved, suggesting a high redshift ($z > 0.6$). But
otherwise this object remains one of only a few XBLs whose optical spectrum is
entirely featureless.

MS 1133.7+1618: While the cross-correlation procedure resulted in a
firm redshift values for this source of $z=0.574$, this value places CaII H\&K quite close to
night sky 6300\AA\ and G-band near the atmospheric B-band. Therefore, this redshift must be
viewed as tentative. An unpublished radio
map obtained for this source using a snapshot observation in A-
configuration at 6 cm reveals a small, slightly bent triple source,
typical of a wide-angle-tail at $z \sim 0.5$.

MS 1205.7-2921:  Our spectrum of this object in Figure~\ref{fig-4.1}
shows a rather large H\&K break, however the spectrum is poor
and $B_{4000}$ is poorly constrained.

MS 1207.9+3945: A VLA snapshot of this source is presented in Figure~\ref{fig-4.3}.
Its extended radio power is high (log $P_{ext}$ = 25.53 W Hz$^{-1}$), but not 
inconsistent with an FR--1.

MS 1219.9+7542: This source has been identified as a cluster in the Bright SHARC
survey (Romer et al. 1999); however our analysis of archival ROSAT PSPC and HRI data
indicate that the X-ray source is unresolved at the location of the optical and
radio counterpart.  Further, multiple HRI observations show that the source is variable
by a factor of 2.  A cluster may be present but the unresolved X-ray source can account for
the entire IPC discovery flux. 

MS 1221.8+2452: The M91 spectrum of this BL Lac showed H\&K quite close to the spectrograph
join, so that a new spectrum was obtained at the MMTO, which confirms the redshift obtained
by M91 (see Figure~\ref{fig-4.1}).

MS 1229.2+6430: While only a modest SNR spectrum is available for this source,
H\&K, G-band and Mg ``b" are all clearly present so that the redshift is firm. A
higher resolution spectrum obtained at the MMTO for this object (not
shown) confirms H\&K at the same redshift.

MS 1256.3+0151: While our modest SNR spectrum of this source obtained at the MMTO
confirms its BL Lac nature, no redshift is obvious, nor was one obtained using
the cross-correlation procedure with an absorption-line template. However, weak
emission lines are potentially present at 4587\AA\ and 6735\AA, which, if real,
yield a redshift match for C III] 1909$\lambda$ and MgII 2798$\lambda$ at
$z=1.40$. 
The rest-frame equivalent widths of these low contrast features
are 4.2\AA\ and 5.8\AA\ respectively.  Better spectroscopy is required to be
certain of these identifications; however if these lines are real, MS 1256.3+0151
would be the only object in the EMSS XBL sample to show broad, luminous (log
$L_{line} \approx$ 42--43 erg s$^{-1}$) emission lines typical of FR--2s, quasars
and some RBLs (RS00).  No deep, 20cm VLA observation is available for this
source.

MS 1258.4+6401: The low-SNR MMTO spectrum confirms the BL Lac nature of this faint
source but yields no redshift information.

MS 1312.1--4221: A low resolution spectrum of this source was obtained at the LCO
100" telescope during the initial source identification process. We show here both
this modest SNR spectrum and a later, red spectrum also obtained at the LCO 100",
from which the firm redshift is derived.

MS 1332.6--2935: We show our new MMTO spectrum of this source in Figure~\ref{fig-4.1}. The
cross-correlation of this spectrum yields a firm redshift of $z=0.513$, which
corrects the previously published value of $z=0.256$, which was based on a low-SNR
spectrum obtained at the LCO 100".

MS 1333.3+1725: This new BL Lac Object was identified using the new ROSAT HRI
position from Hattori \& Morikawa (private communication) of 
$\alpha = 13^h35^m46.9^s$, $\delta = +17\arcdeg09\arcmin41.5\arcsec$ (J2000). 
Analysis of their HRI image in the manner of RSP99 confirms the location of an
X-ray point source at the location of a
20th mag. galaxy with $z=0.465$, the object at (0,0) in the image of Gioia \&
Luppino (1994). Our spectrum of this object in Figure~\ref{fig-4.1}
shows a rather large H\&K break, however the spectrum is poor
and $B_{4000}$ is poorly constrained. A weak radio
source ($0.9$ mJy)  is detected at the location of the optical counterpart. The image of
this field  shown in Gioia \& Luppino (1994) reveals no obvious cluster of galaxies and so
despite  the somewhat large H\&K break, this source is 
classified as a BL Lac.

MS 1402.3+0416: In M91 we reported no redshift for this object. Despite a
substantial red component seen in the Palomar spectrum, no absorption features were
seen. A newer spectrum of this source was obtained at the MMTO when the source was
brighter and the red component not so obvious. A cross-correlation analysis of this
spectrum yields a good cross-correlation peak at $z=0.343$. However, the line
identification at this redshift is not obvious in Figure~\ref{fig-4.1}. We,
therefore, list this redshift as tentative. This modest redshift is consistent with
the well-resolved FR--1 type radio structure seen by PS93.

MS 1458.8+2249: The M91 spectrum of this BL Lac showed H\&K quite close to the
spectrograph join, so that a new spectrum was obtained at the MMTO, which confirms
the redshift obtained by M91 (see Figure~\ref{fig-4.1}).  A possible line is seen at
4230\AA\ in the MMTO spectrum, which matches the position of [Ne V] $\lambda$3426 at
the the redshift of this object.  However, this line is not seen in the Palomar
spectrum, suggesting that either the line and/or continuum flux varied or that the line is
not real.

MS 1552.1+2020: The original EMSS survey identification spectrum is all that is
available for this BL Lac (Figure~\ref{fig-4.1}) but it clearly shows CaII H\&K, so the
redshift is firm.  A previous redshift of $z=0.222$ was erroneously reported for
this object in M91.

MS 2306.1--2236: This source is identified with a well-resolved galaxy at low-$z$.
Despite the substantial H\&K break, we identify it as a BL Lac given its steep
X-ray spectrum, variable X-ray flux and absence of any obvious cluster on the POSS
(see finding chart in Maccacaro et al. 1994).

MS 2316.3--4222: This source is identified as a BL Lac based upon the work presented
in Crawford \& Fabian (1994).

MS 2336.5+0517: Despite a modest SNR spectrum of this object obtained at the MMTO,
we find no firm redshift. A very tentative $z=0.74$ is suggested based upon a
slight increase in continuum level redward of the atmospheric B-band, which could
be the H\&K break.  This redshift is consistent with the weak extended radio
structure seen in Figure~\ref{fig-4.3}, but given the significant flux
and extent to the 20cm emission, this redshift is probably best treated
as an upper limit for the distance to this object.

MS 2342.7--1531: A modest SNR MMTO spectrum is shown in Figure~\ref{fig-4.1}, which 
confirms the
BL Lac classification of this source but offers no redshift information. 

MS 2347.4+1924: The CFHT spectrum of this faint source shown in Figure~\ref{fig-4.1} yields a
firm detection of the H\&K break at 6010\AA. A deep image of this source obtained
by us shows a resolved elliptical galaxy, consistent with $z=0.515$.  The slightly
resolved radio emission (see Figure~\ref{fig-4.2}) is also consistent with this redshift.

\section{The Completeness of the EMSS BL Lac sample}

Due to the manner in which the radio observations were made and in which the
optical identification program was carried out, it is appropriate to address the
completeness for the M91 and D40 sub-samples as well as for the
entire EMSS BL Lac sample.


The M91 sample has been scrutinized most carefully over the years due to its
importance in determining the luminosity function, cosmological evolution and
\vvmax\ value for XBLs. 
All EMSS sources bounded by $\delta \geq -20$\arcdeg\ and $f_x \geq 5$ x $10^{-13}$
ergs cm$^{-2}$ s$^{-1}$ have been identified so that the only additions to the M91
sample would come either from the correction of a previously incorrect ID or
from a revised X-ray spectral index or Galactic $N_H$ value, which raises the flux of the
source to greater than the flux limit. This can occur because the conversion
from IPC counts s$^{-1}$ to flux units is dependent upon the assumed spectral
index and a large Galactic absorption value can diminish the observed flux
substantially. Of the known EMSS BL Lacs, two have been added to the M91 sample
by this effect (see Section 2.1), while one other (MS 1704.9+6046) remains just
under the flux limit using a ROSAT PSPC spectral index. There are no unidentified
sources (or suspected misidentifications) just below the M91 flux limit which could
become sample members due to this effect (see below).

While it is somewhat less scrutinized and not completely identified, the D40
sample is virtually complete.  All extragalactic and unidentified sources north
of $\delta  = -40$\arcdeg\ have been observed with the VLA to a $5\sigma$ limit of
$\sim 1$ mJy.  Assuming $\alpha_{ro} \geq 0.25$, the observed lower 
limit within the EMSS, and $V \leq 20$, this should be sufficient to detect any
BL Lac which meets the D40 sample X-ray flux criteria ($f_x \geq 2$ x
$10^{-13}$ ergs cm$^{-2}$ s$^{-1}$).
Within this sky area only four X-ray sources which meets the D40 sample X-ray flux criteria
remain unidentified or are poorly identified.  And only two of these four X-ray sources are
detected radio sources.  These four sources are listed in Table~\ref{tbl-4.4}. Also listed
in Table~\ref{tbl-4.4} are four other unidentified sources which are possible BL Lacs. 
Those included are: (1) sources which have X-ray flux densities close enough to the D40 
flux limit such that a steeper spectral index than the assumed value of $\alpha_x
= -1$ could bring them into the D40 sample; and (2) fainter X-ray sources which have
detected radio sources in their error circles.

Of the currently unidentified sources in Table~\ref{tbl-4.4}, the only sources which
are plausibly BL Lac objects are: (1) MS 0134.4+2043, wherein a 22nd mag. red object
at the radio position could be the X-ray emitter; (2) MS 1317.0--2111, wherein a
non-detection by the ROSAT HRI leaves the identification unknown (see RSP99); and (3) MS
2225.7--2100 where only a poor optical spectrum exists for the X-ray/radio source.  In the
other case with a radio detection (MS 0501.0--2237), the radio source is too far away from
both the IPC and PSPC positions to be considered a plausible ID. However, we caution that our
VLA snapshot survey may not have been quite deep enough  
to detect all the EMSS BL Lac objects in the sample down to the faintest
X-ray flux densities.

South of $\delta  = -40$\arcdeg\ not all of the EMSS sources have been observed
at radio frequencies and so we claim no completeness level for the EMSS sample of
BL Lac objects south of $\delta  = -40$\arcdeg.

Therefore, the current status of the EMSS identification process yields two
complete samples to slightly different degrees of confidence: the M91 defined sample of
26 objects, which has been completely identified and thoroughly scrutinized for
misidentifications using new ROSAT HRI observations (RSP99); and the larger
D40 sample of 41 BL Lacs, which includes the M91 sample as a subset, bounded by
the VLA observing area of $\delta  = -40$\arcdeg\ and by a flux above which almost all
sources are plausibly identified ($f_x \geq 2$ x $10^{-13}$ ergs cm$^{-2}$ s$^{-1}$;
see above). The latter sample has not been scrutinized in detail for misidentified
sources. Given the analysis of March\~a \& Browne (1995) and the results of RSP99,
which found two additional sources in the M91 area, we expect that
there may be a few
misidentified sources absent from the D40 sample. Since we have already reobserved with
the ROSAT HRI several of the most obvious misidentifications in the D40
sample,
we expect that the remaining numbers of misidentifications are small and that the
D40 sample is largely complete.




\section{The X-Ray Luminosity Function and Cosmological Evolution of XBLs}

Here we present the updated XLF and \vvmax\ for the M91 sample, which was first
presented in M91, and for the new D40 sample.  In addition to the newly added
objects we make the following refinement.  Whereas M91 assumed an X-ray spectral
index $\alpha_x = -1$ for all EMSS BL Lacs, here we use the $\alpha_x$ as
determined for each source via ROSAT PSPC observations (see \S 2) or by the
$\alpha_x - \alpha_{ox}$ correlation in Padovani \& Giommi (1996). The
tentative redshifts listed in Table~\ref{tbl-4.1} are assumed to be correct for the
results presented in this Section.  For objects
without redshifts we assume $z=0.3$, the median redshift for both the M91 and D40
samples.  This assumption is reasonable as the distribution of redshifts for the
EMSS sample is narrow ($0.1 < z < 0.7$), with a sharp peak at $z \approx 0.3$
(Figure~\ref{fig-4.7}).  For MS 0205.7+3509, the only EMSS XBL for which the quoted
redshift is plausibly a lower limit from an intervening galaxy (Stocke et al. 1995),
we assume that the quoted redshift is for the BL Lac, regardless of this
uncertainty. As 
discussed in M91, uncertain redshift information for some of the EMSS XBLs does not
significantly affect the \vvmax\ test, as it is almost independent of redshift for
the low redshift range of the EMSS BL Lac sample ($z < 0.7$ for all but one 
XBL).  The area surveyed by the EMSS is a function of the flux limit; hence the
volume function for the EMSS is complex.  For this reason we use the \veva\ test of
Avni \& Bahcall (1980).  M91 details the application of the
\veva\ test to the EMSS and the same procedure is followed herein.  

Table~\ref{tbl-4.6} presents the results of the \veva\ test as it is applied to
several X-ray flux-limited subsamples of EMSS BL Lacs.  By column: (1) is the sample; (2) is
the number of objects in the sample; (3) the \veva\ and its error; the error is
$\sqrt{1/12N}$; (4) is the probability that BL Lac subsample is consistent with
no evolution, as determined by a K-S test.  M91 first reported \vvmax\ $= 0.329\pm0.062$
for the M91 sample.  With the updated redshifts and X-ray flux densities presented in
Table~\ref{tbl-4.1} and the addition of four new BL Lacs, the
\vvmax\ of the complete M91 sample is raised to \vvmax\ $= 0.399\pm0.057$; and
\vvmax\ $= 0.427\pm0.045$ for the new D40 sample.  In addition to the M91 and D40 
samples we determined the \veva\ value for two other high-flux EMSS subsamples.  

It is not clear what, if any, trends exist in Table~\ref{tbl-4.6}.  The \vvmax\ for the
smaller samples are consistent with no evolution, but as the sample size increases this is
less so.  
Della Ceca (1993) first noted this result, suggesting that the EMSS sample
was  incomplete at lower flux levels.  However in RSP99 and this paper we have shown that the 
EMSS samples, and in particular the M91 sample, are now virtually complete.
Further, the negative evolution result persists in the larger D40 sample.
Even though the D40 sample has a larger \vvmax, it is 
less consistent (only a 4\% probability) with a no evolution  result than the M91 sample. 
As all the EMSS sources bounded by M91 sample X-ray flux and declination criteria have been
identified, the only possible additions to the M91 sample would come either from the
correction of a previously incorrect identification or from a revised X-ray spectral index or
Galactic $N_H$ value.  As discussed in \S 4, we believe no BL Lac objects are excluded due to
the modified EMSS BL Lac selection criteria; and only one or two objects could possibly be
moved into the M91 sample for the latter two reasons.
While the \vvmax\  values for the M91 and D40 samples are indicative of negative
evolution, both results are consistent with a no evolution scenario for
XBLs at the $2\sigma$ level.  

Table~\ref{tbl-4.7} presents the results of the \veva\ test as it is applied to subsamples of
the D40 divided by their \fxfr\ ratios.  The column identifications are the same as
Table~\ref{tbl-4.6}.  We find that only the more extreme HBLs (log(\fxfr) $\geq -4.5$) clearly
show negative evolution, whereas the less extreme HBLs and the intermediate BL Lacs are
compatible with a no evolution result.  BL Lac samples drawn from the ROSAT All-Sky Survey
also show this characteristic (Bade et al. 1998; Giommi, Menna \& Padovani 1999). 

The negative evolution of EMSS XBLs is modeled as both density evolution of the form
$\rho(z) = \rho(0)(1+z)^{\beta}$ and luminosity evolution of the form $L_x(z) =
L_x(0)(1+z)^{\gamma}$.  Figure~\ref{fig-4.4} shows the variation of \veva\ with the
evolution parameters $\beta$ and $\gamma$. Three cases are shown in this
Figure: (1) open circles: original M91 result; (2) closed circles:
revised M91 sample using revised X-ray spectral index data from ROSAT;
(3) closed triangles: D40 sample with ROSAT spectral indices. The new \veva\ best-fit
values for the M91 sample (i.e., the values for which the evolved \veva\ = 0.5) are $\beta
= -3.5$ with $\pm2\sigma$ limits of -7.0 to +0.5 and $\gamma = -4.0$ with $\pm2\sigma$
limits of -12.0 to -0.25.  Because of the slight increase in the unevolved \veva\ the
measured  evolution is not as strongly negative as first reported in M91; $\beta$ has
increased from -5.5 to -3.5 and $\gamma$ has increased from -7.0 to -4.0.  These values are
slightly higher for the D40 sample, $\beta = -2.8$ and $\gamma = -3.0$. 
Figure~\ref{fig-4.5} shows the integral X-ray luminosity functions evolved to $z=0$ for the same three cases as in Figure~\ref{fig-4.4}. 
Both the derived, best-fit luminosity- and density-evolved XLFs are shown for the
original M91 sample (open circles), the updated M91 sample (closed
circles) and the new D40 sample (closed triangles). 
Figure~\ref{fig-4.6} shows the differential form of the XLF; each bin is $10^{44}$
erg s$^{-1}$ in size.  As can be seen in both Figures~\ref{fig-4.5} and \ref{fig-4.6}
the XLF is not significantly changed from our first determinations using
the M91 sample.  The new objects in the M91 and D40 samples
extend the XLF down to somewhat lower luminosities (log $L_x \sim 43.5$ erg s$^{-1}$);
and, as was first reported in RSP99, we see no evidence of flattening of the XLF at
log $L_x < 45.0$ erg s$^{-1}$.

\section{Impact on Unified Schemes}

Here we discuss the relationship between EMSS XBLs and FR--1 radio galaxies.  The relationship
between XBLs in the EMSS sample to RBLs in the 1Jy survey will be discussed in RS00.

\subsection{Extended Radio Luminosities \& Core Dominance Values}

If low-luminosity, FR--1 radio galaxies are the parent population of BL Lacs, the
distribution of their emission line luminosities and extended radio power levels 
should be
comparable.   In addition to those presented in PS93, the new VLA maps presented here
complete deep 20cm mapping of all the BL Lacs in the M91 sample.  
All of the EMSS XBLs have measured or upper limits to their extended radio powers which
are consistent with FR--1s, as defined by Owen \& Laing (1989).
Three objects have extended radio powers which are somewhat high for 
FR--1s (log $L_{ext} > 25$ W Hz$^{-1}$), one of which is based upon a very tentative 
redshift ($z=0.74$; MS 2336.5+0517).  

Due to the number of unresolved objects the distribution of
extended radio luminosities for XBLs is not well defined, making a detailed
comparison between FR--1s and XBLs difficult; e.g., the extended luminosity limit
for MS 1019.0+5139, while consistent with being an FR--1, is nonetheless very low. 
The extended flux limits placed upon MS 1019.0+5139 and MS 1235.4+6315 by our VLA
observations are considerably lower than other EMSS BL Lacs observed (PS93 
and here) because they were observed with the B-array, which has a larger beamsize
that is better suited for detecting faint, extended radio flux at the redshifts of
most EMSS BL Lacs ($z < 0.5$).  Thus, deeper 20cm B- and C-array observations should
be obtained of these sources to determine whether or not the distribution of
extended radio flux densities for XBLs matches FR--1s precisely or not.

In the orientation hypothesis, XBLs are believed to be FR--1s viewed at intermediate
angles to the line of sight 
(Urry \& Padovani 1995).  
If so, the core-to-extended flux ratios ($f$ values) should indicate
the degree of Doppler boosting of the core and hence the range of orientation angles for 
XBLs.  To test this hypothesis, the $f$ values for XBLs in the M91 sample
are compared to FR--1s from the B2 sample (Ulrich 1989; as presented in PS93) and RBLs
from the 1Jy sample (Stickel et al. 1991; RS00) in
Figure~\ref{fig-4.9}.  Padovani \& Giommi (1995) argued against the unified model on
the basis that the radio structures of XBLs shown in PS93 are more core-dominated
than expected.  Based upon derived beaming parameters for BL Lacs (e.g., Urry et al. 1991,
Celotti et al. 1993), they argue that most XBLs should be lobe-dominated radio
sources ($f < 1$).  This situation is analogous to steep-spectrum radio quasars,
lobe-dominated sources which are believed to be seen at angles
intermediate to flat-spectrum radio quasars and FR--2s.  However,
the Padovani \& Giommi (1995) conclusion
is based upon the brighter radio EMSS XBLs observed by PS93,  
possibly introducing a bias towards more core-dominated sources via Doppler
boosting.  So, it was necessary to complete the radio mapping of the M91
sample in order to remove any possible bias that the PS93 radio flux-limited 
sample may have created.  
However, as shown in Section 3.2, the
$f$ values for the newly mapped sources are quite high as well; and most of these
$f$ values are lower limits because all but three of the newly mapped XBLs were
unresolved by these new observations.  With these new observations the \avgf\ for EMSS XBLs has actually
increased such that \avgf\ $\geq 4.2$ for the M91 sample.  No XBLs with $f \approx
0.02$ are seen, which are predicted by the unified model.  Thus these observations 
do not ameliorate the excessive core-dominance values of XBLs. 

However, an orientation scheme is still possible if XBL jet Lorentz factors are
moderately higher ($\gamma > 5$) and the jet is confined laterally.  In the Padovani \&
Giommi (1995) model the jet outflow is assumed to be well collimated such that the observed
Doppler boosting is dependent only upon the beam pattern, which, for low $\gamma$ is quite
broad (i.e., the Doppler factor $\delta > 1$ for $\theta < $ cos$^{-1}
\sqrt{(\gamma-1)/(\gamma+1)}$; which, for $\gamma = 5$, gives $\delta > 1$ for $\theta <
35$\arcdeg).  If $\gamma > 5$ and the X-ray and radio-emitting jet is laterally confined, the
observed beam pattern would primarily depend on the physical opening angle of the jet,
$\theta_{jet}$, with an abrupt ``edge."  In this case, because the beam pattern of the jet
falls off dramatically for viewing angles $\theta > \theta_{jet}$, XBLs could still have
large radio core-dominance values and there would be an absence of $f$ values intermediate
between that of XBLs and FR--1s (i.e., a lack of objects with $f \approx 0.02$). This model
can also explain the observed absence of a large population of low-luminosity BL Lacs (RSP99;
Owen, Ledlow \& Keel 1996).  

We note that the problem of unification via the orientation hypothesis is exacerbated
if X-ray emitting regions have greater Lorentz factors than the radio emitting regions
($\gamma_x > \gamma_r$; e.g., Georganopoulos \& Marscher 1998).  
If we assume EMSS XBLs are seen at orientation angles $\theta \leq 1/\gamma_x$, the
radio emission will be less Doppler boosted than if $\gamma_x = \gamma_r$, resulting in the
expected distribution of core dominance values to be pushed towards even more lobe-dominated
sources.

\subsection{Other Impacts on Unified Schemes}

\subsubsection{Emission Line Properties}

In the orientation hypothesis, BL Lacs should have emission line luminosities similar to
their parent population.  Obscuration is not believed to be important as BL Lac sightlines
are believed to be at small angles to the jet axis (e.g., Chiaberge et al. 1999; Fabbiano et
al. 1984; Worrall \& Birkinshaw 1994).  Only a few EMSS XBLs have any observed emission
lines,  all of which are quite weak (log $L_{em} \leq 40$ erg s$^{-1}$).  Despite the moderate
to poor quality of some spectra, upper limits of log $L_{em}  =$ 40-41 erg s$^{-1}$ are set
for most emission lines.  The lack of observed emission lines in XBLs is completely
consistent with FR--1s being the  parent population for XBLs; none conclusively show the
luminous emission lines typically seen in FR--2s (e.g., Zirbel \& Baum 1995).

\subsubsection{Absorption Line Properties}

Stocke \& Rector (1997) discovered an excess of MgII absorption systems along the sightlines
of 1Jy BL Lacs, suggesting that some BL Lacs may be gravitationally lensed.  The optical
spectra of many XBLs show the presence of weak stellar absorption lines and a depressed
CaII break, consistent with nonthermal emission from an AGN diluted by a luminous host
elliptical galaxy.  Since most EMSS XBLs lack emission-line redshifts it is possible that
some of the derived redshifts are for an intervening galaxy rather than for the BL Lac
itself.  However, MS 0205.7+3509 is the only EMSS XBL that shows evidence that the observed
galaxy may be intervening rather than the host (Stocke, Wurtz \& Perlman 1995).  The number
of MgII absorption systems along XBL sightlines is not known due to the low source redshift
of XBLs, which places any possible MgII absorption lines in the near-UV.

\subsubsection{Spectral Energy Distributions}

Figure~\ref{fig-4.11} shows the distribution of \fxfr\ values for the M91 and D40 samples. 
The flux densities are $K$-corrected and the X-ray flux values are converted to 1 keV,
assuming the X-ray spectral index determined for each object in \S 3.3.1.  Not surprisingly,
nearly all the EMSS XBLs are HBLs; only a few have log(\fxfr) $< -5.5$.

Fossati et al. (1998) find a trend among blazars wherein the peak frequency \nupeak\ of their
broadband SEDs occurs at lower frequencies for the most luminous sources.  To explain this
result, Ghisellini et al. (1998) have presented a unification model in which the energy
density of the radiating electrons within the jet and their Lorentz factors are correlated,
such that HBLs should have intrinsically lower powers than LBLs.  Thus, their model predicts
a correlation between \nupeak\ and bolometric luminosity \Lbol.

A K--S test indicates a 95\% probability that the \fxfr\ distributions of the M91 and D40
samples are drawn from the same parent population, even though the distribution of X-ray and
radio luminosities are significantly lower in the D40 sample.  If \nupeak\ scales with \fxfr\
and the Ghisellini et al. model is correct, we would expect the \fxfr\ distribution for the
D40 sample to contain a higher fraction of extreme HBLs than the M91 sample because the D40
sample contains XBLs of lower luminosity.
Giommi et al. (1999) also note an apparent lack of correlation between luminosity and
SEDs in a large sample of HBLs.
However, Perlman et al. (1999) find that the correlation between \nupeak\ and \Lbol\ is much
wider than determined by Fossati et al. (1998); and, given that \nupeak\ is poorly
determined by \fxfr\ and that the X-ray flux limit of the D40 sample is only
2.5 times fainter than the M91 sample, we cannot conclude that such a correlation does not
exist among EMSS XBLs.

\subsubsection{Cosmological Evolution}

As is discussed in \S 7, the new D40 and updated M91 samples have significantly larger \vvmax\
values than previously reported for EMSS XBL samples.  While still only
$2\sigma$ below a no-evolution result, negative evolution remains the best fit, especially
for the more extreme HBLs.  The measured \vvmax\ values for the EMSS are
consistent with studies of other complete XBL and HBL samples. It is also consistent with the
observed \vvmax\ for FR--1s from the 2Jy sample (\vvmax\ $=0.42\pm0.05$; Urry \& Padovani
1995),  supporting the notion that XBLs and FR--1s share the same parent population.



\section{Summary \& Conclusions}

We have presented updated and complete radio, optical and X-ray data for BL Lacs in
the EMSS.  Optical spectroscopy is now complete for all known cases of BL Lacs in
the EMSS; and modest dynamic range, arcsecond-resolution VLA maps are now available
for the complete M91 sample. 
We
have also defined a new, virtually complete sample of 41 XBLs in the EMSS with $f_x
\geq 2$ x $10^{-13}$ ergs cm$^{-2}$ s$^{-1}$ and $\delta \geq -40$\arcdeg. 
Redshift information is now available for all but four of the 44 BL Lacs in the
sample; five other objects have tentative redshifts, two of which are highly uncertain. 

Updated redshifts and X-ray flux densities, as well as the addition of four objects, has
increased the \vvmax\ value for the M91 sample to $0.399\pm0.057$; and \vvmax\ $=
0.427\pm0.045$ for the newly defined D40 sample. Thus, all current data continue to
support negative evolution for XBLs, especially the more extreme HBLs.  XBL evolution is
comparable to FR--1s (\vvmax $= 0.42\pm0.05$ for FR--1s in the 2Jy sample; Urry \& Padovani
1995).   The observed spectral and radio properties of XBLs are also completely consistent 
with an FR--1 parent population.  But the radio core-dominance values 
for the M91 sample are likely too high for XBLs to be seen
at intermediate angles as suggested by the simplest unified model, which
needs to be modified significantly to still be viable.
Our results may also contradict the unification model of Ghisellini et al. (1998), which
predict an anticorrelation between peak frequency and bolometric luminosity.


Deep 20cm VLA observations of the D40 sample should be obtained to complete the
observations for this new, nearly complete XBL sample.  Based upon the number of
unresolved sources seen in the M91 sample, B- and C-array observations are more
likely to detect extended radio flux.  Similar observations should be
obtained for all M91 XBLs which were not resolved by the A- and B-arrays.  It is
important to detect extended radio flux for these sources so that the
distribution of extended radio luminosities for XBLs can be accurately compared to
FR--1s.

\acknowledgments

Research on BL Lac objects at the University of Colorado was supported by NASA grant
NAGW-2675.  This research has made use of data obtained through the High Energy
Astrophysics Science Archive Research Center Online Service, provided by the
NASA/Goddard Space Flight Center as well as NASA's Astrophysics Data System (ADS)
Abstract Service.

\clearpage

\begin{figure}
\plottwo{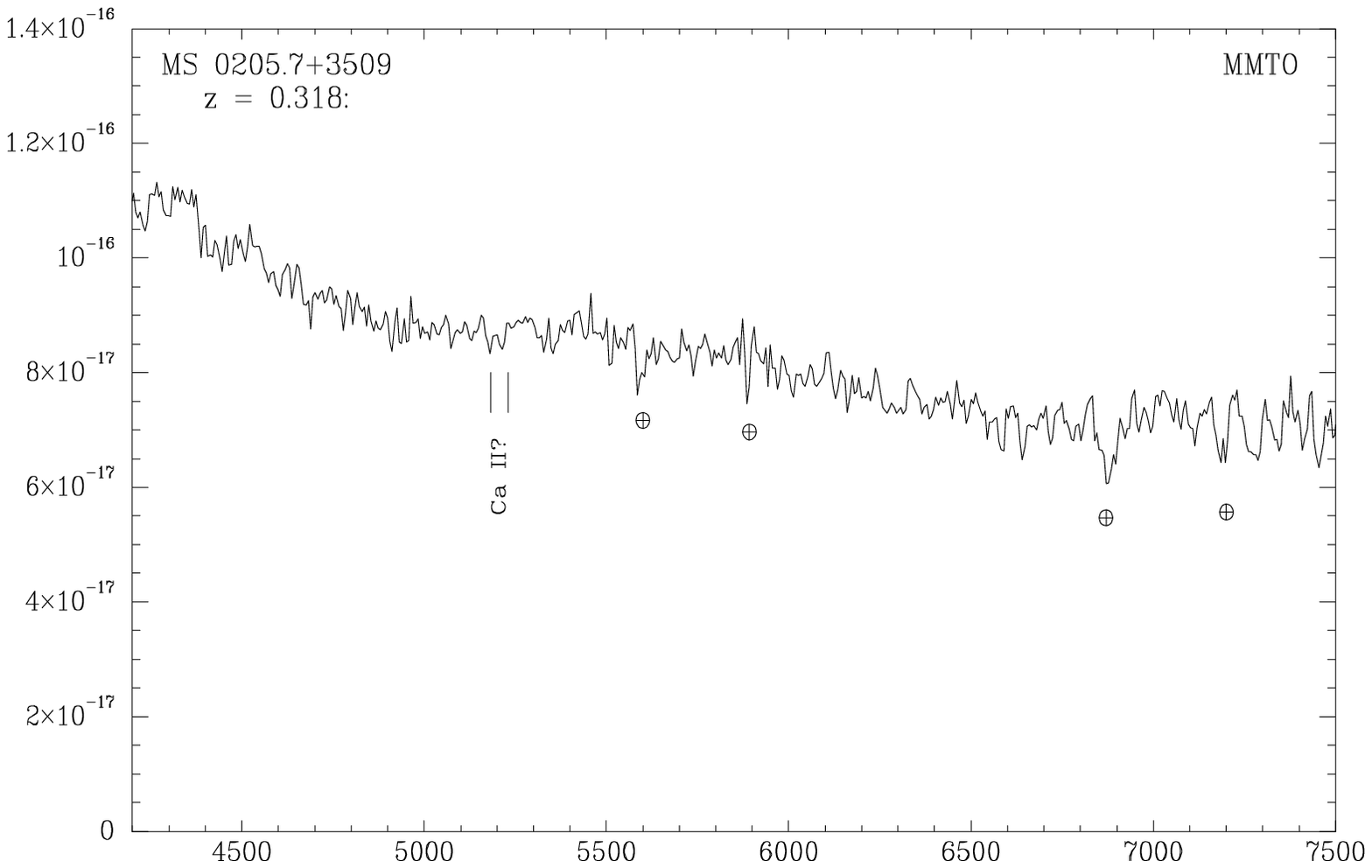}{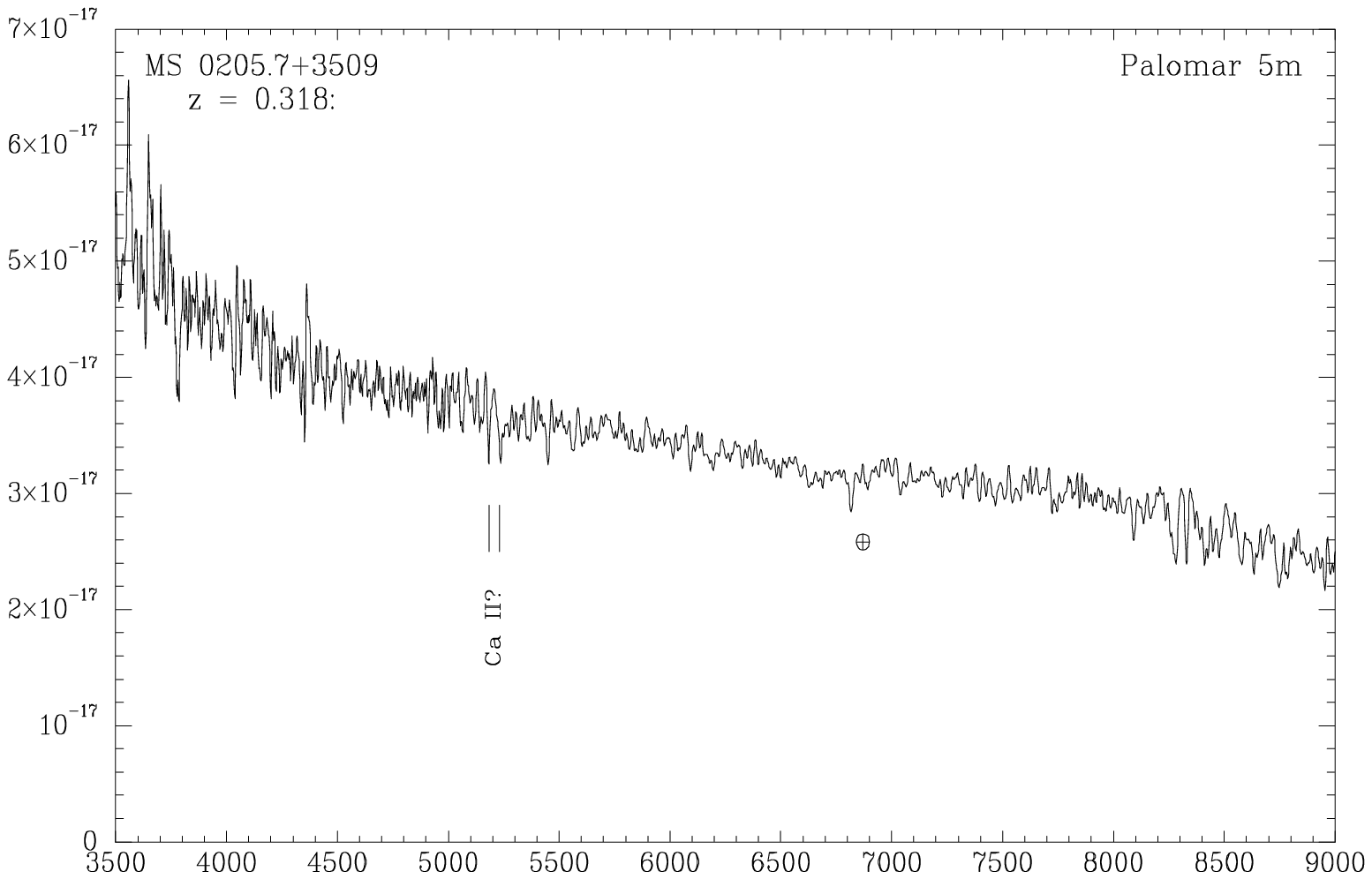}
\end{figure}
\begin{figure}
\plottwo{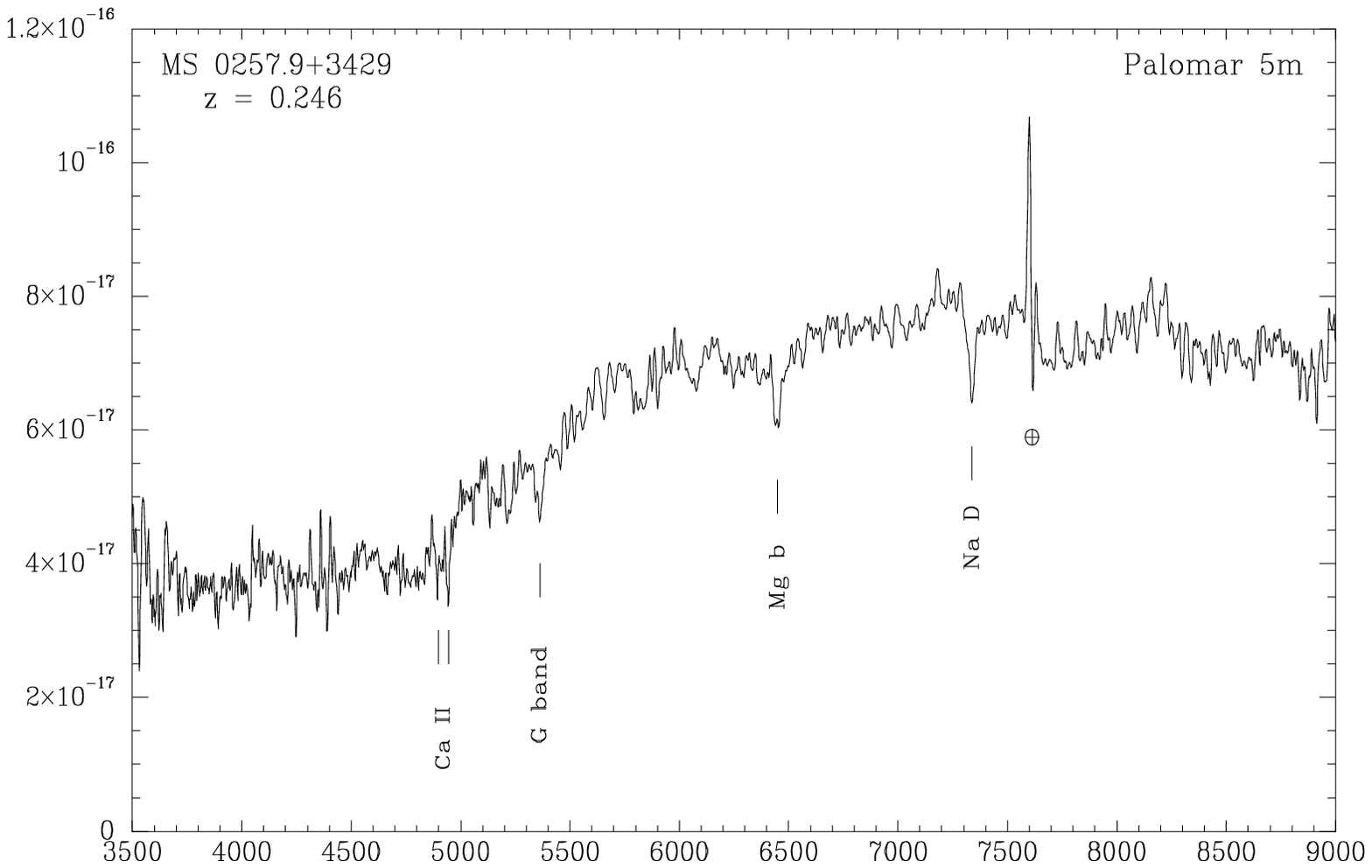}{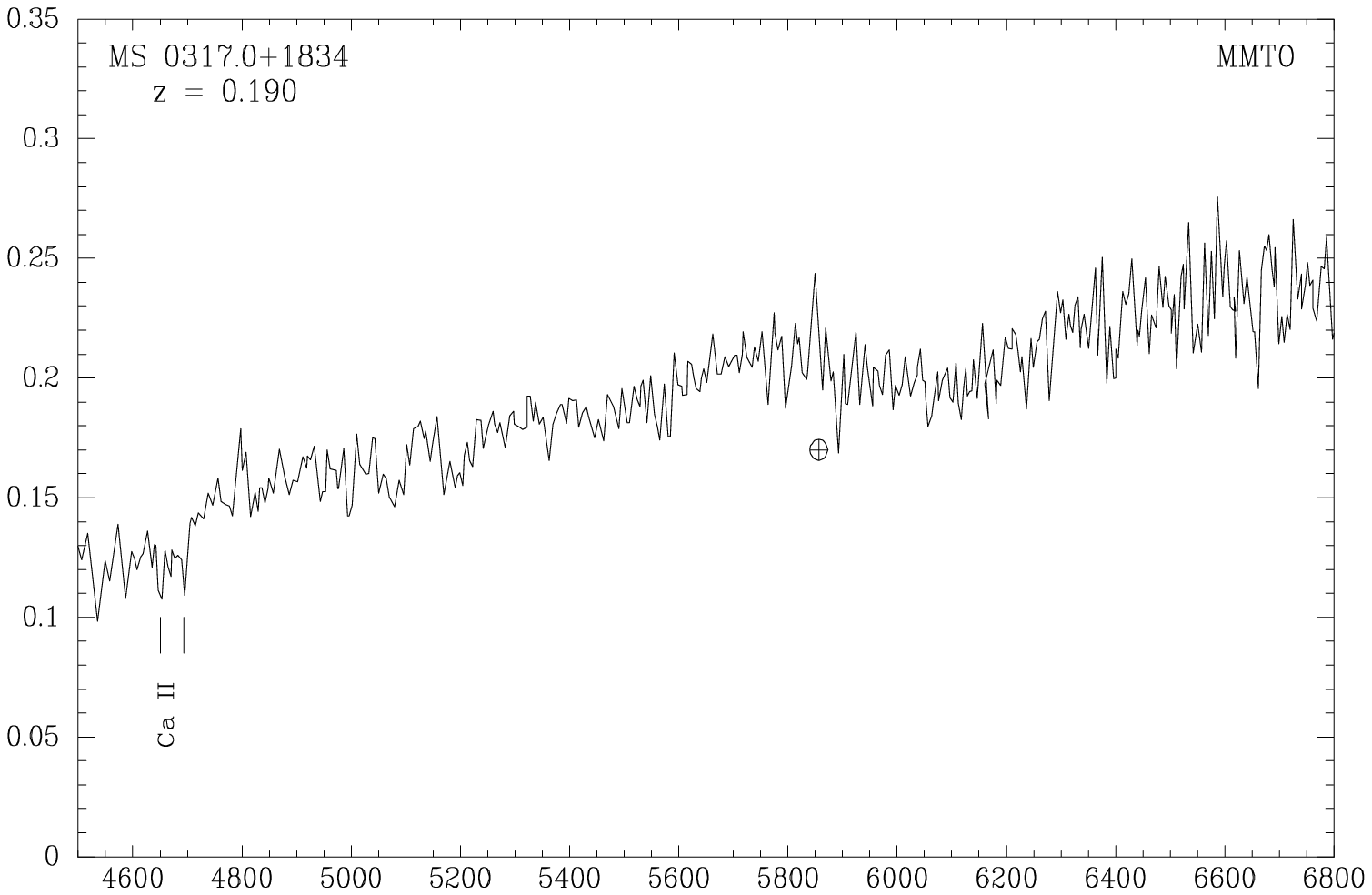}
\end{figure}

\begin{figure}
\plottwo{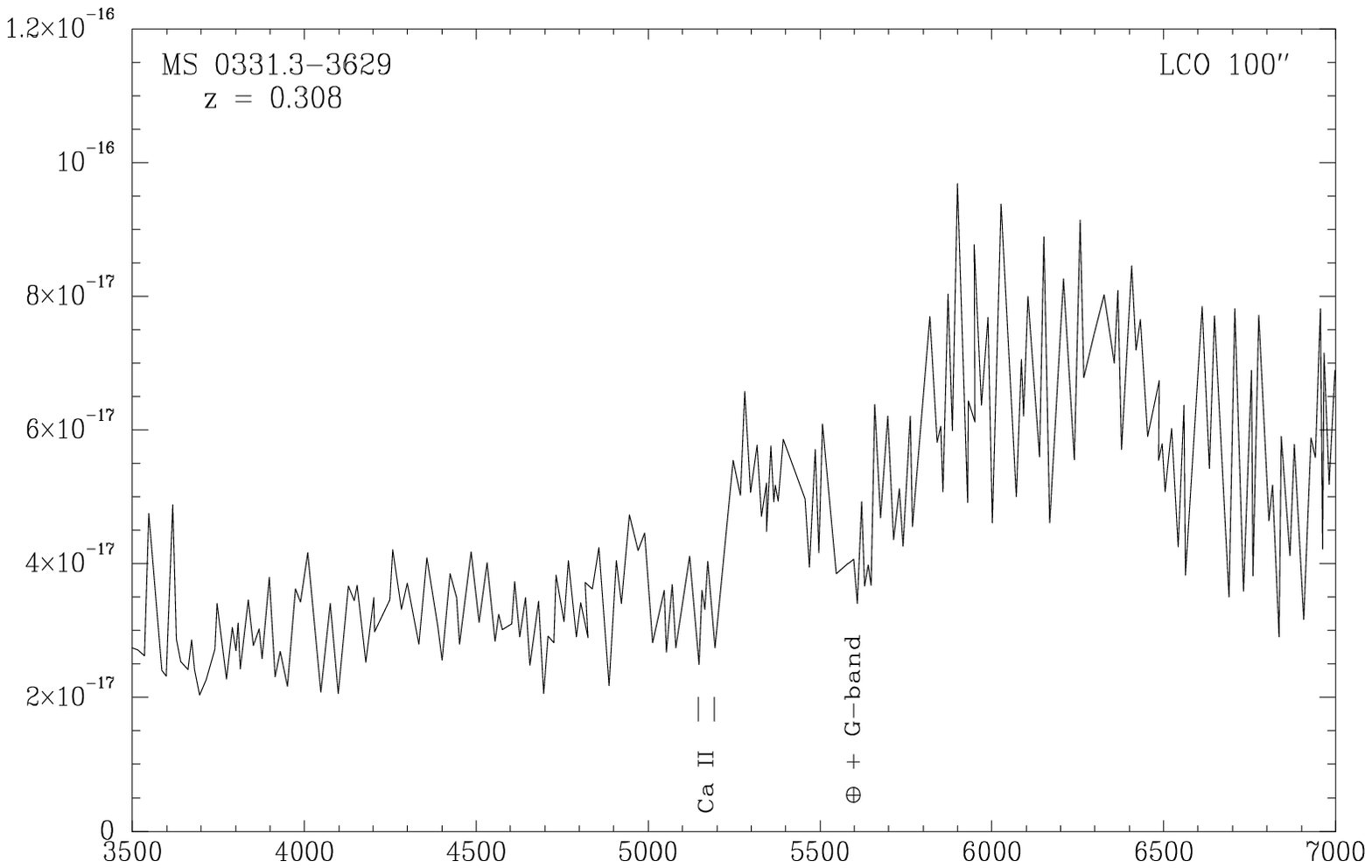}{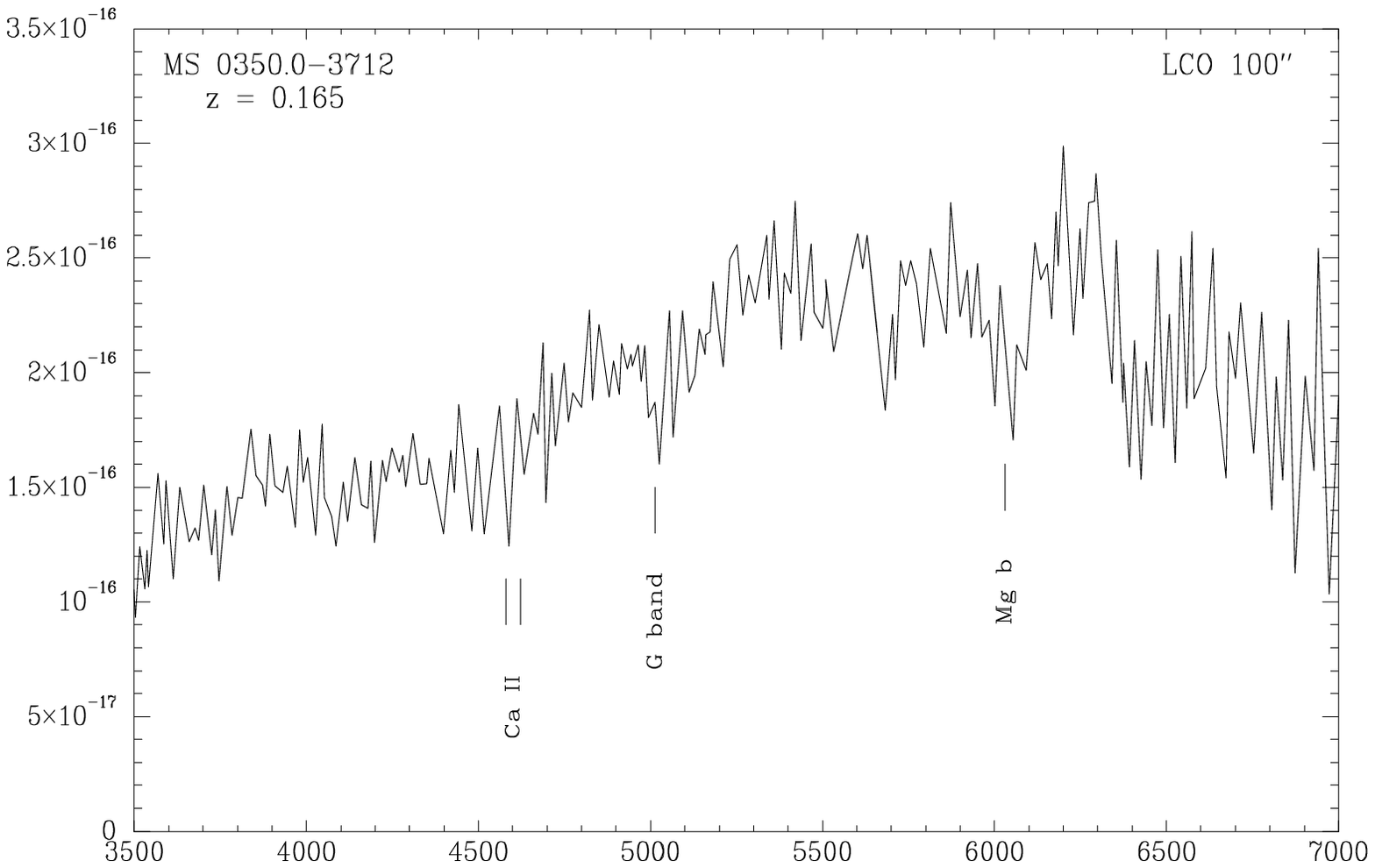}
\end{figure}
\begin{figure}
\plottwo{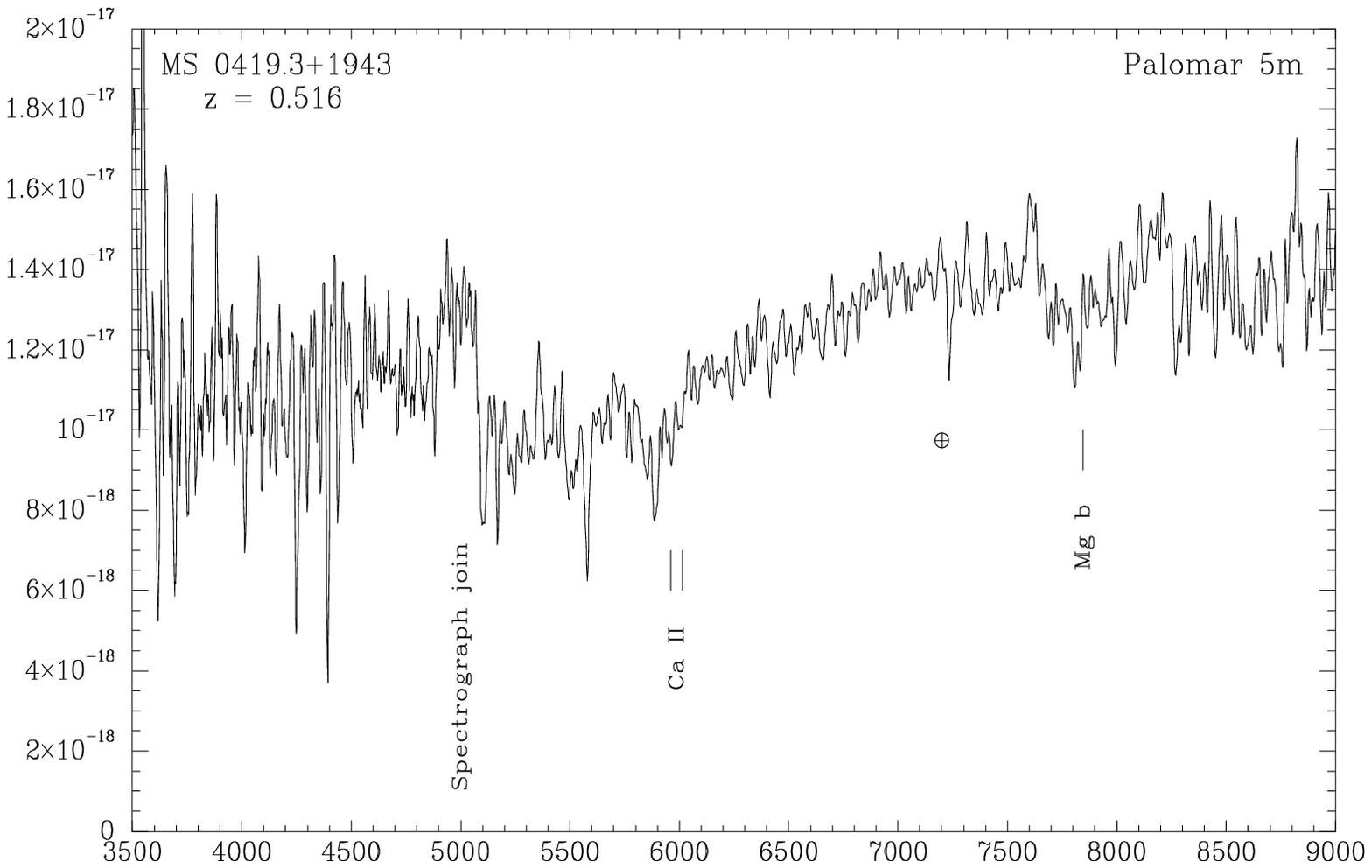}{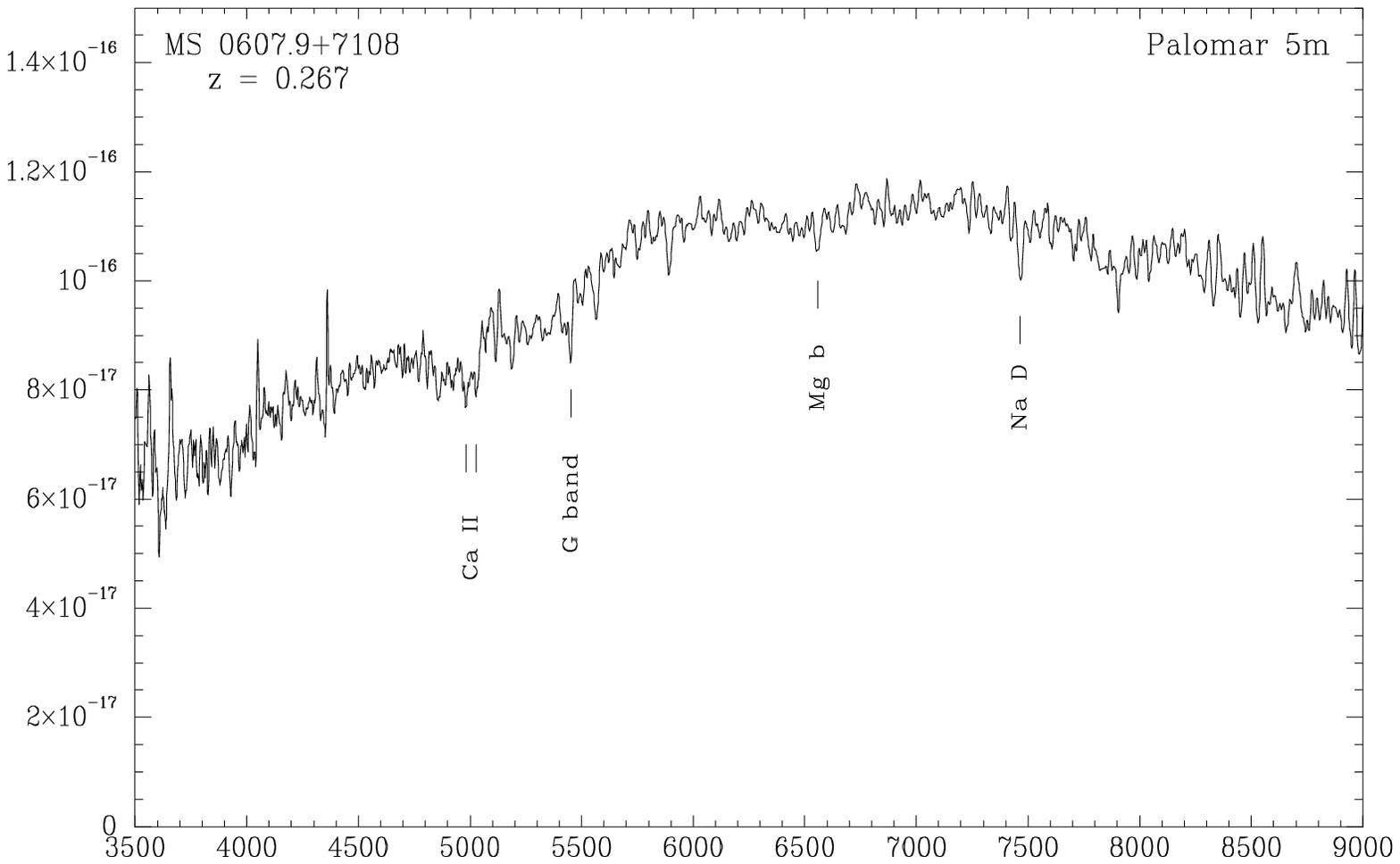}
\end{figure}
\begin{figure}
\plottwo{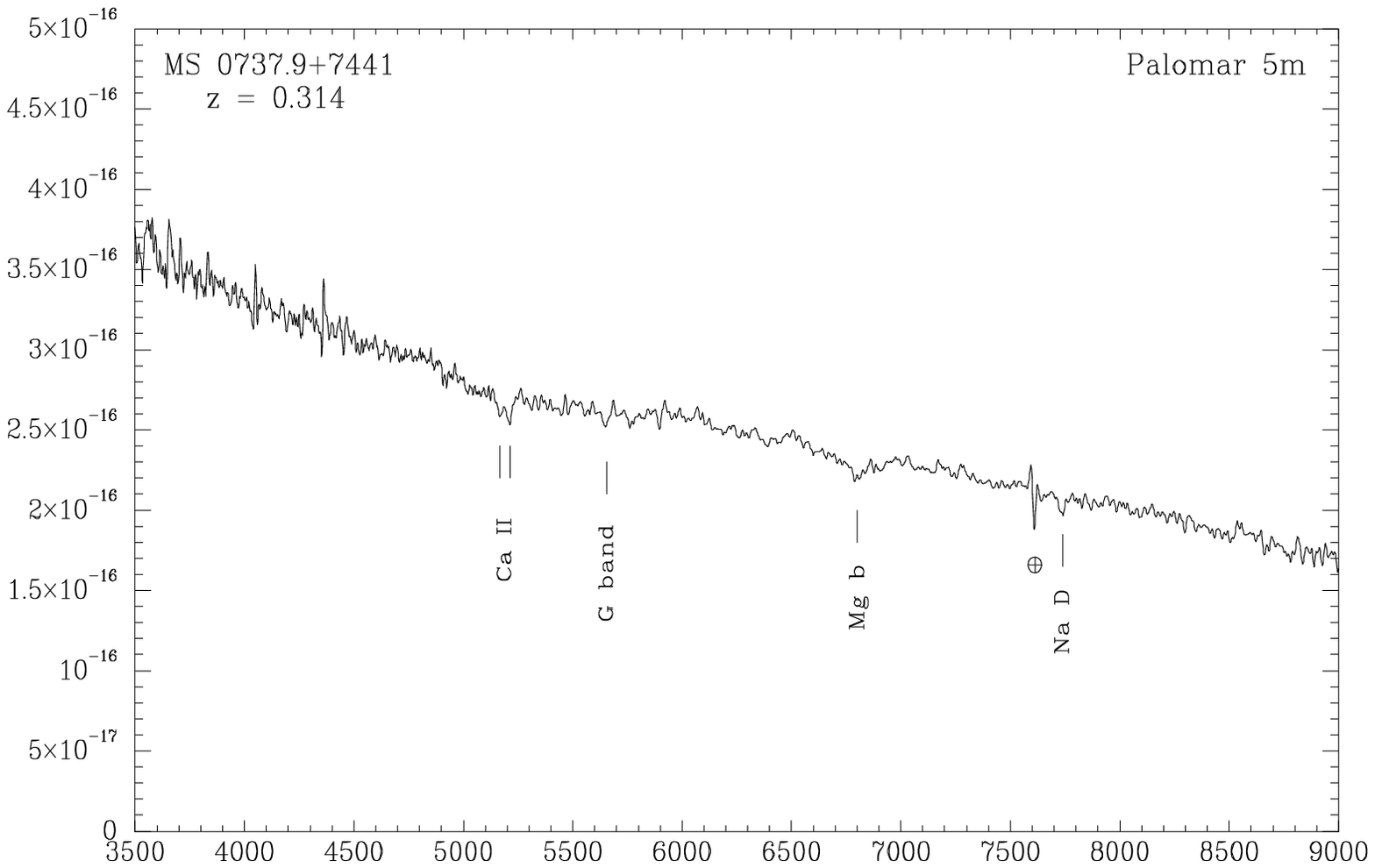}{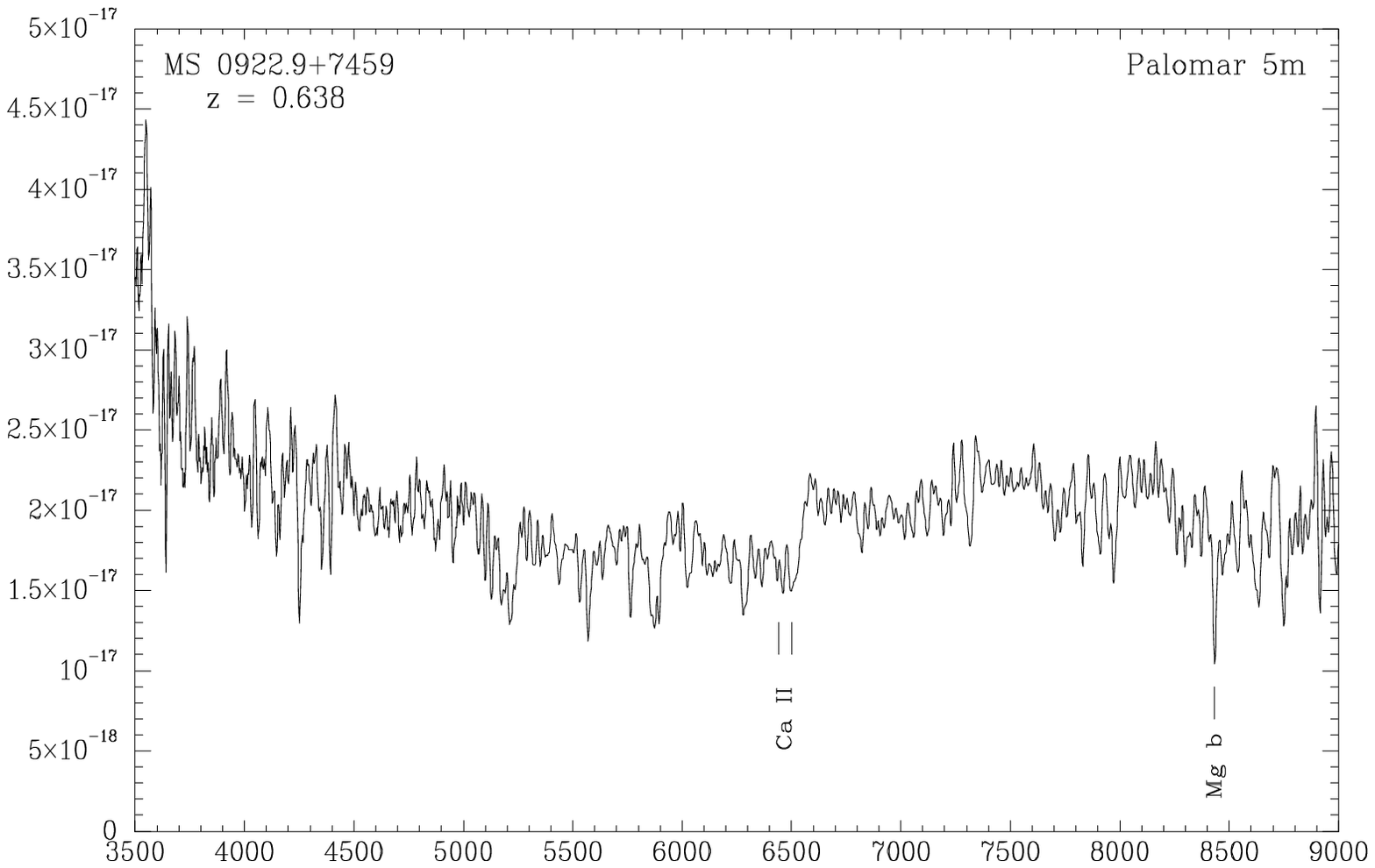}
\end{figure}
\begin{figure}
\plottwo{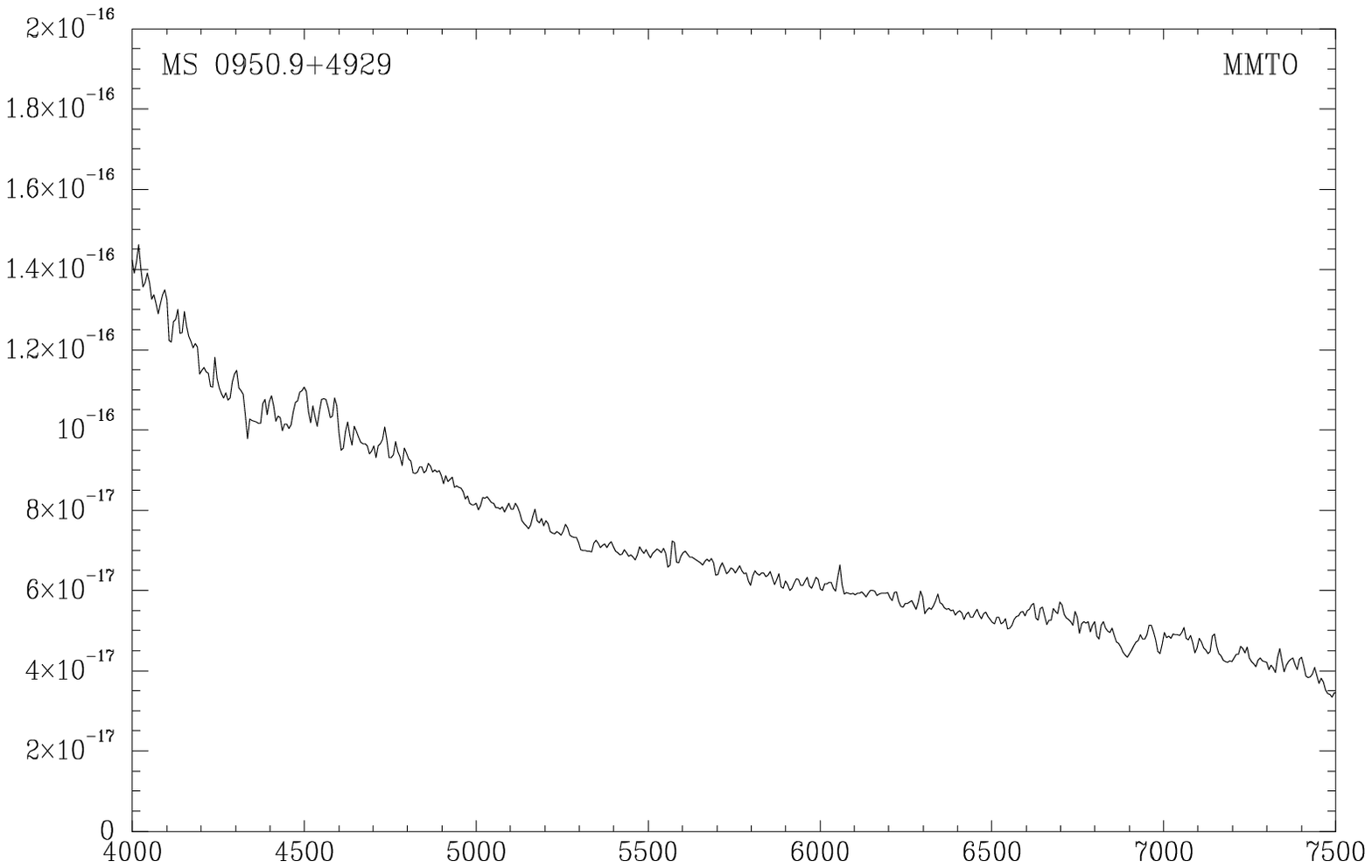}{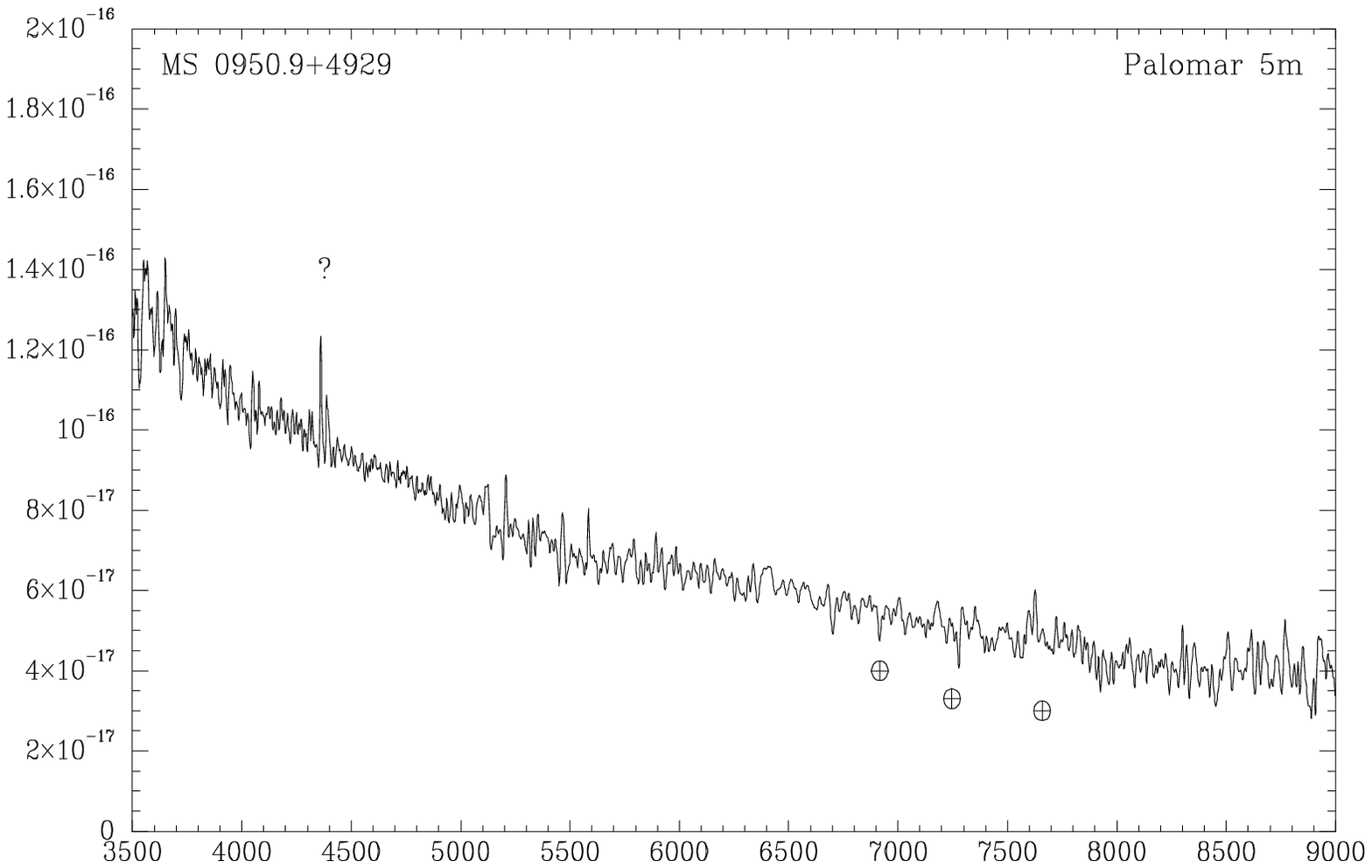}
\end{figure}

\begin{figure}
\plottwo{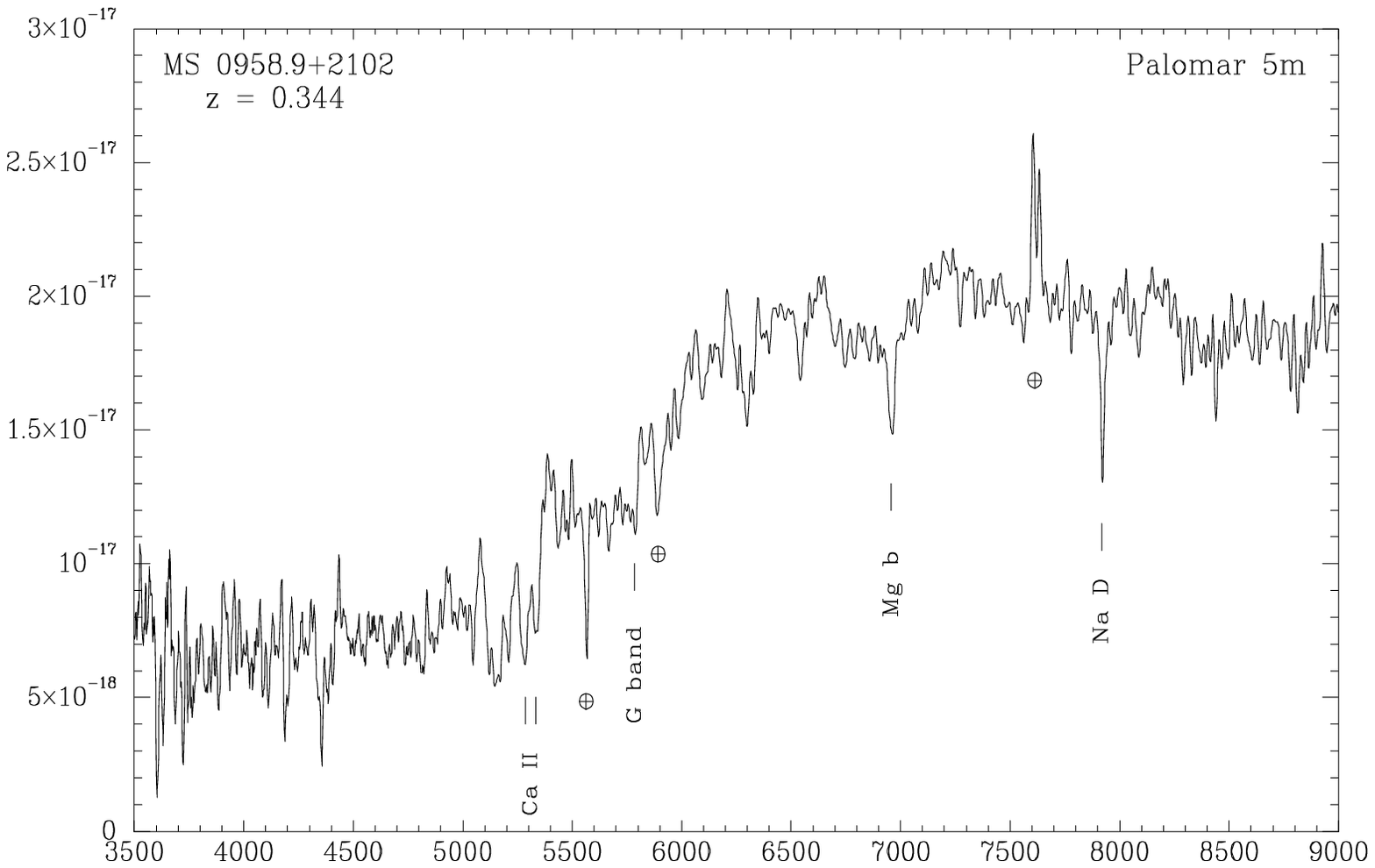}{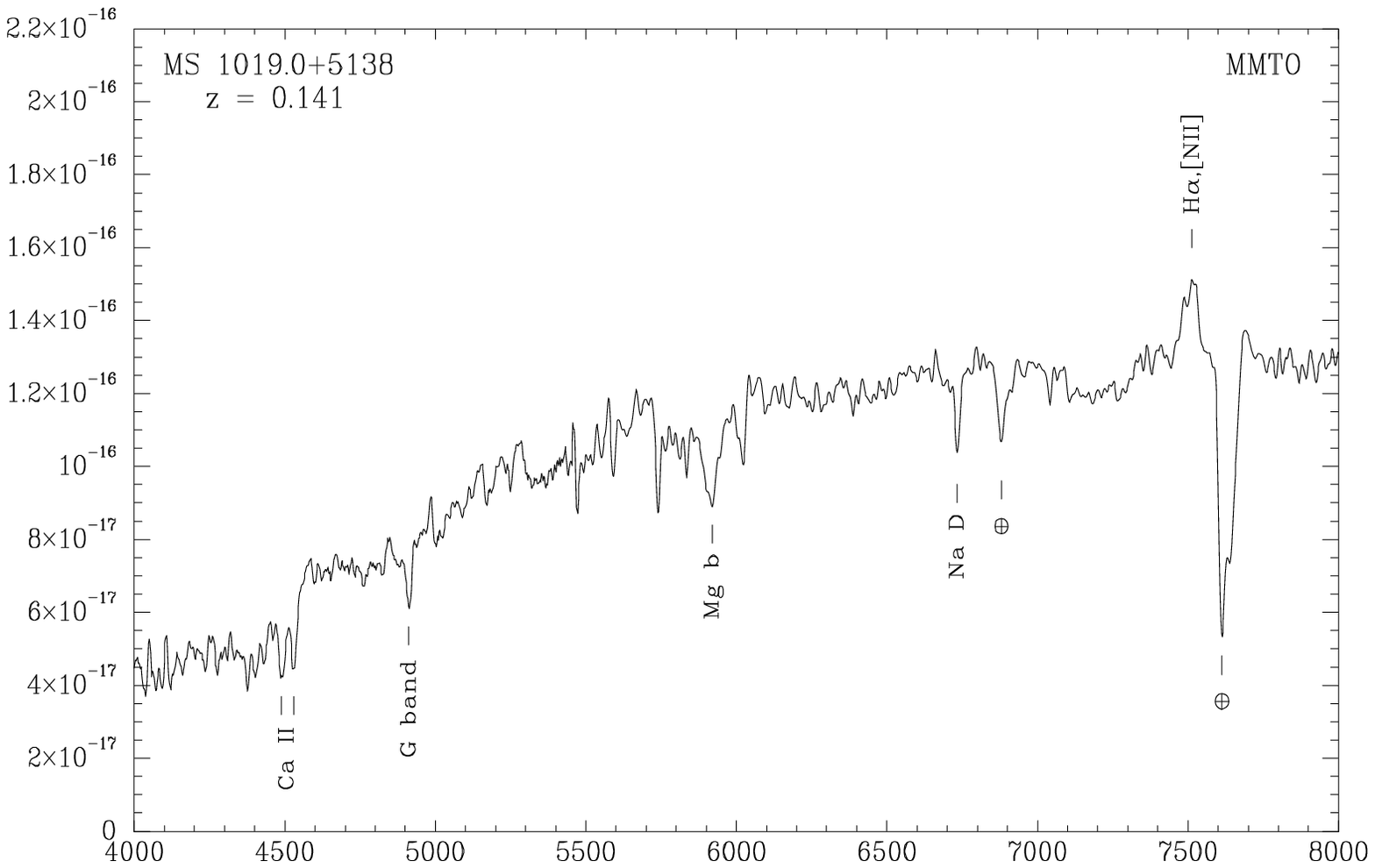}
\end{figure}
\begin{figure}
\plottwo{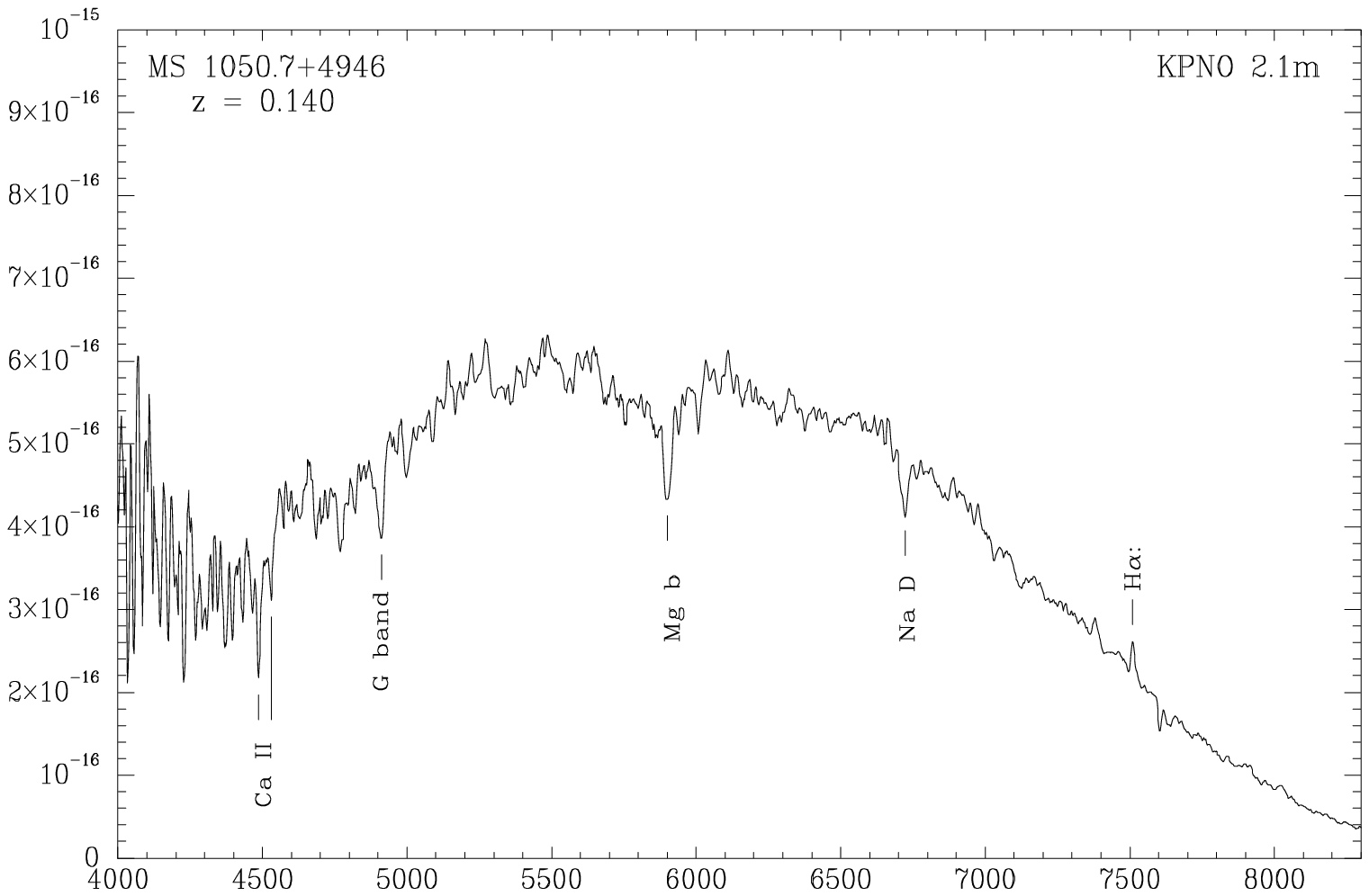}{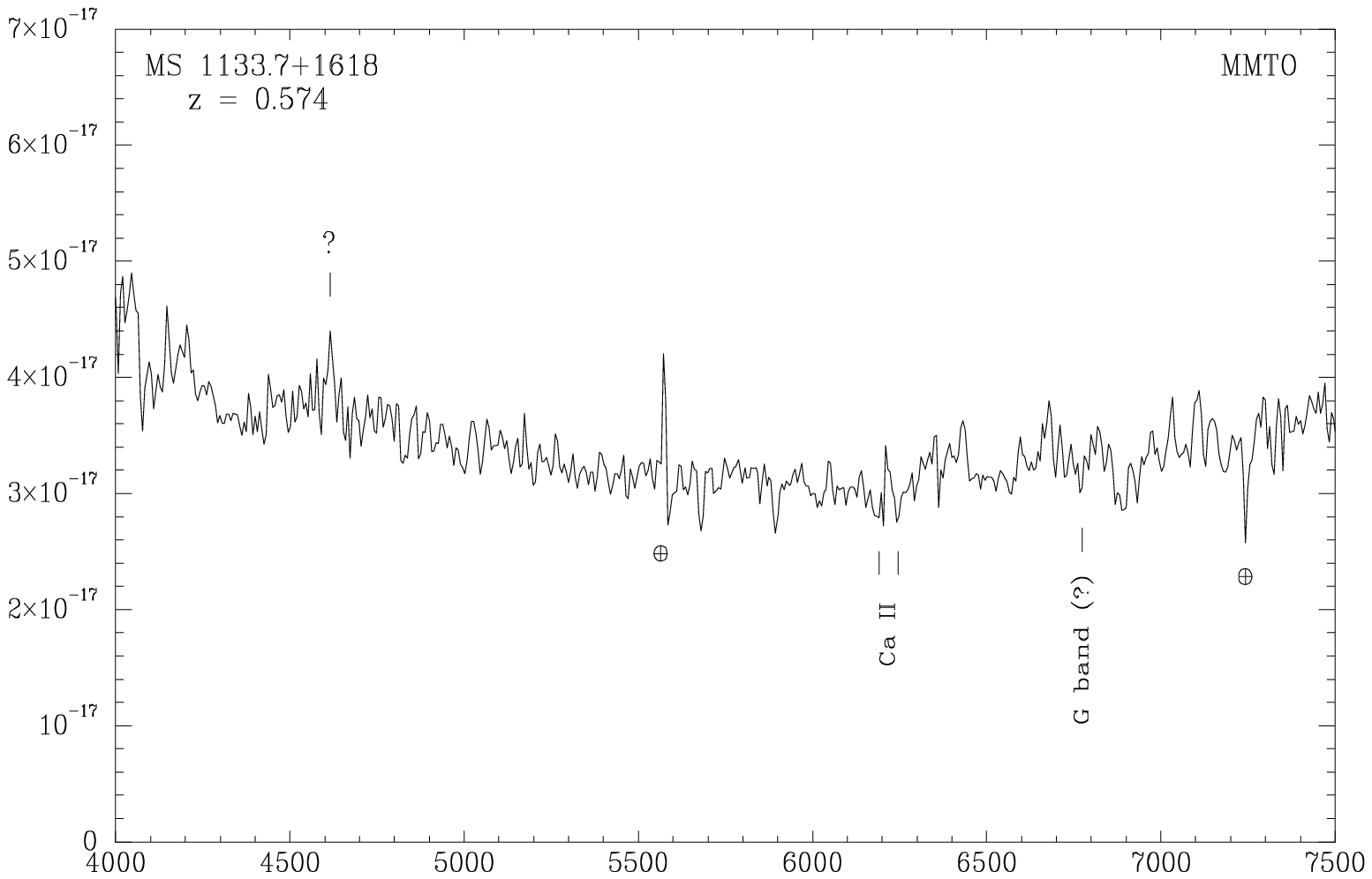}
\end{figure}
\begin{figure}
\plottwo{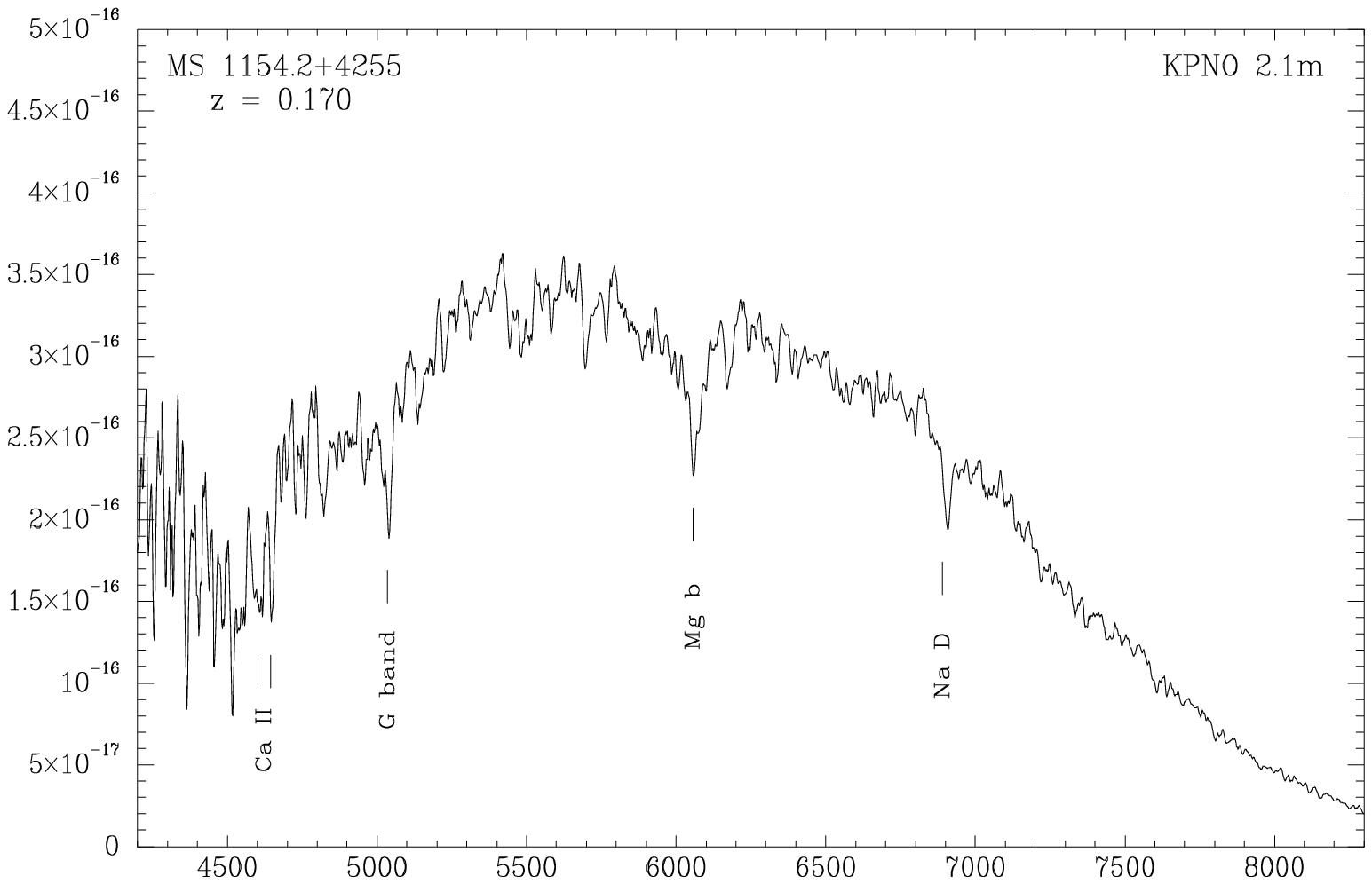}{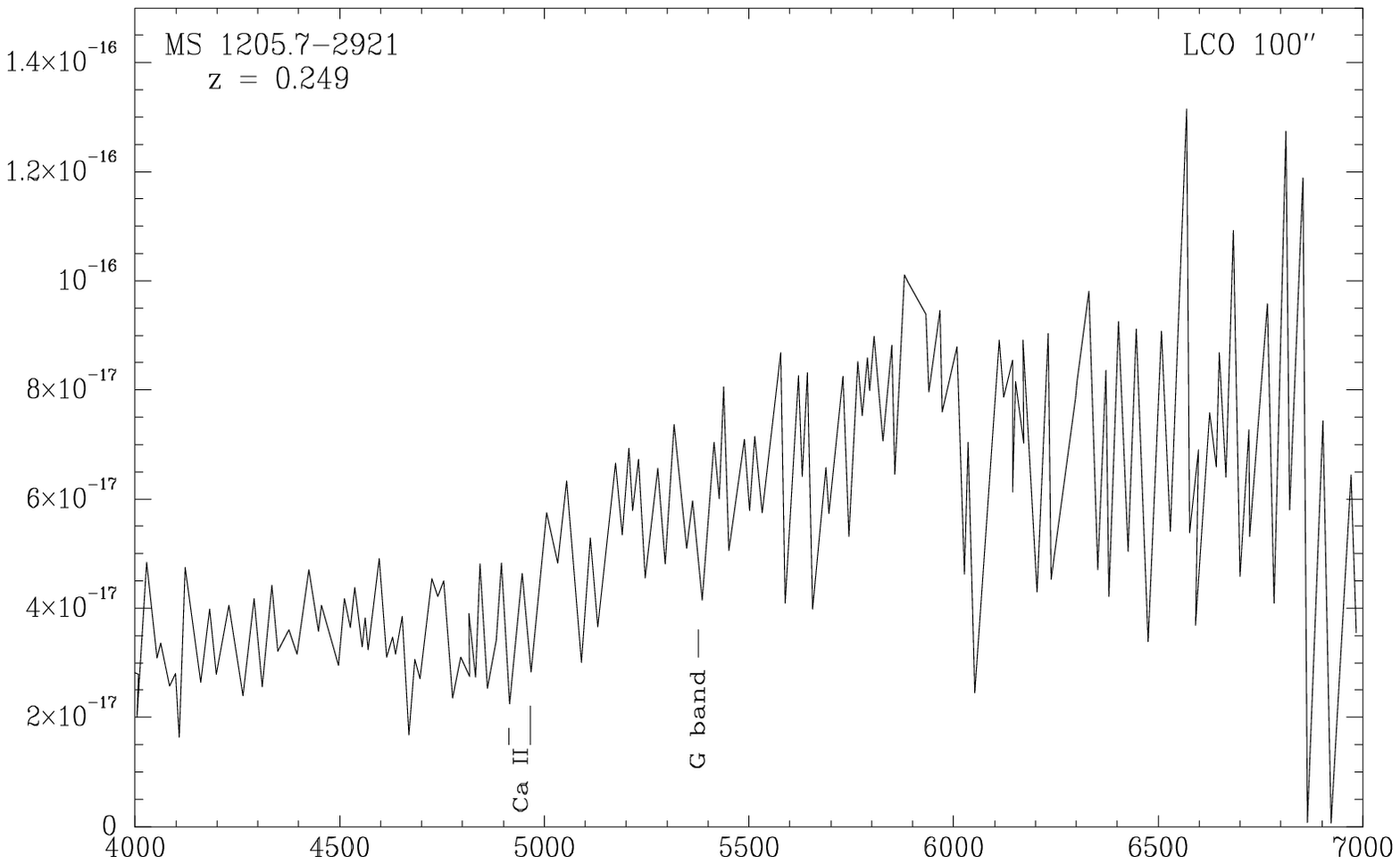}
\end{figure}
\begin{figure}
\plottwo{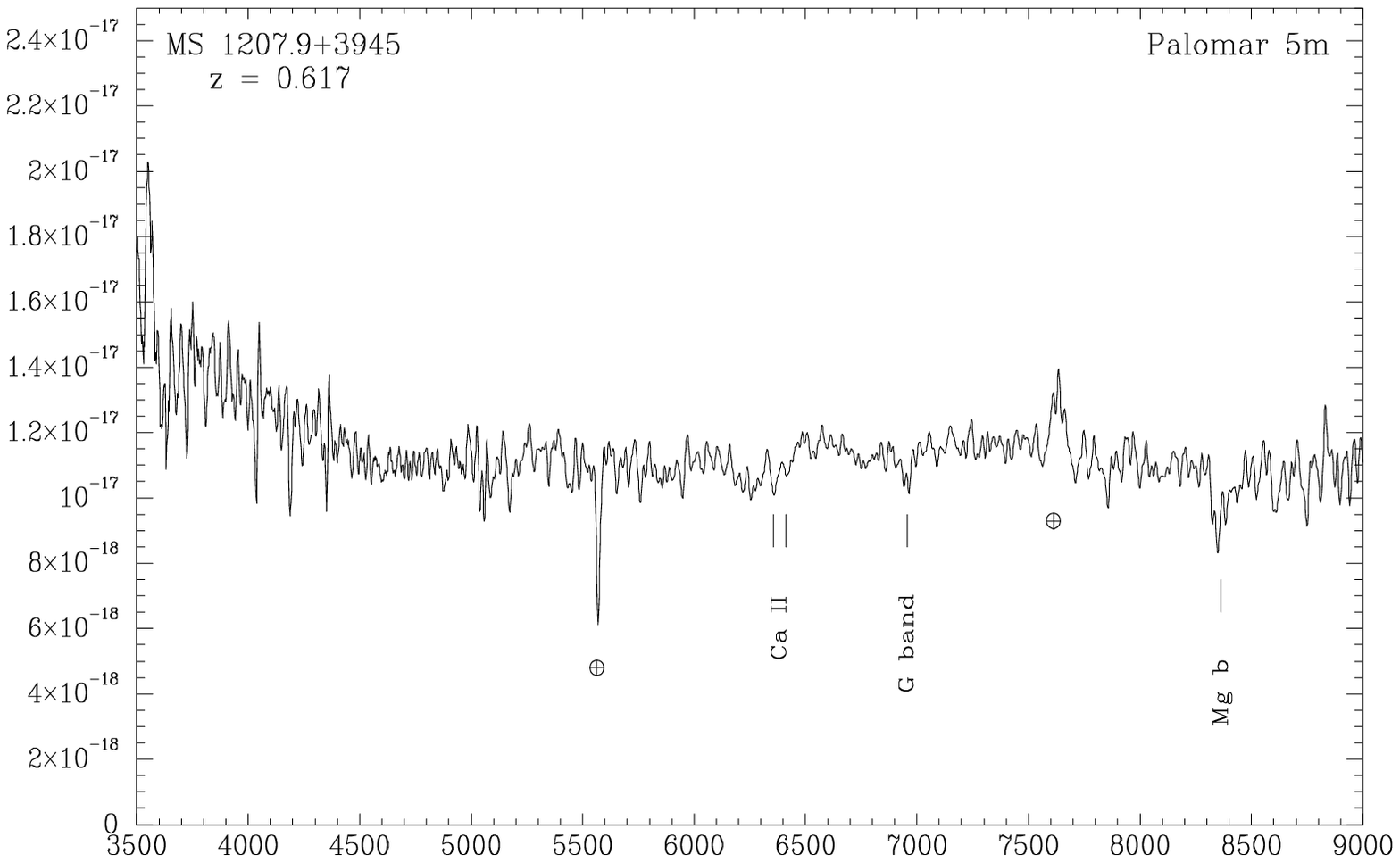}{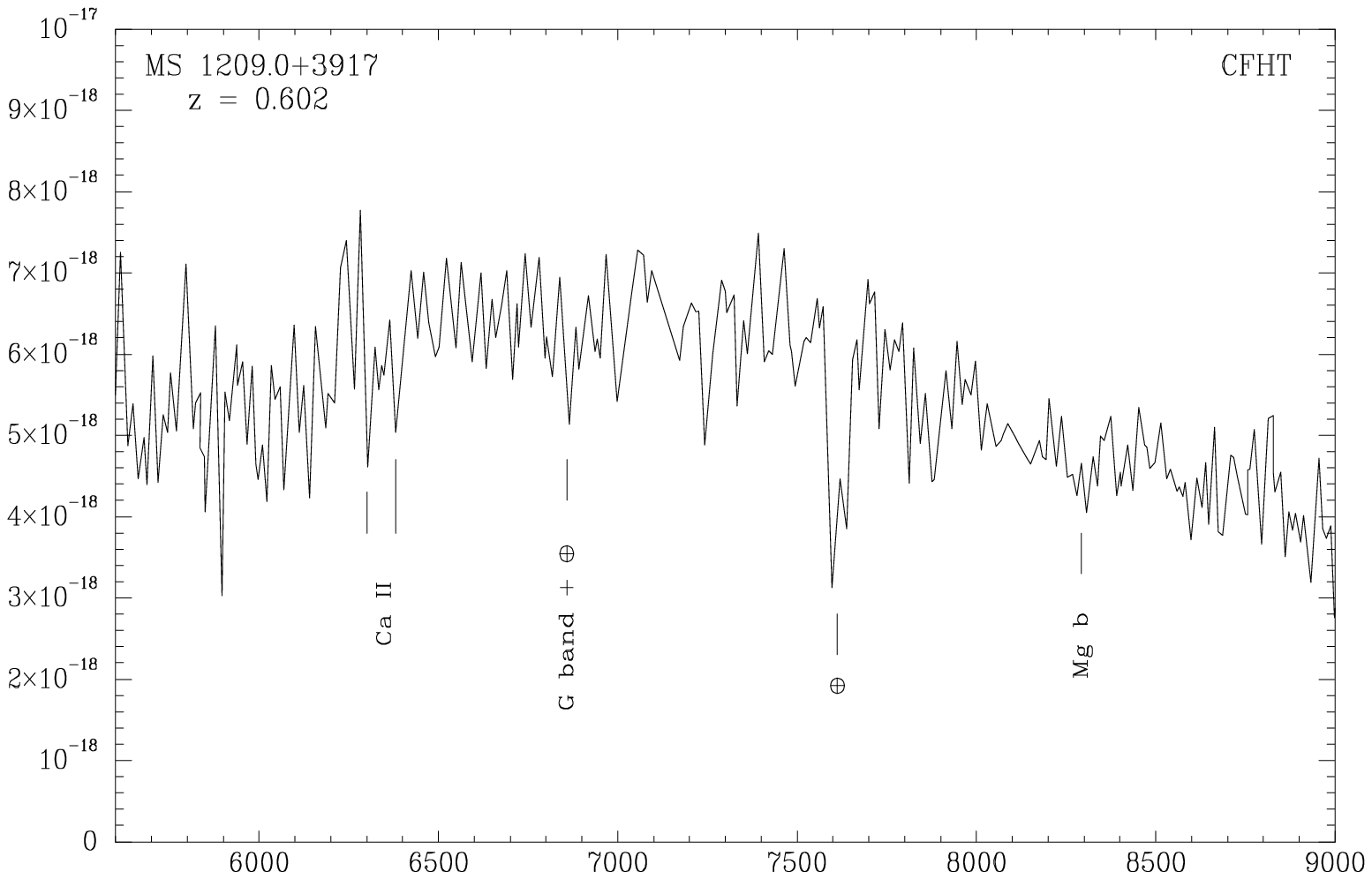}
\end{figure}

\begin{figure}
\plottwo{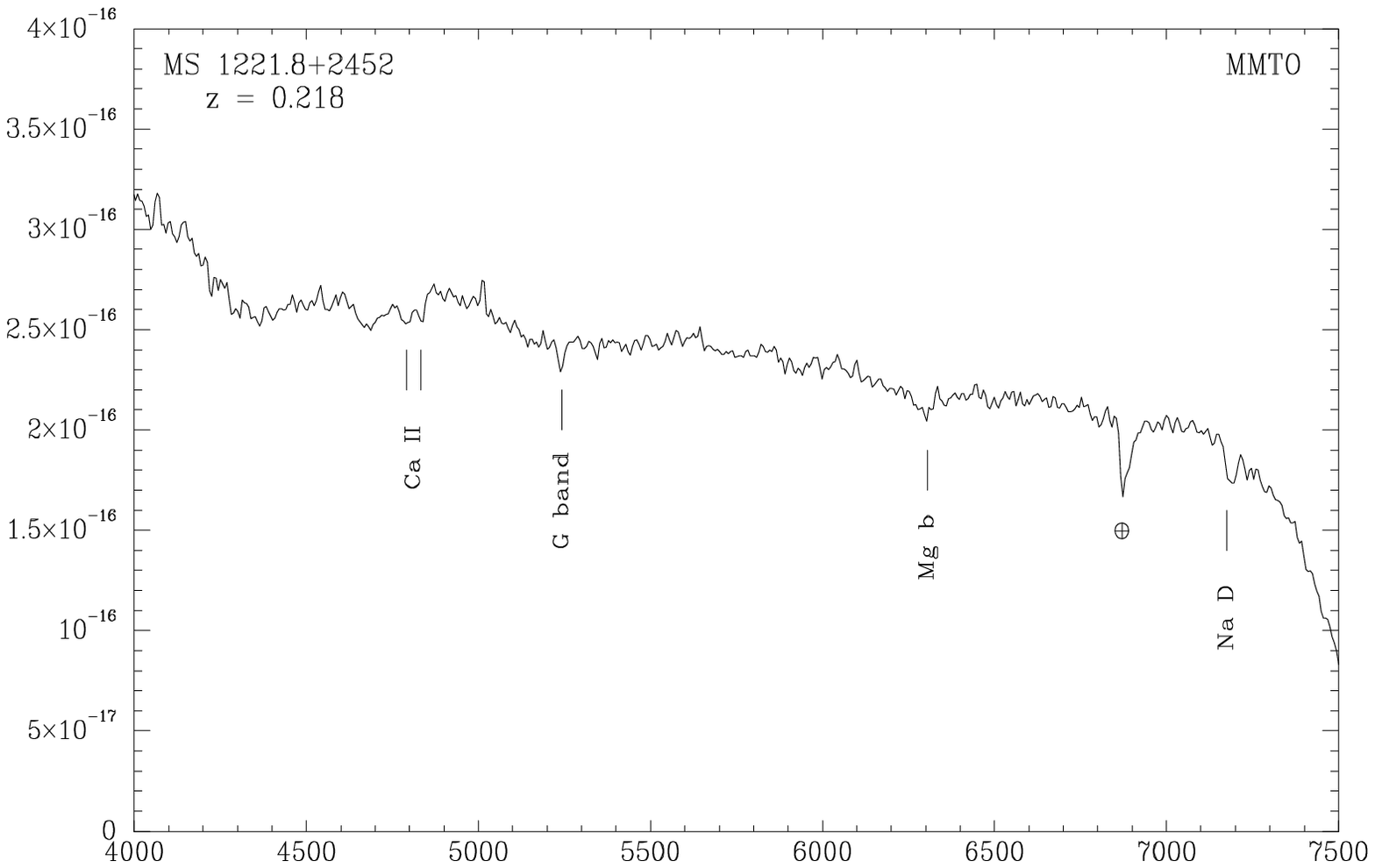}{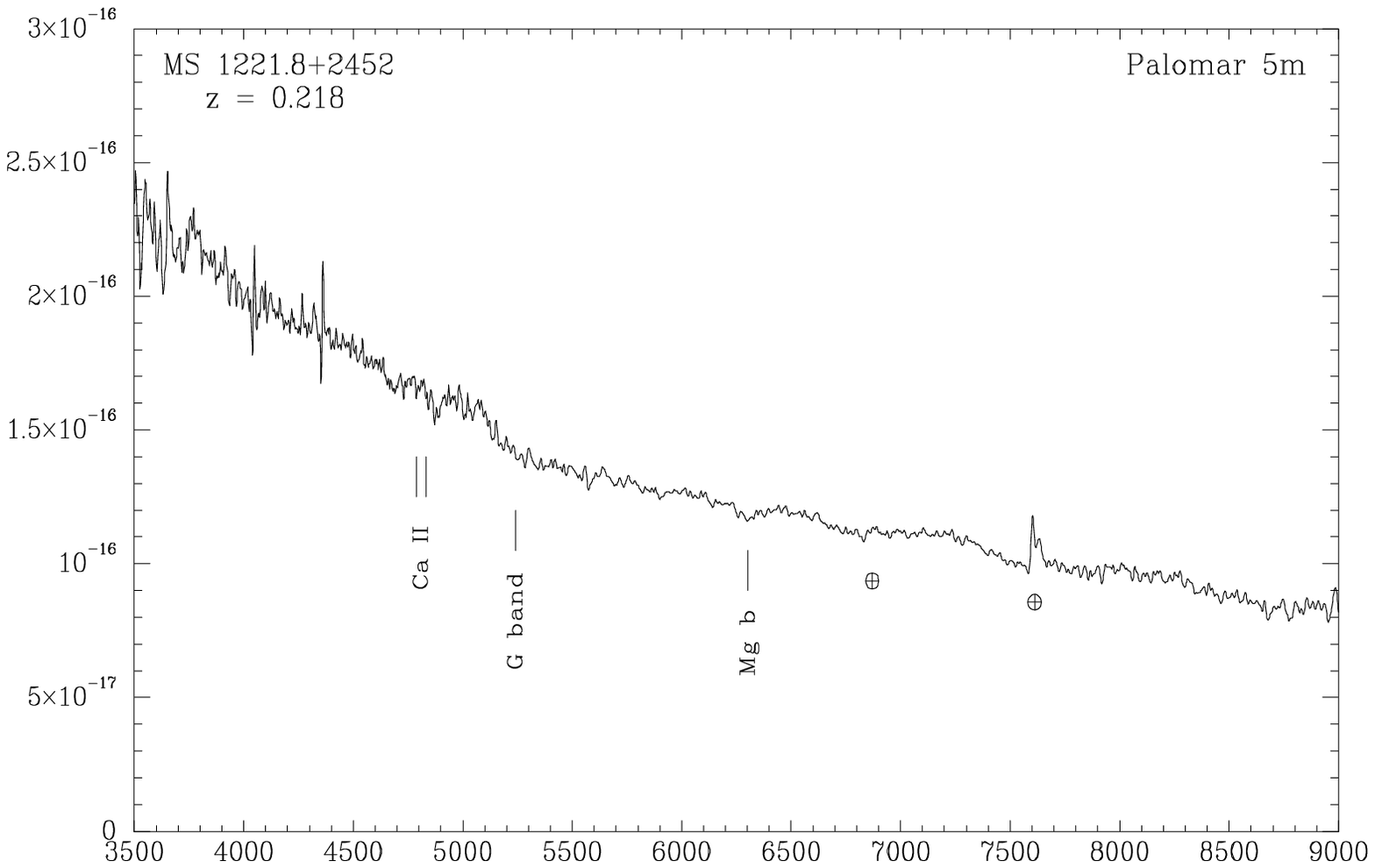}
\end{figure}
\begin{figure}
\plottwo{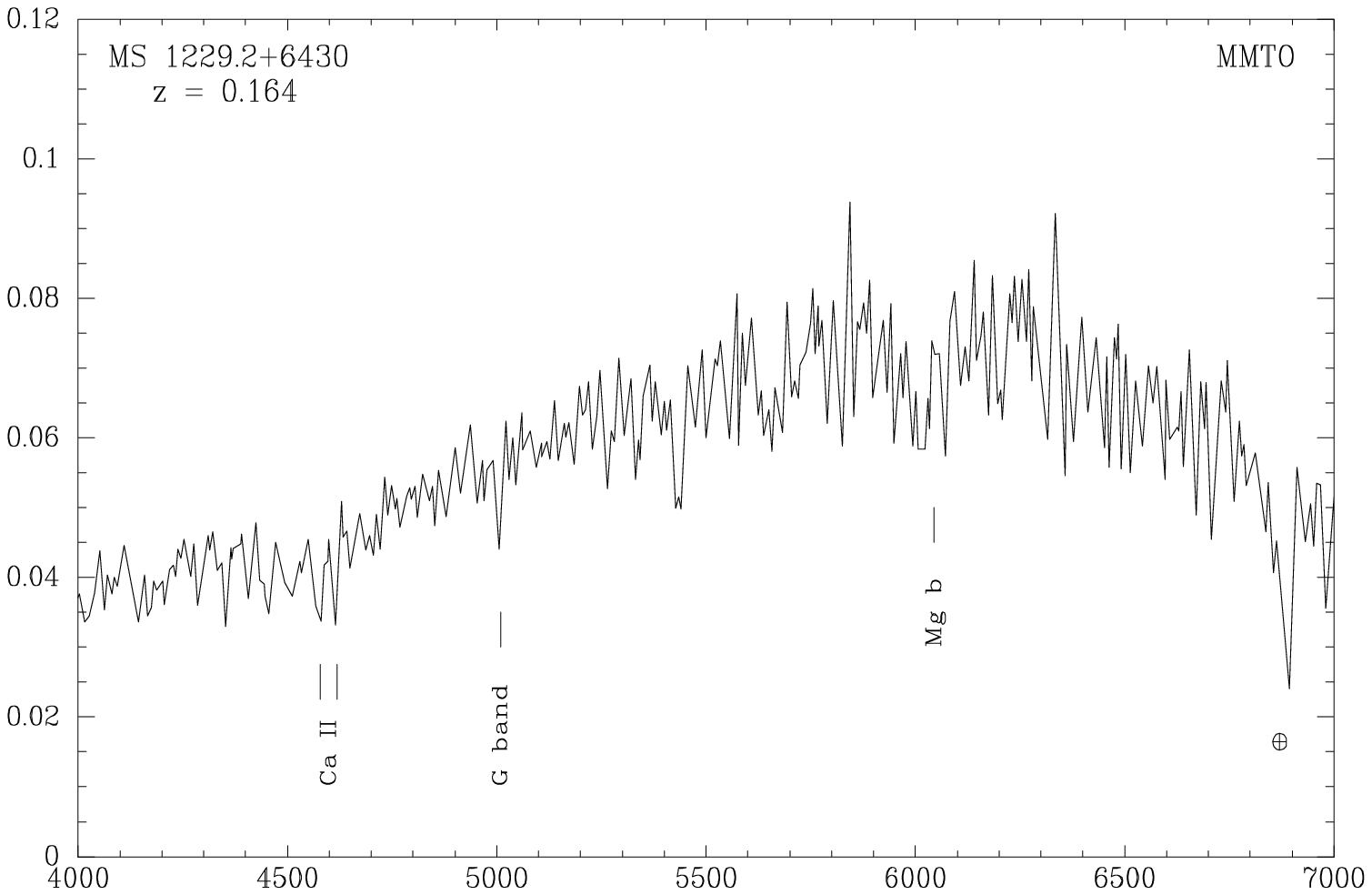}{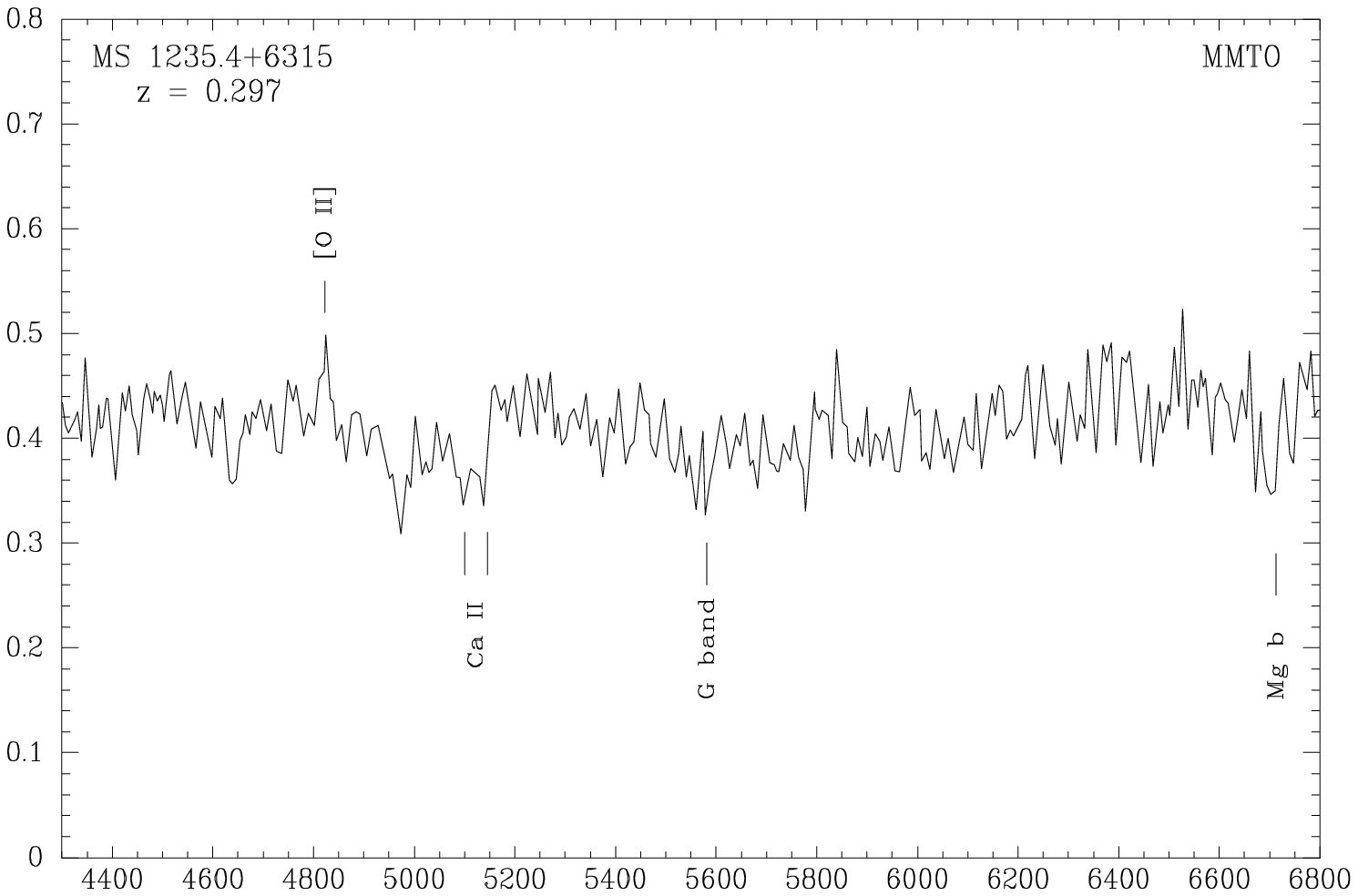}
\end{figure}
\begin{figure}
\plottwo{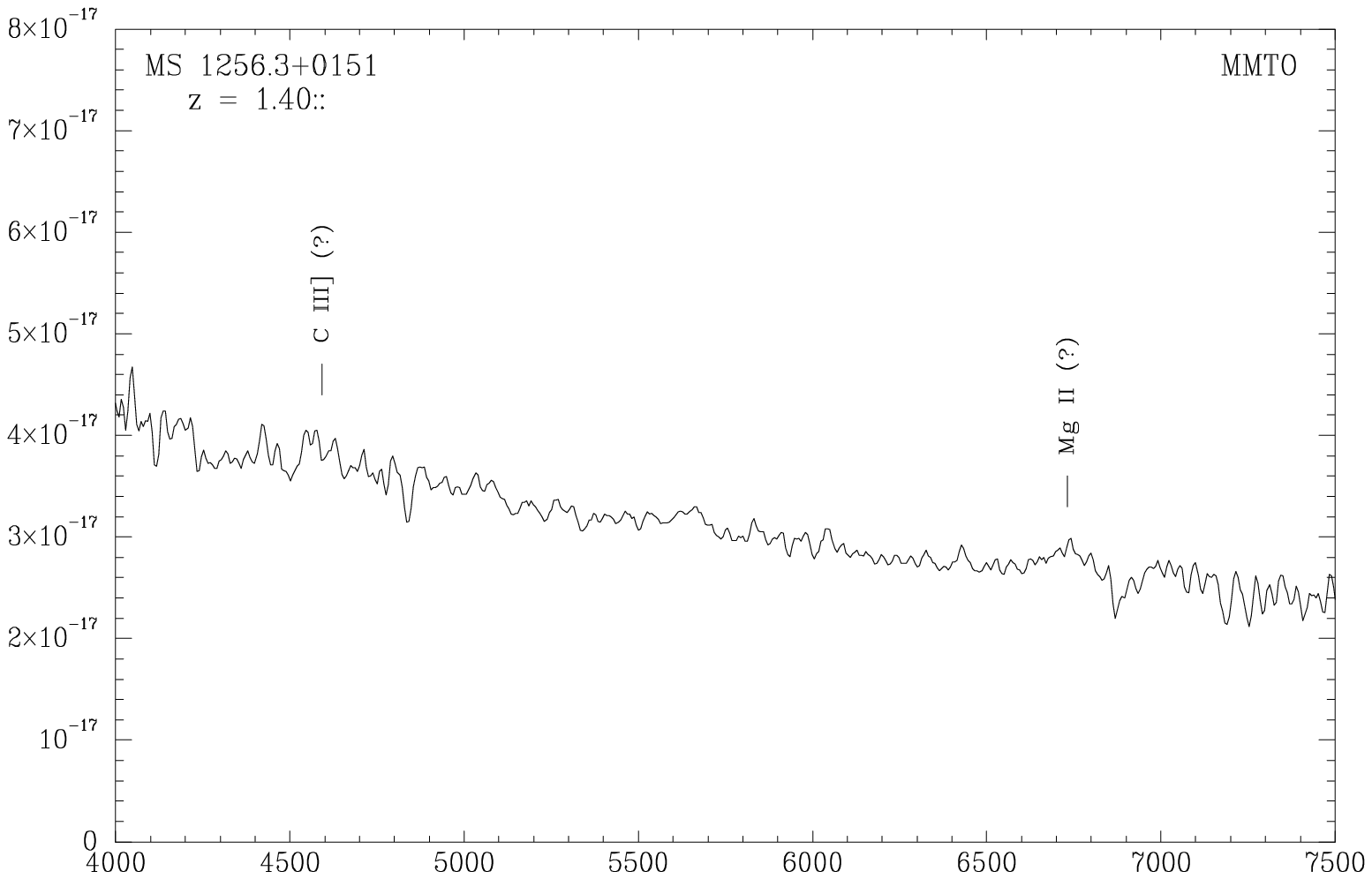}{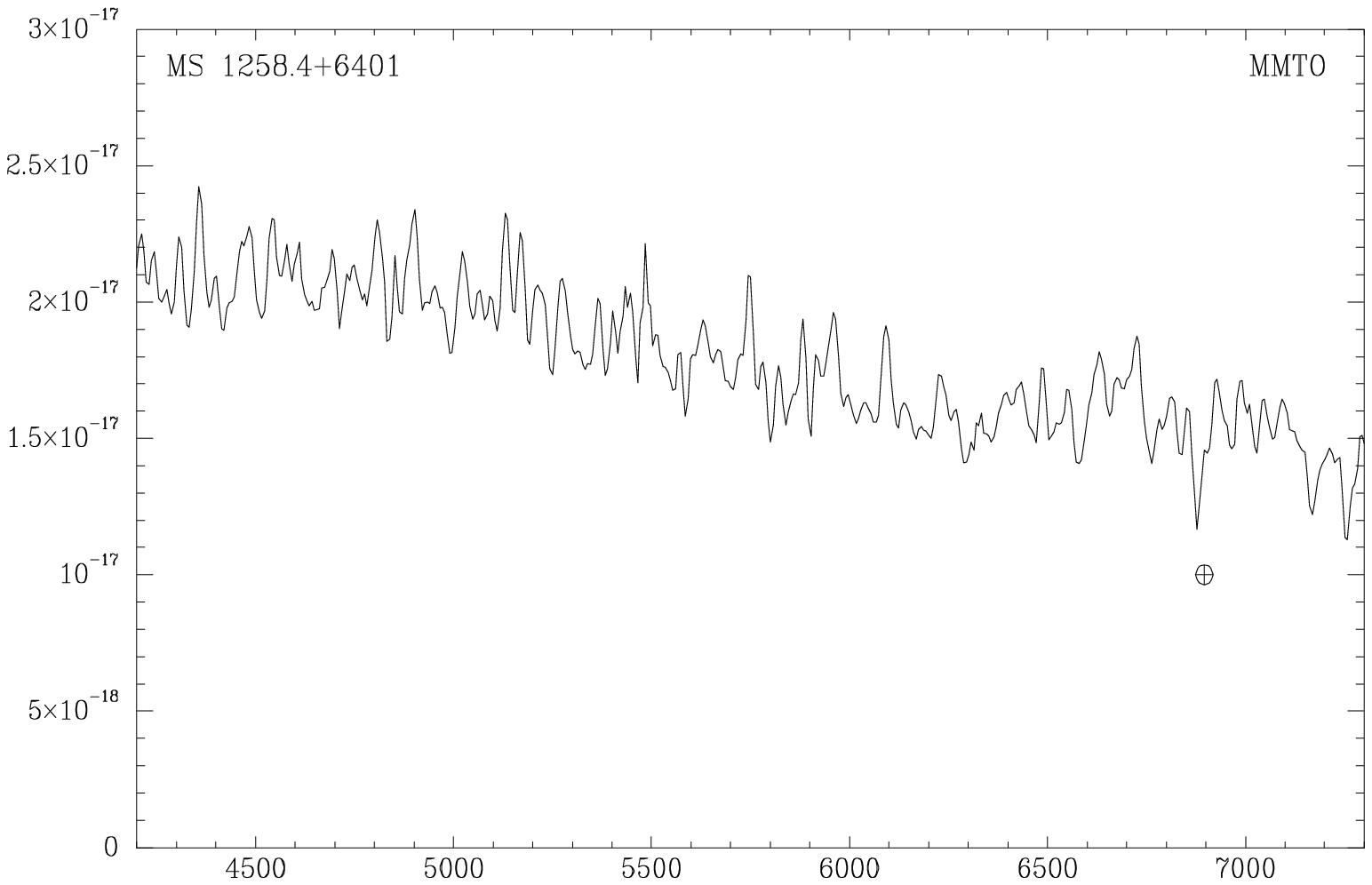}
\end{figure}
\begin{figure}
\plottwo{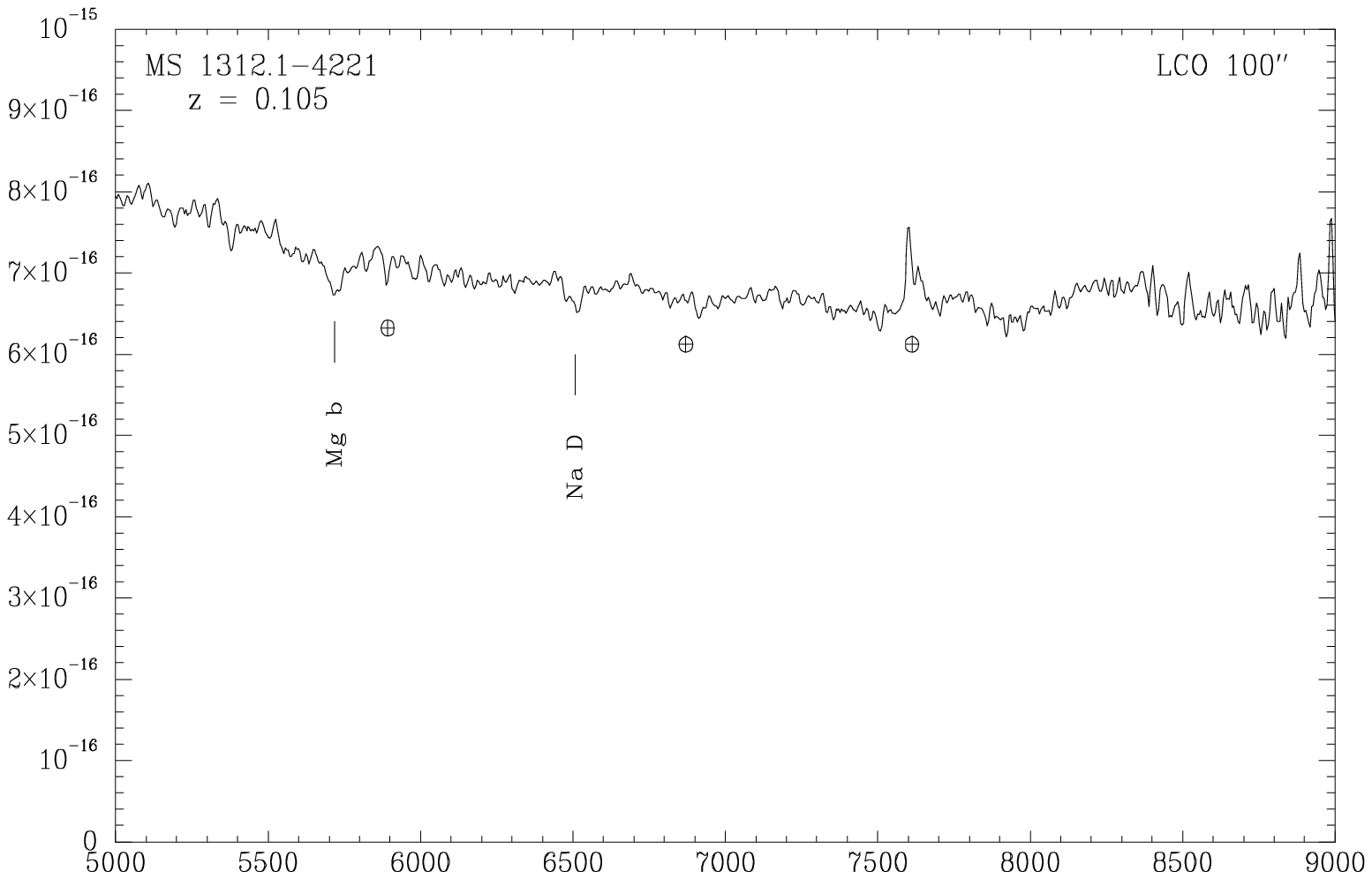}{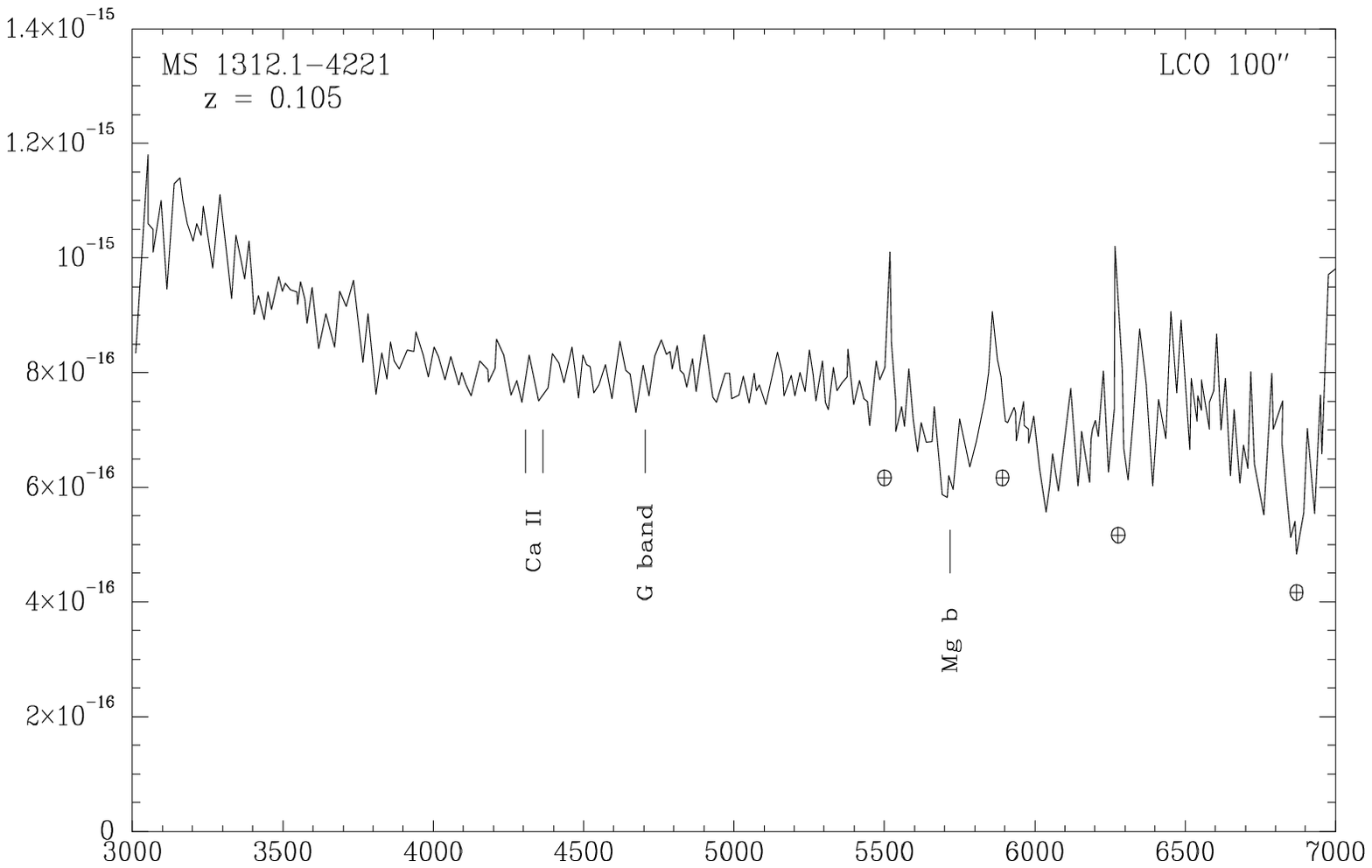}
\end{figure}
\clearpage
\begin{figure}
\plottwo{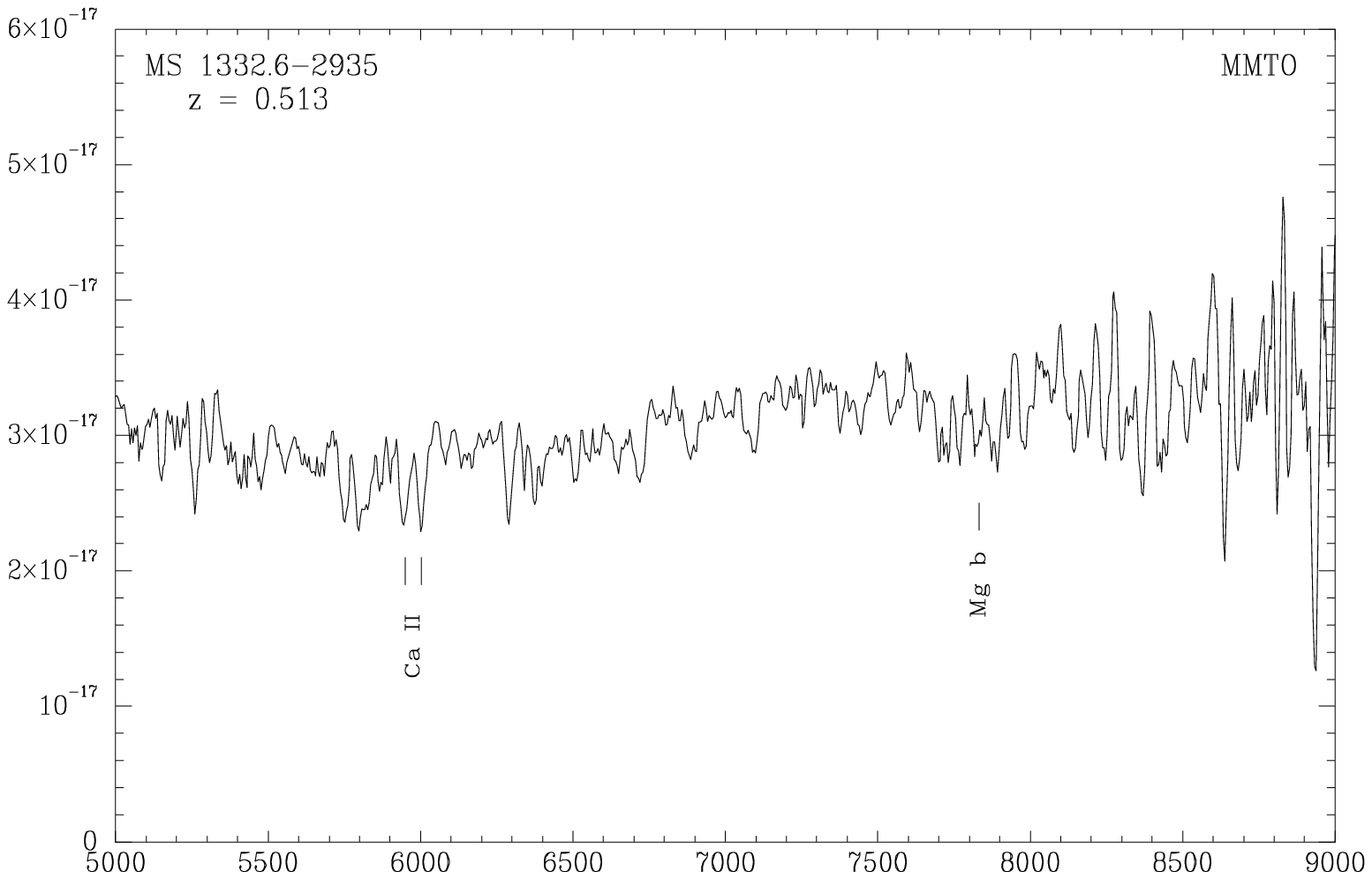}{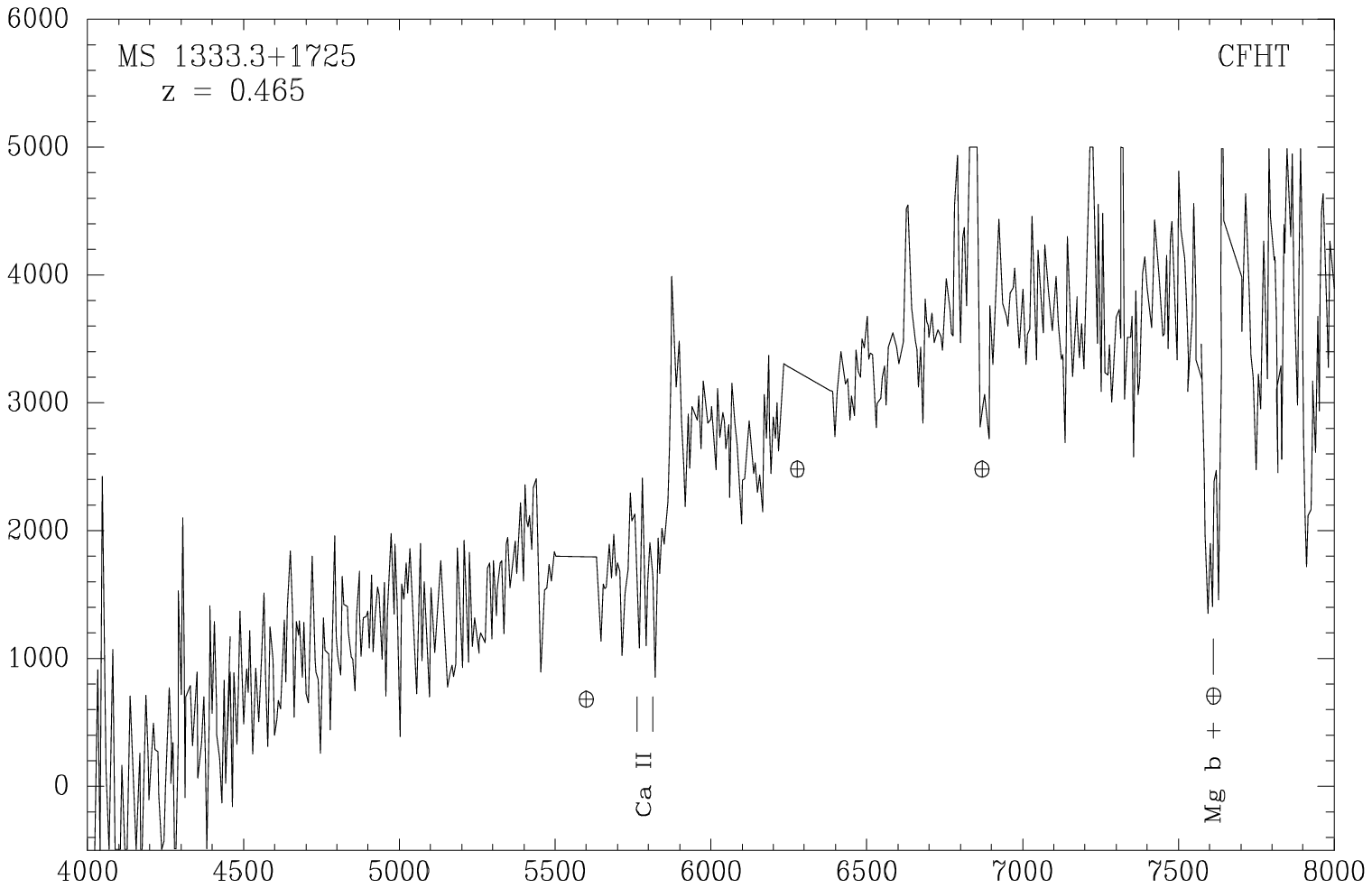}
\end{figure}
\begin{figure}
\plottwo{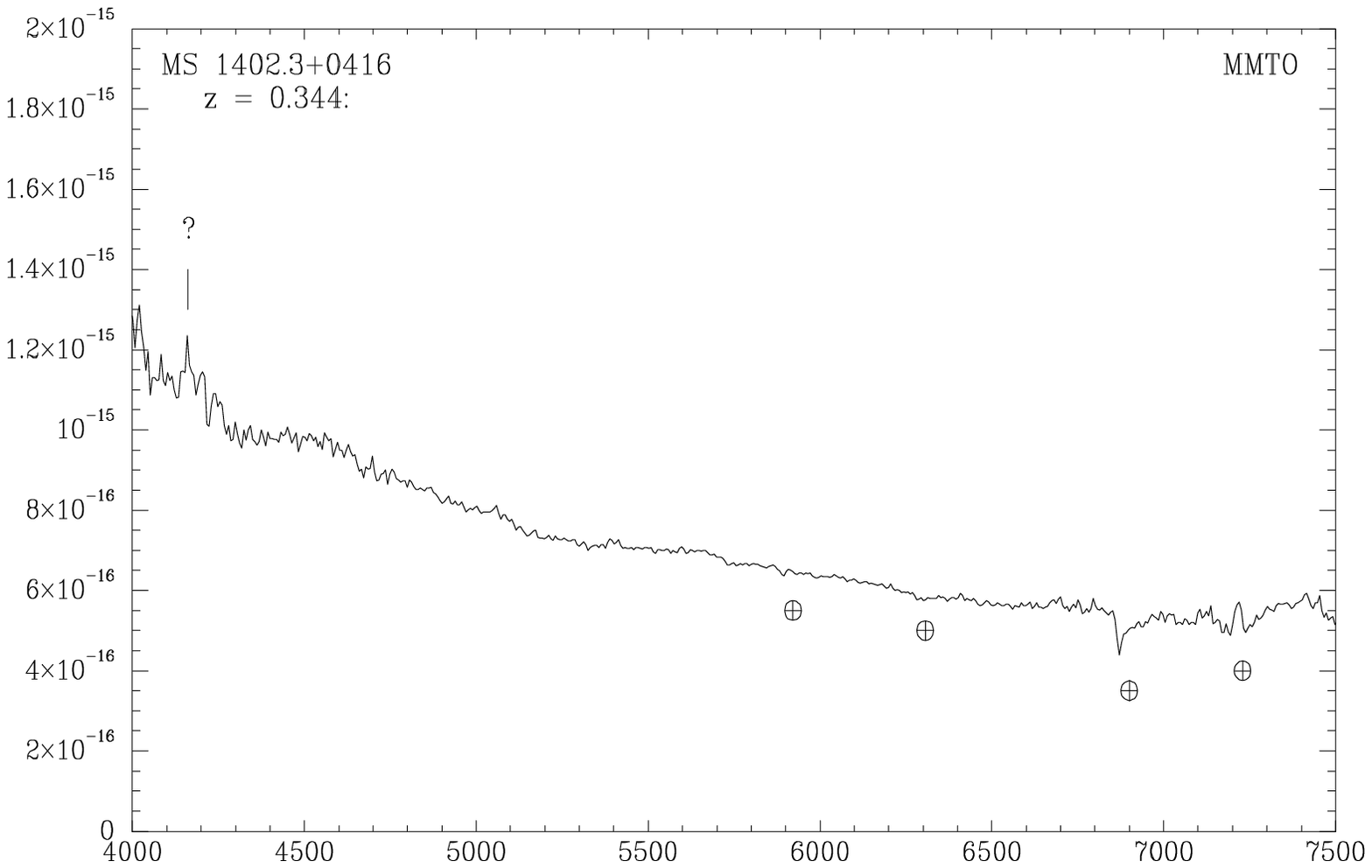}{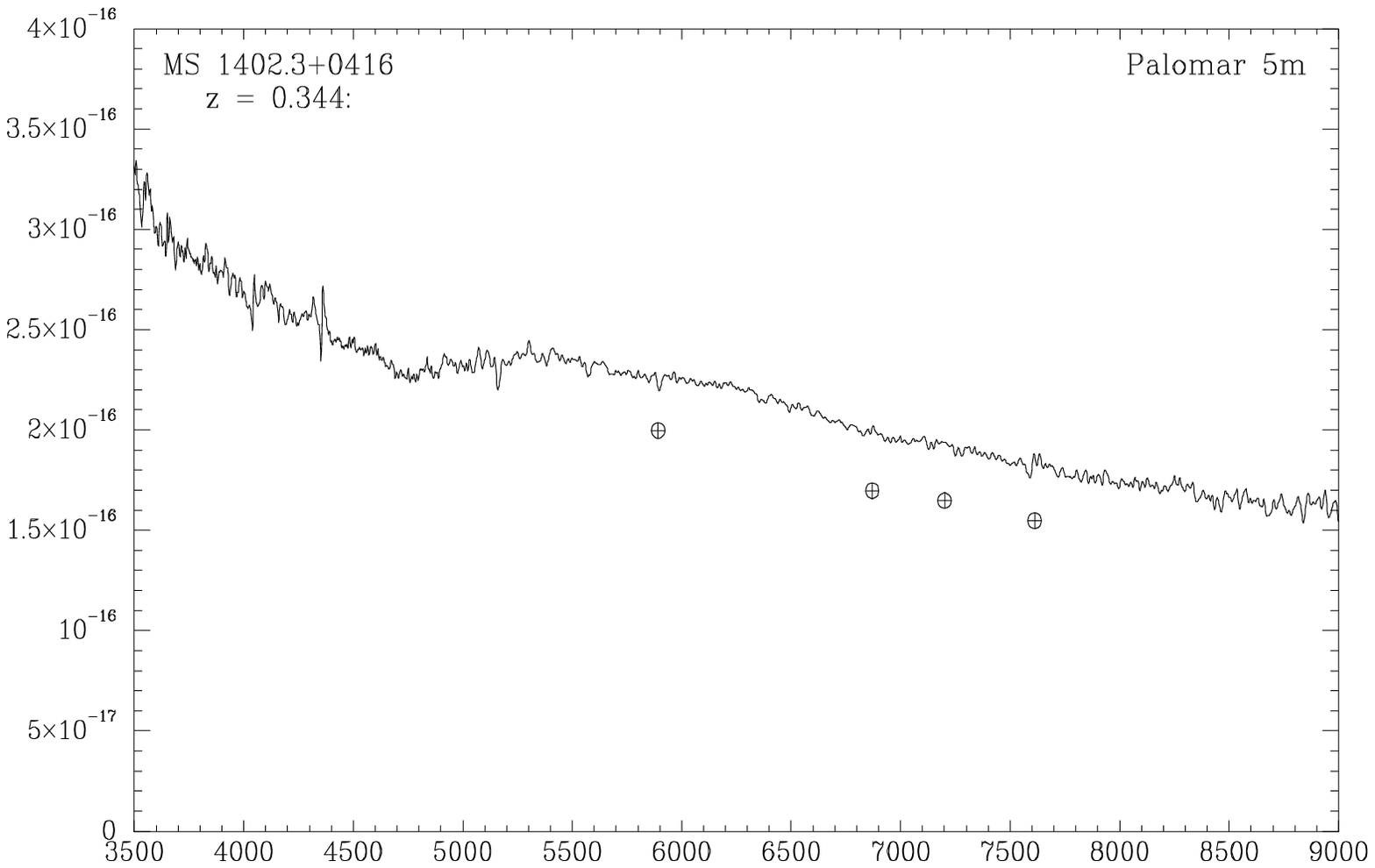}
\end{figure}
\begin{figure}
\plottwo{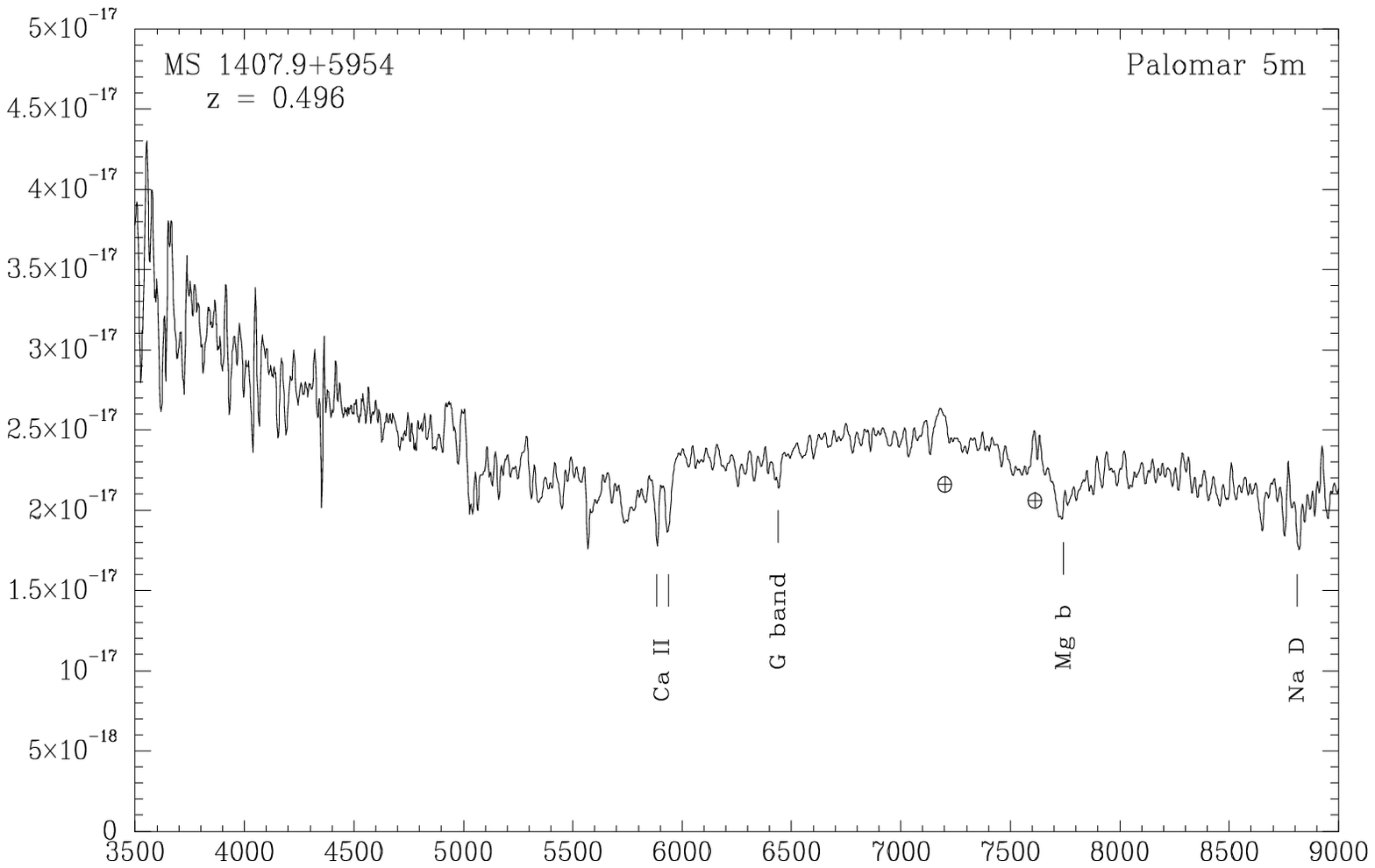}{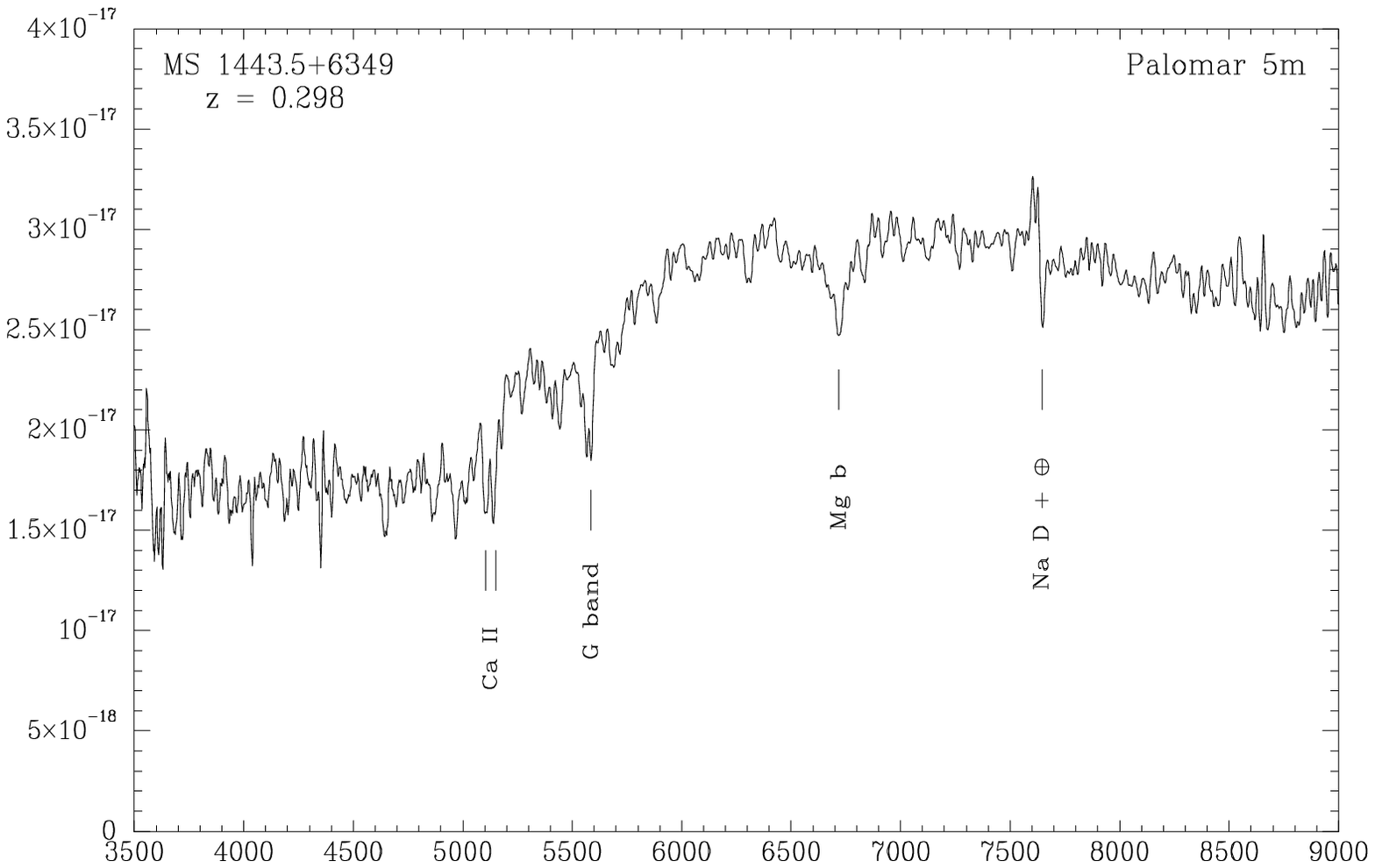}
\end{figure}
\begin{figure}
\plottwo{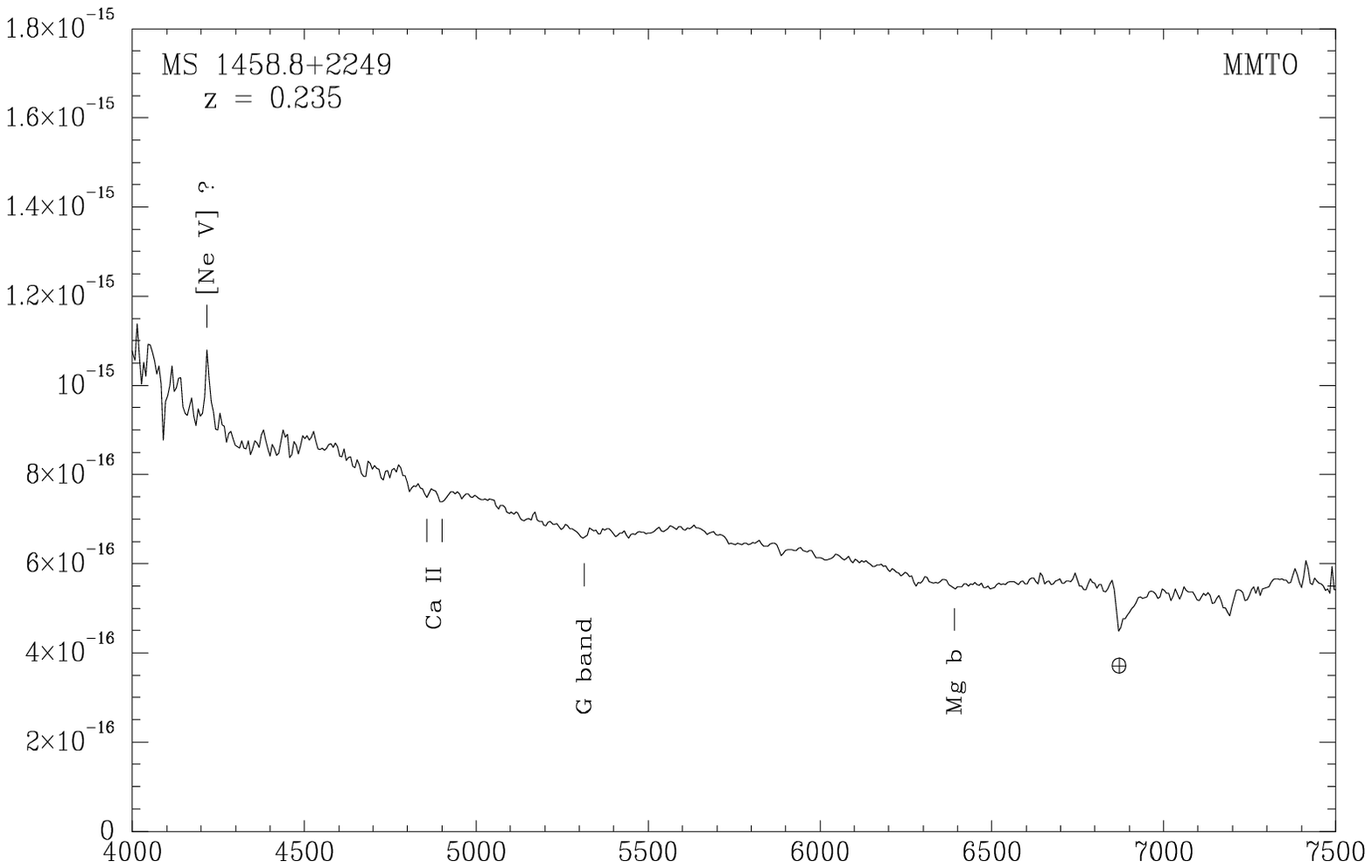}{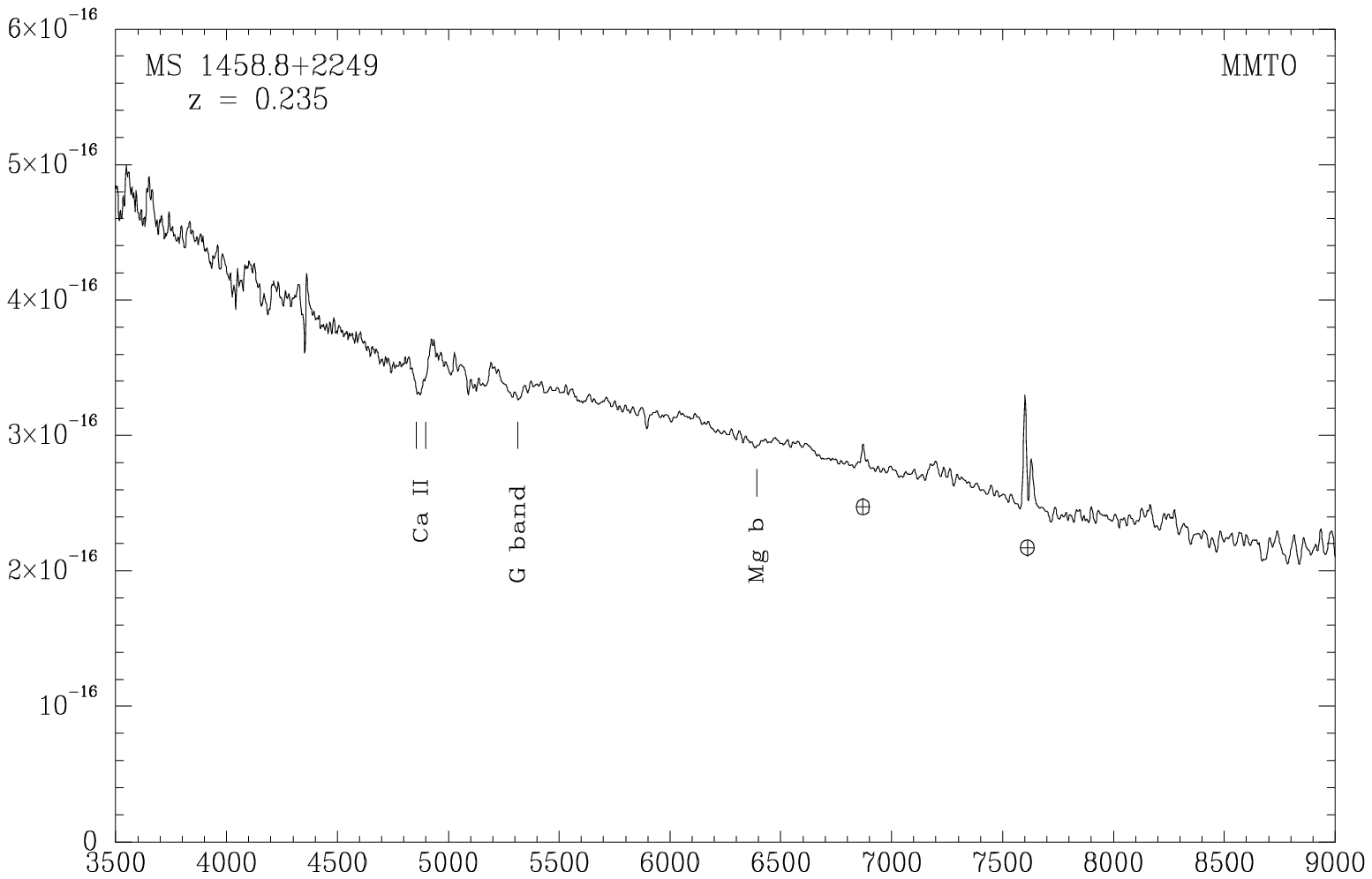}
\end{figure}

\begin{figure}
\plottwo{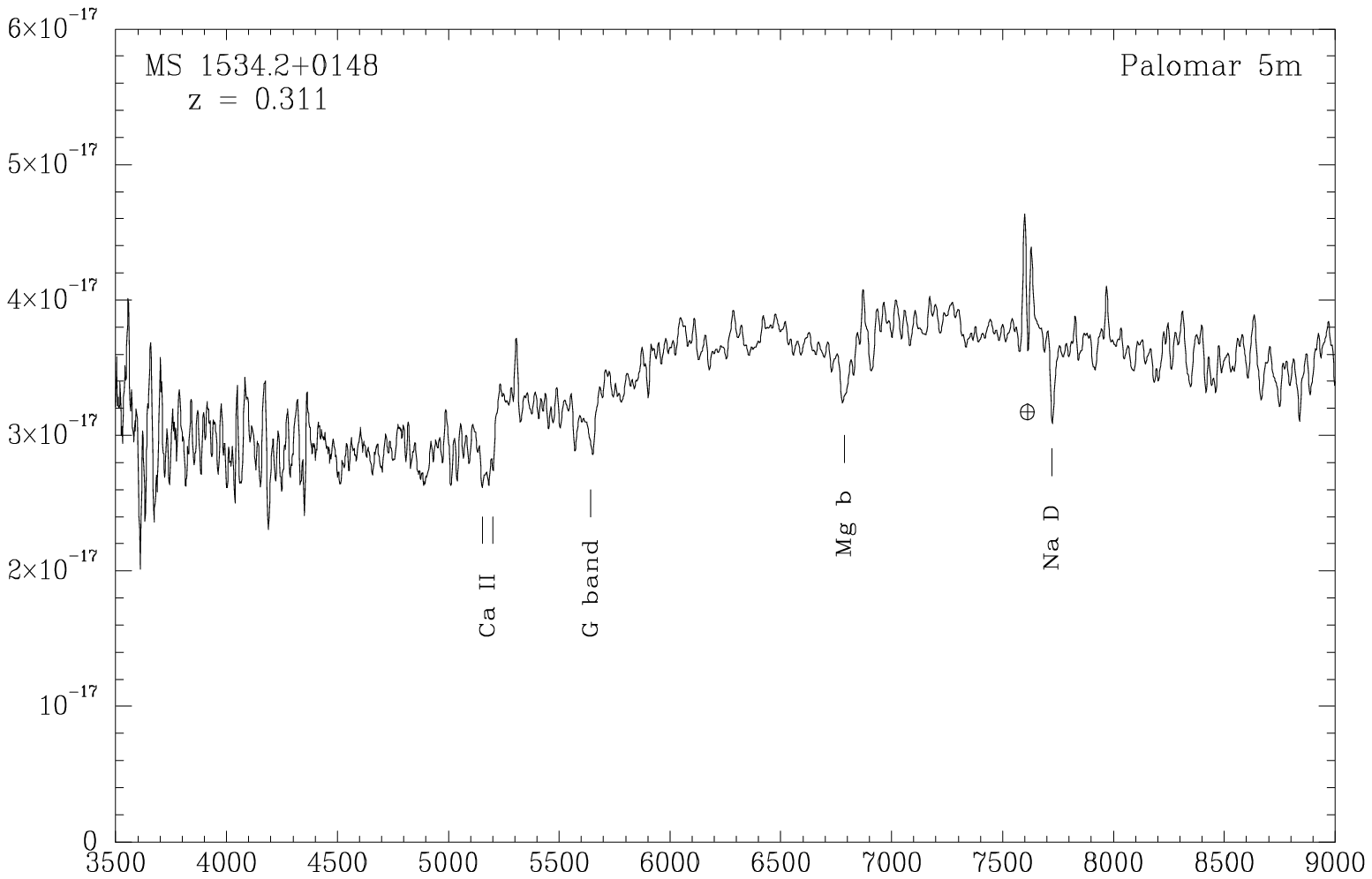}{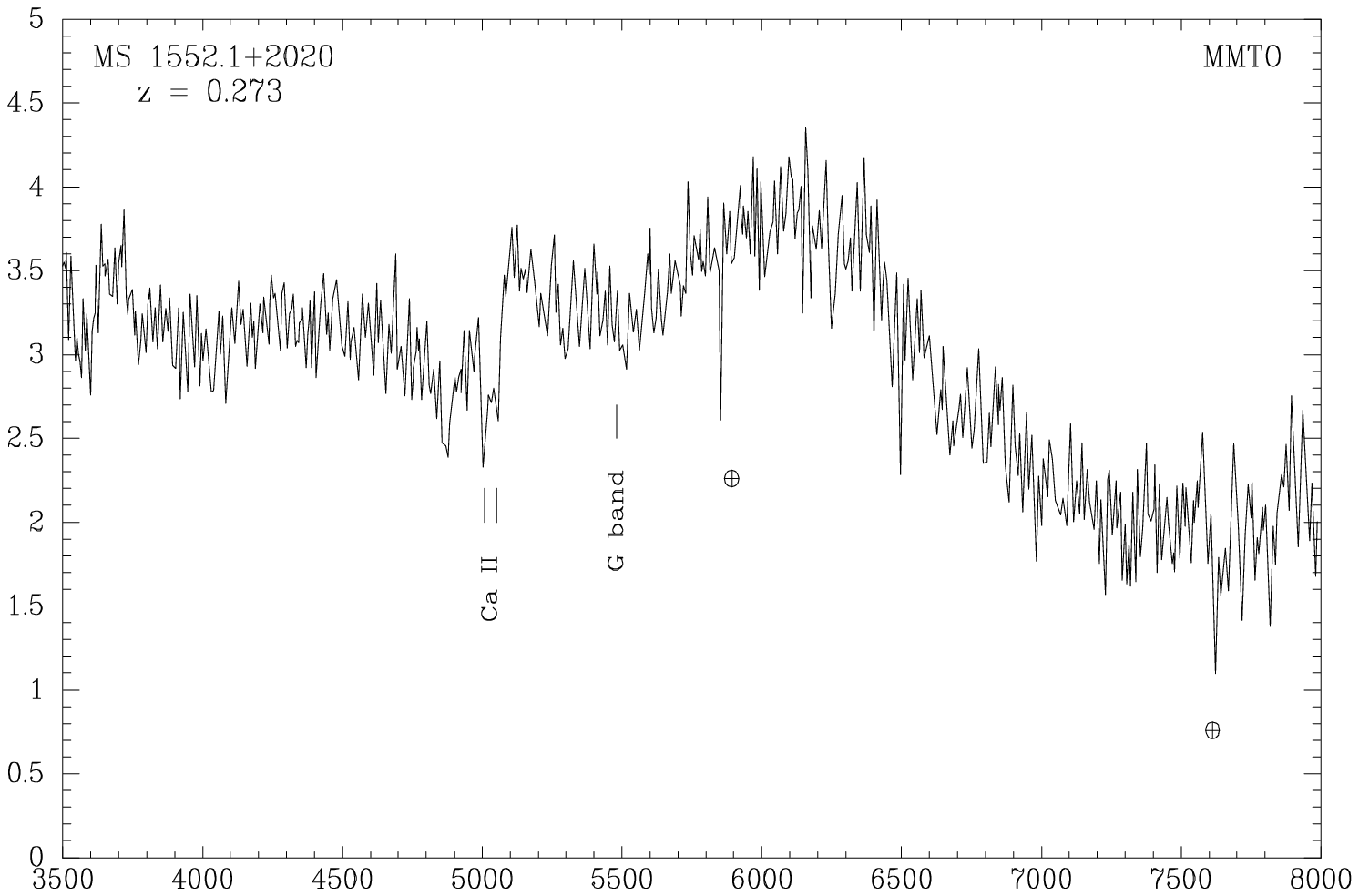}
\end{figure}
\begin{figure}
\plottwo{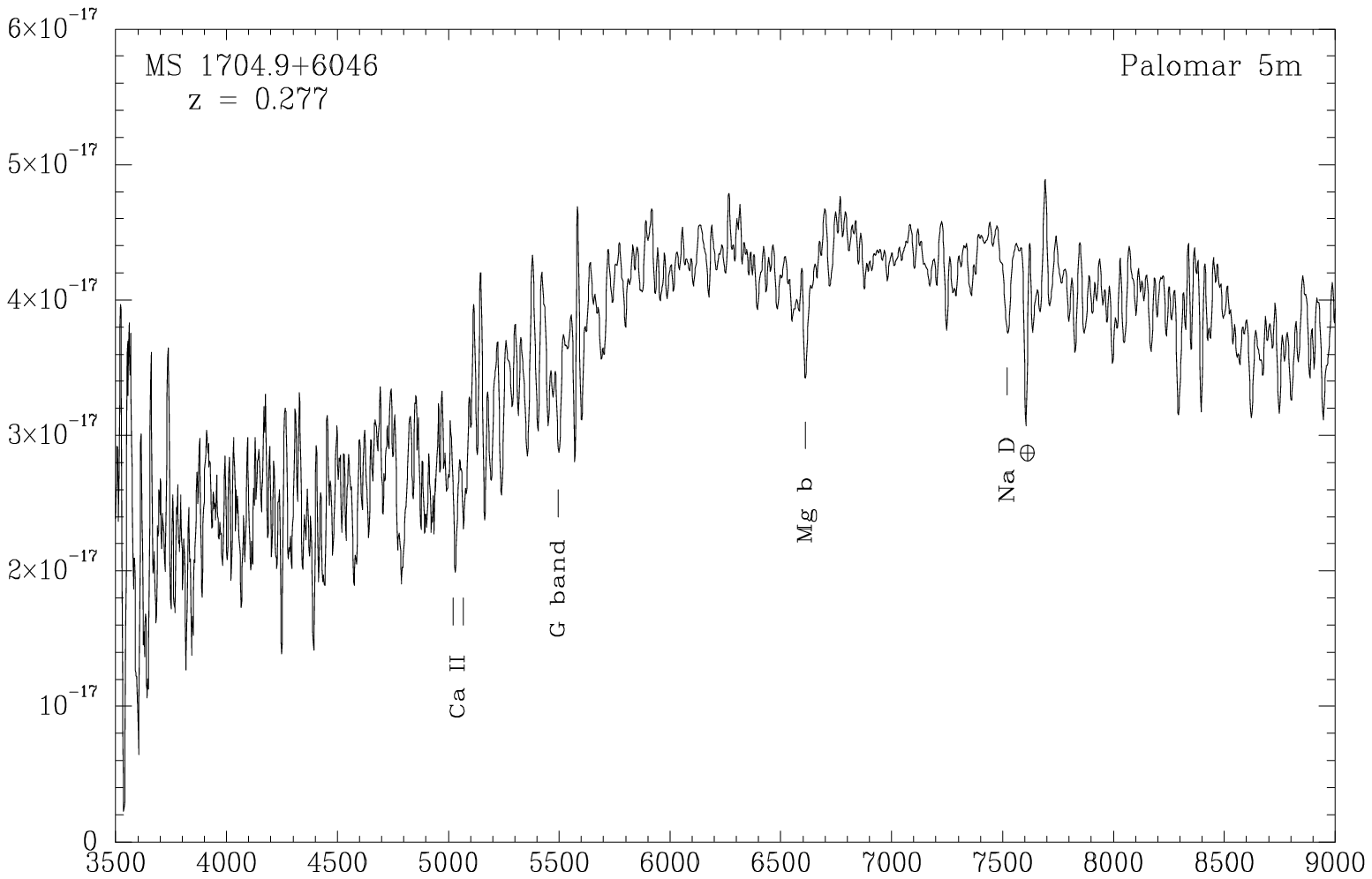}{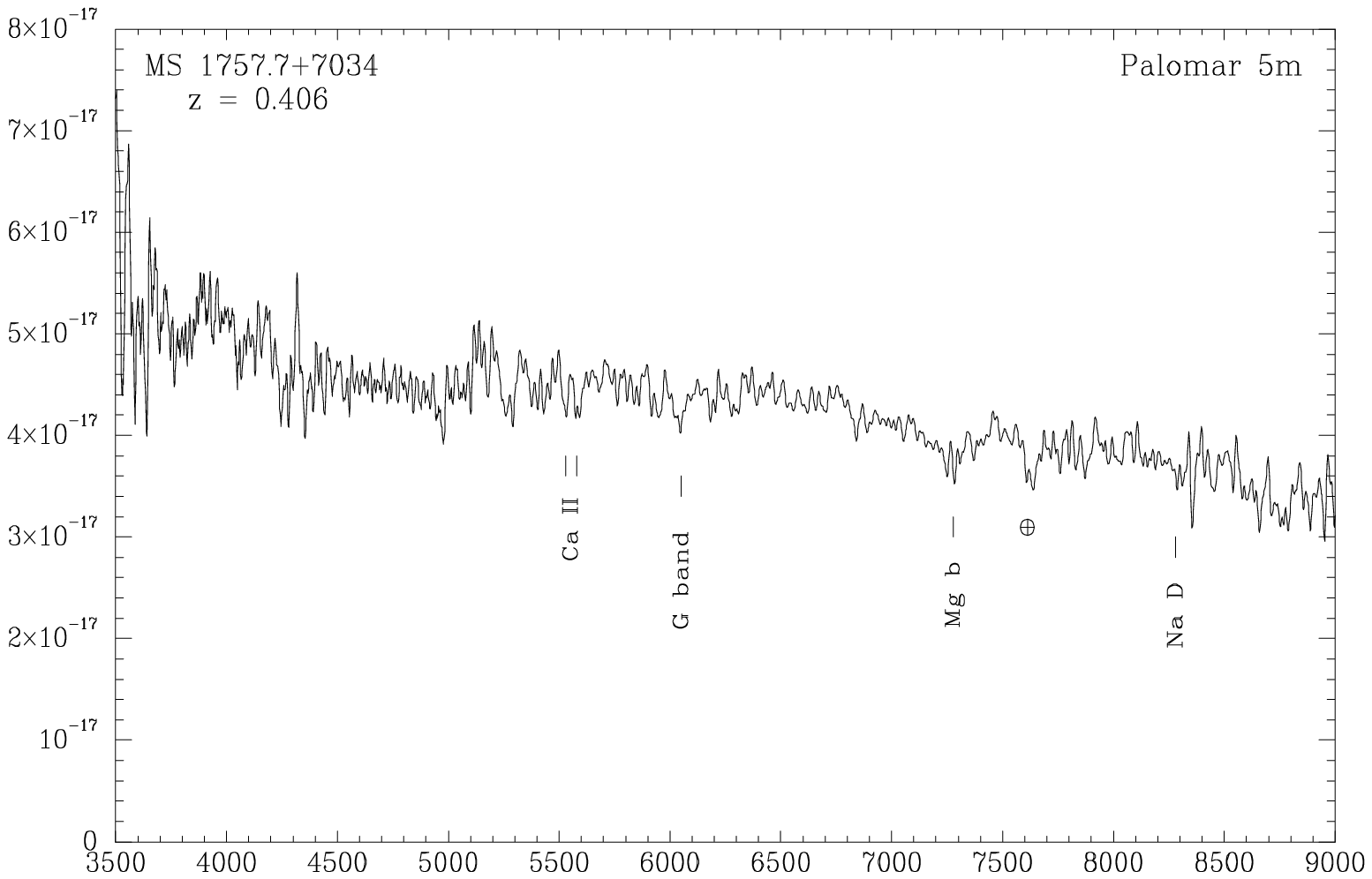}
\end{figure}
\begin{figure}
\plottwo{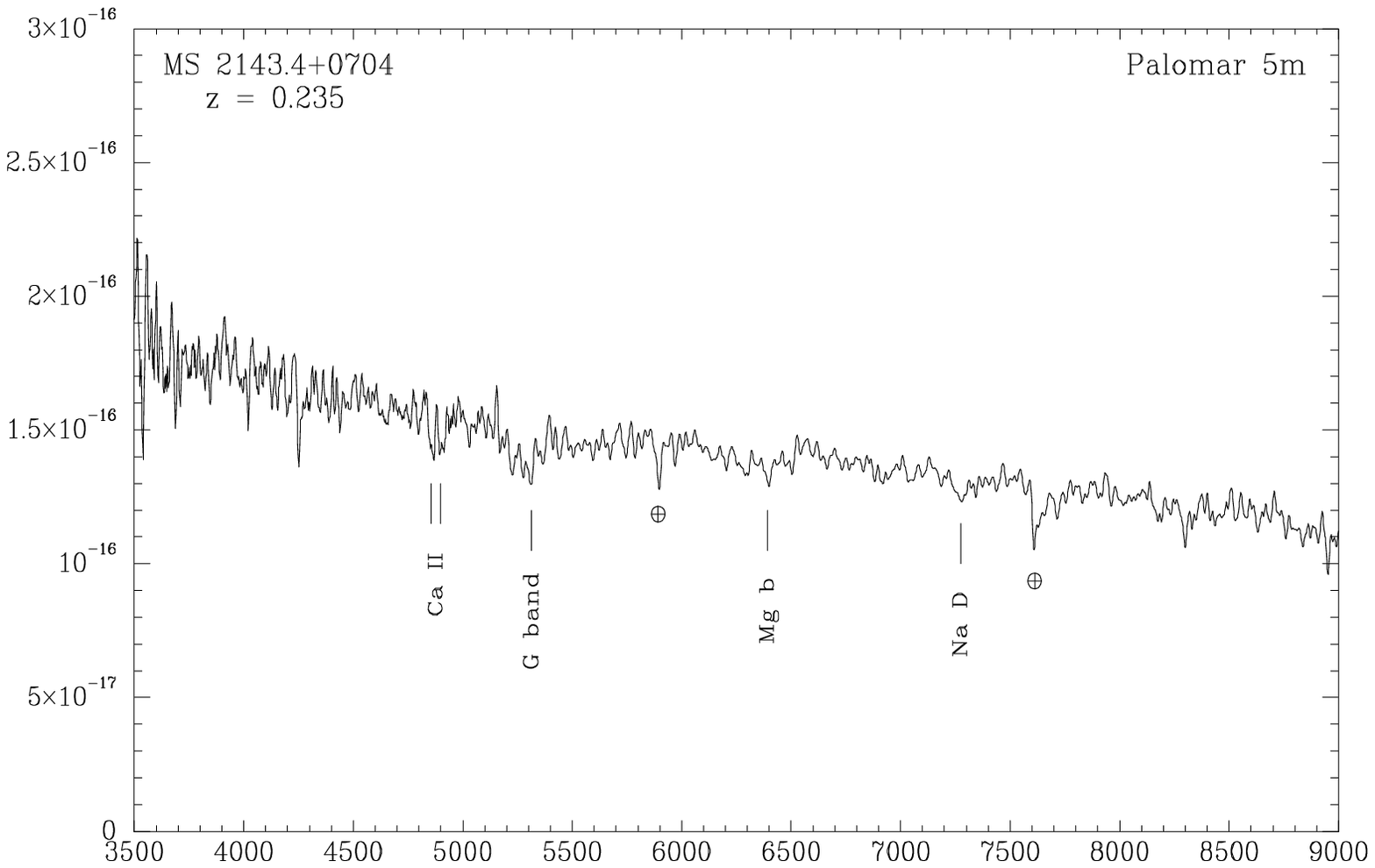}{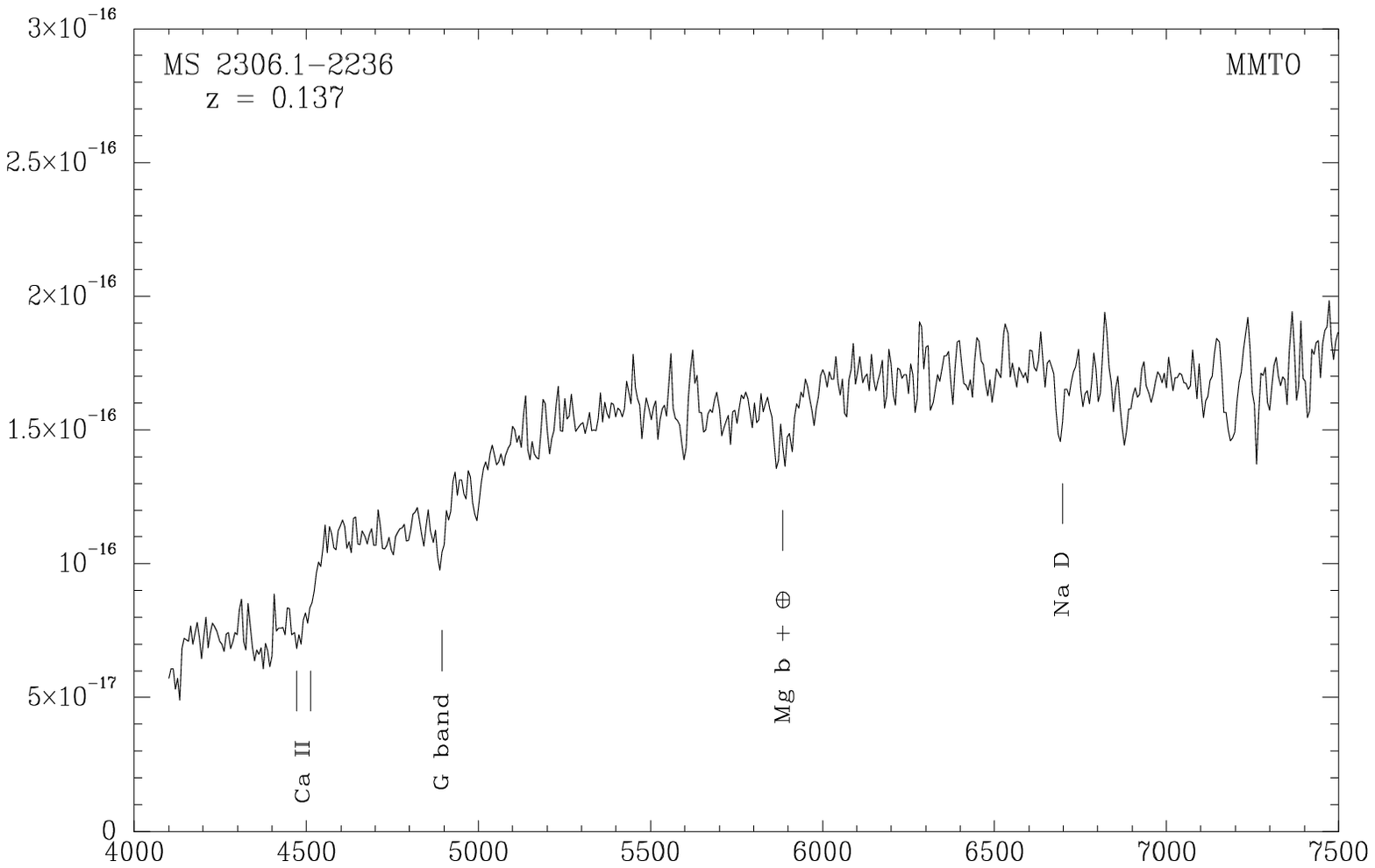}
\end{figure}
\begin{figure}
\plottwo{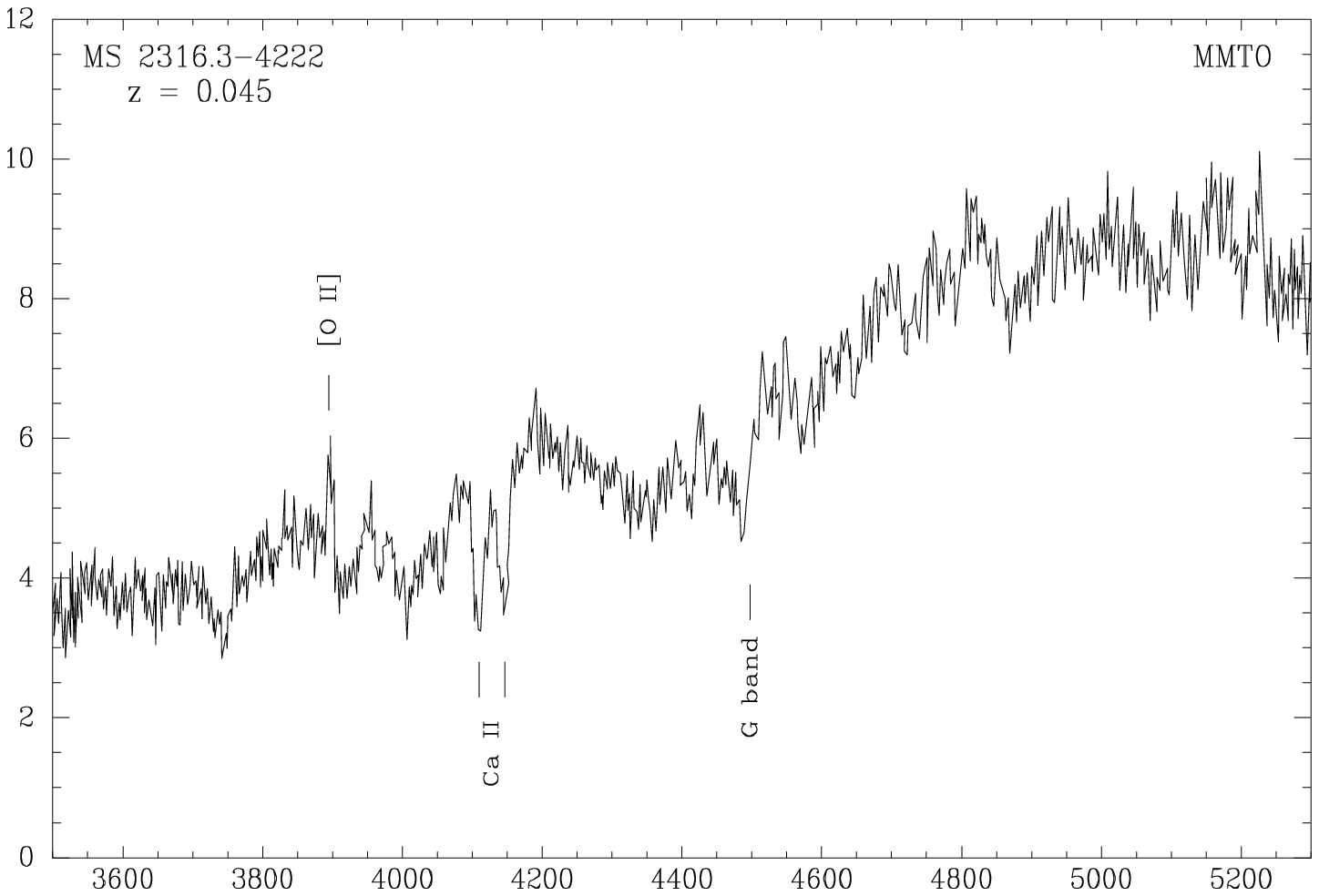}{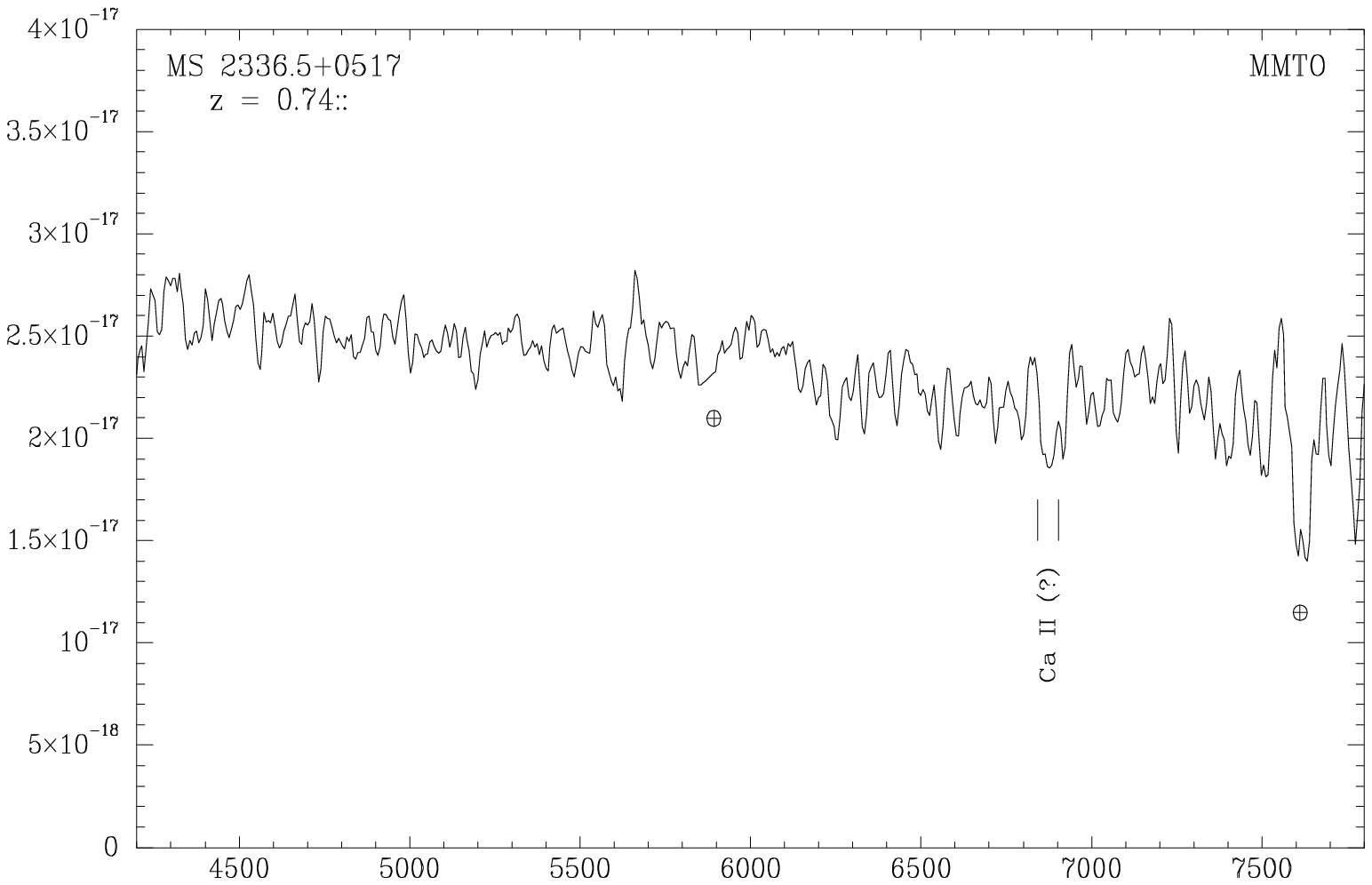}
\end{figure}

\begin{figure}
\plottwo{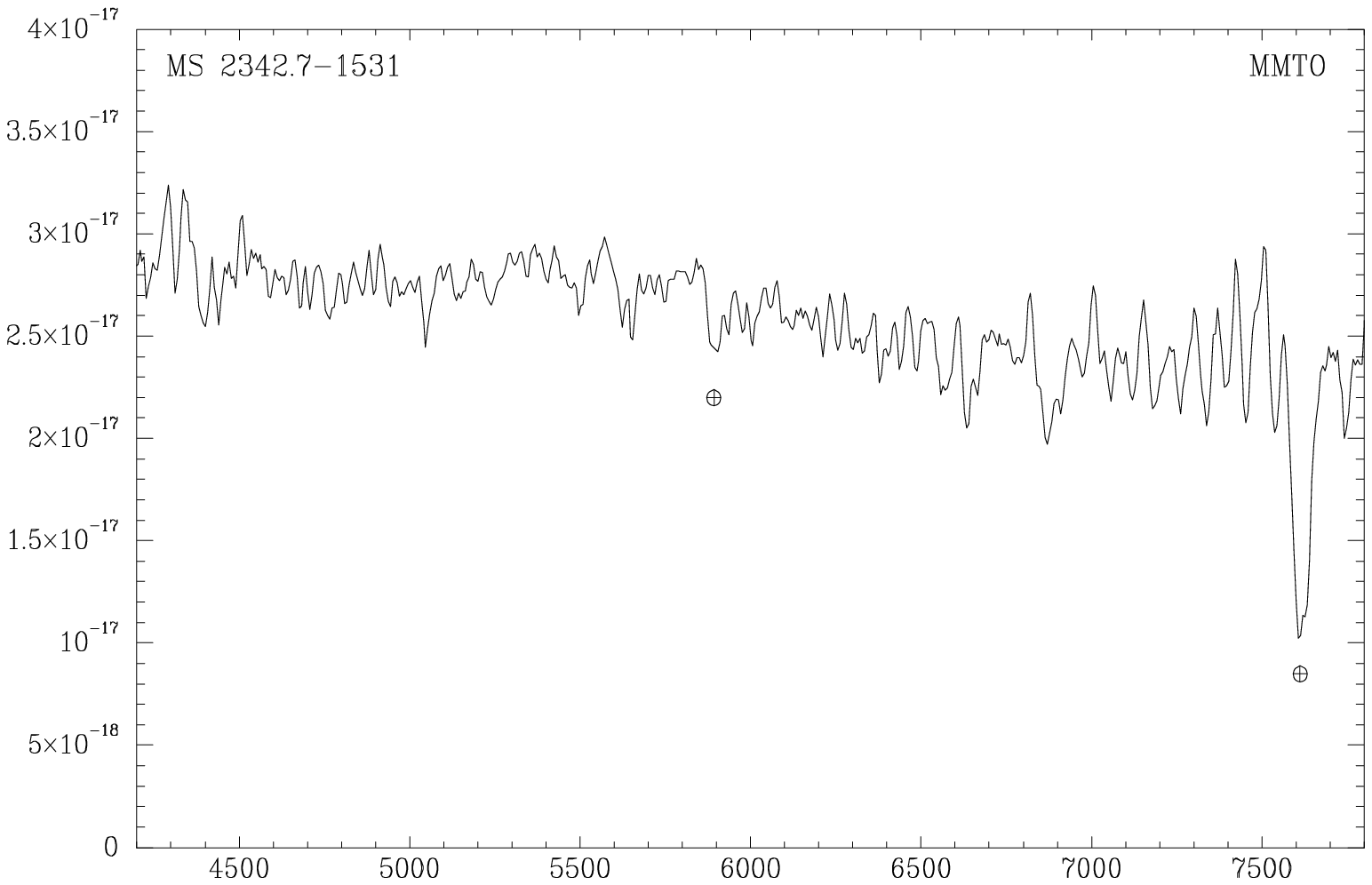}{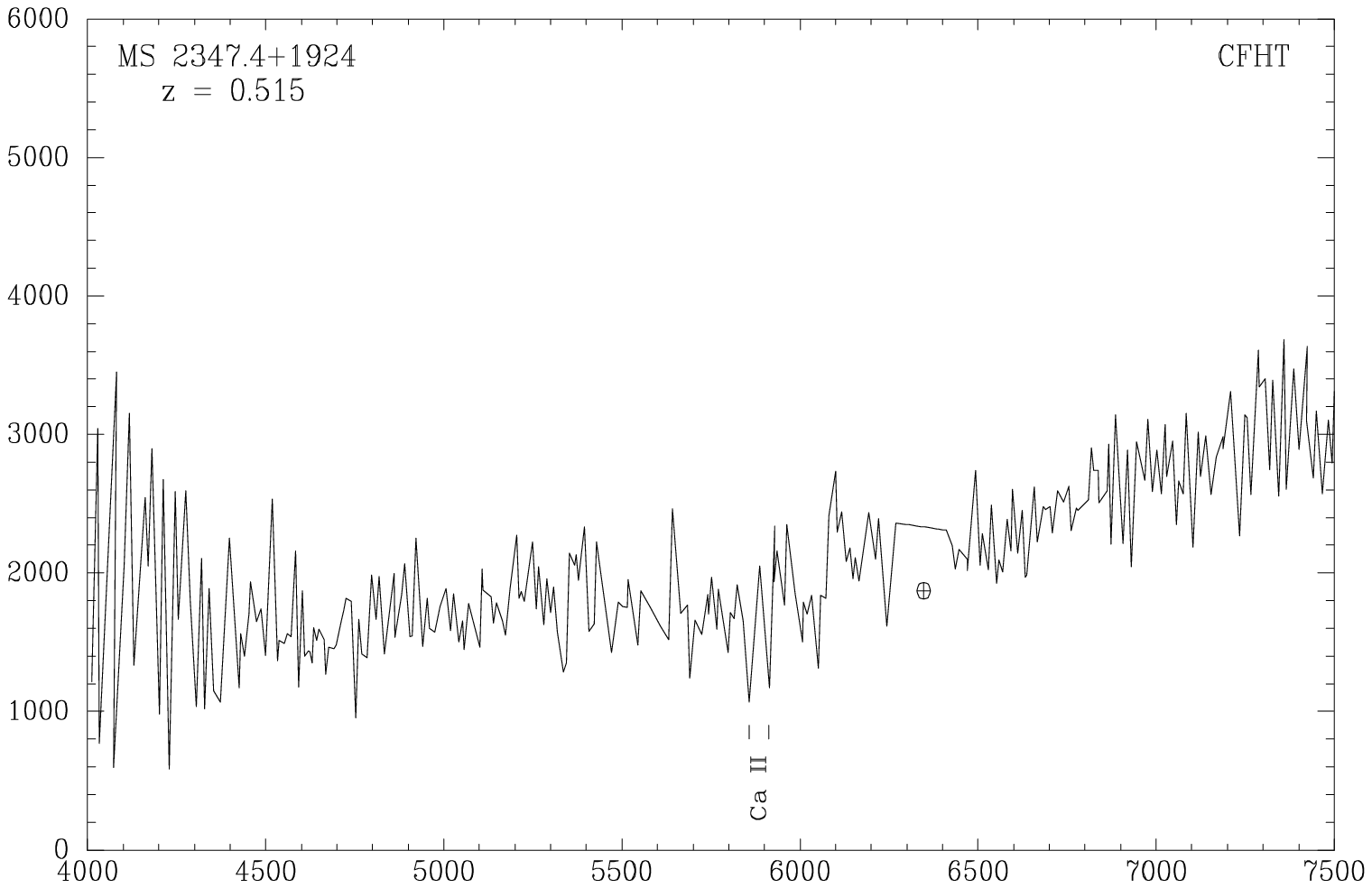}
\end{figure}
\clearpage
\begin{figure}
\plottwo{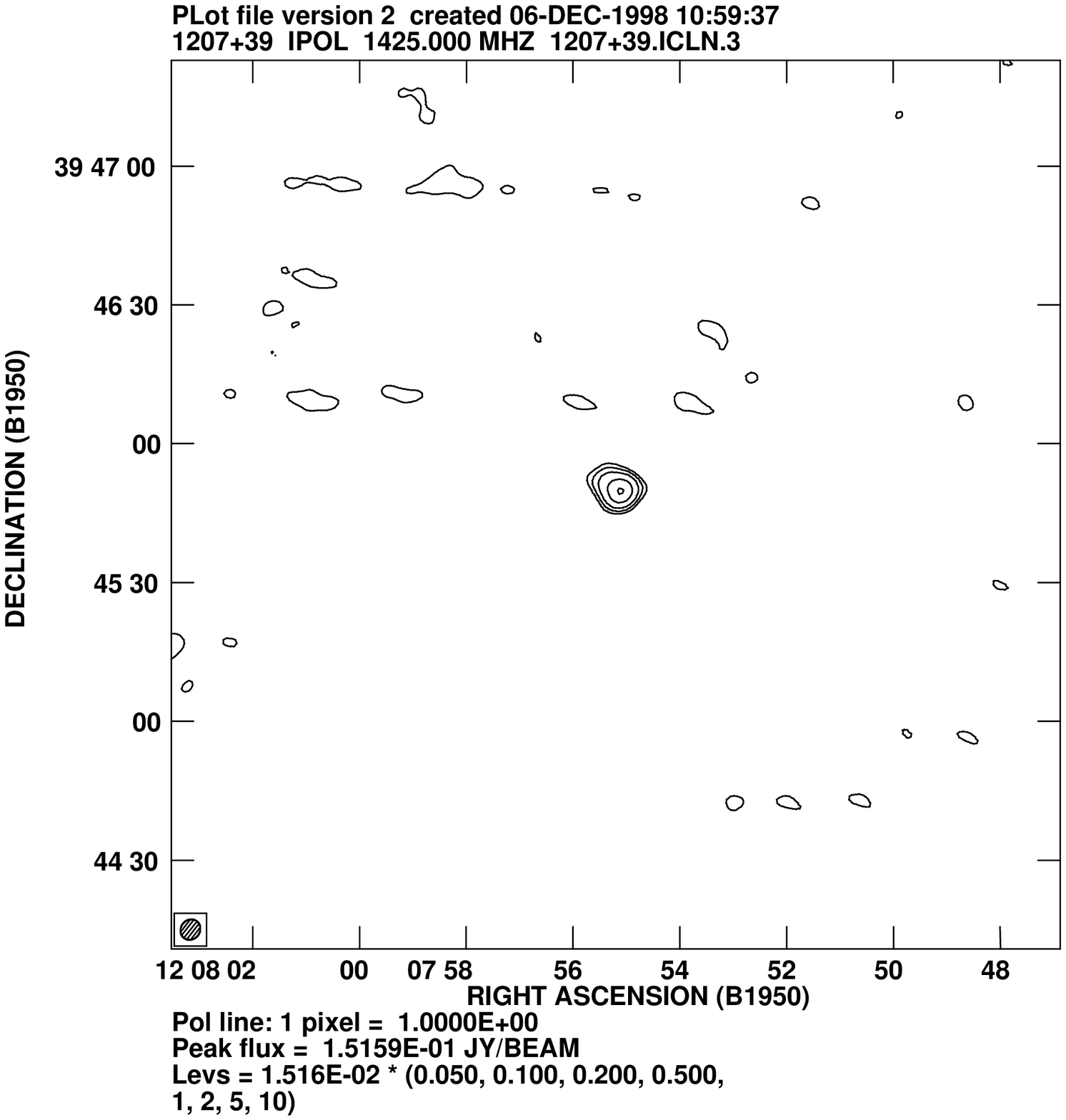}{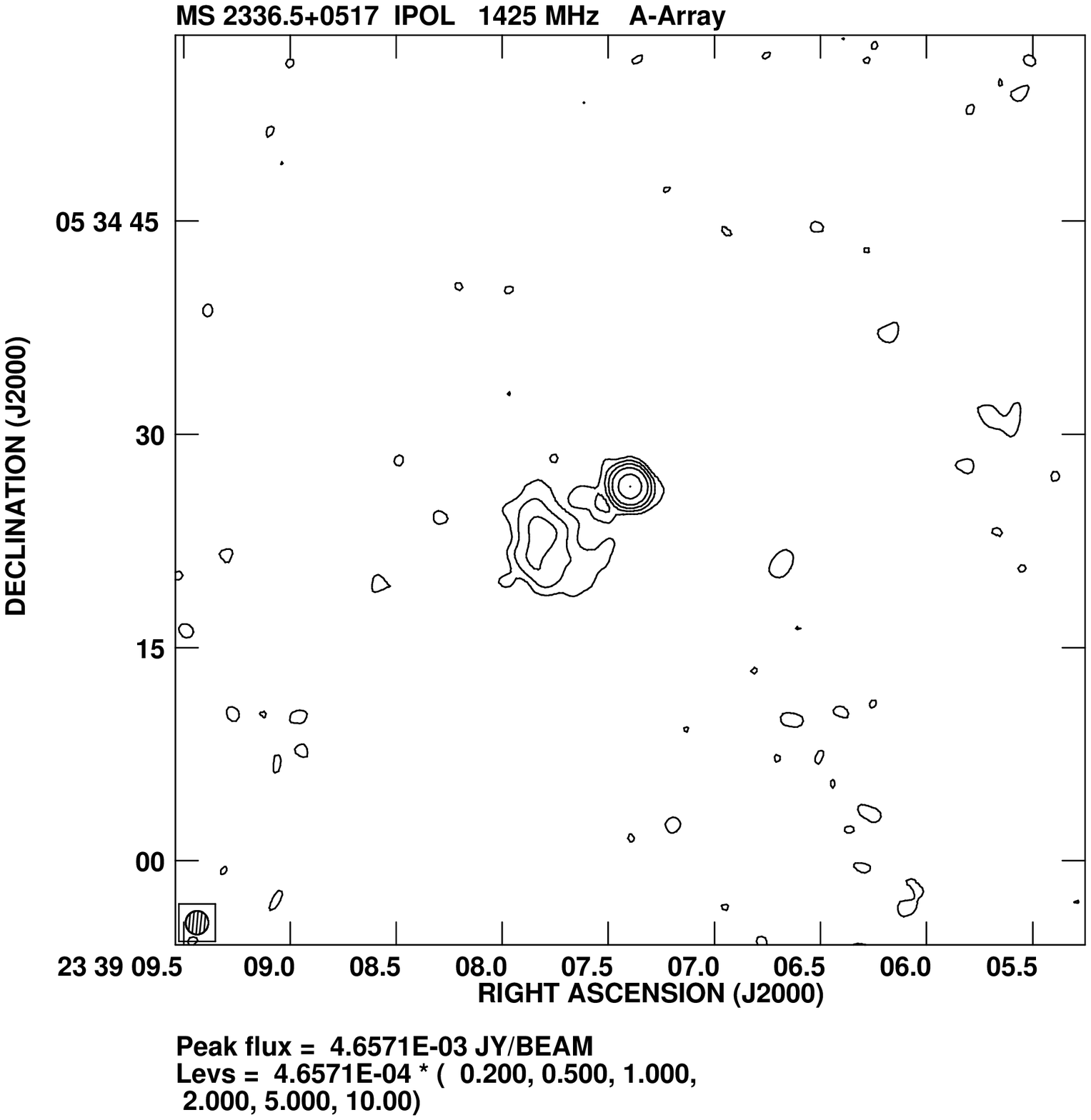} \\
\plotfiddle{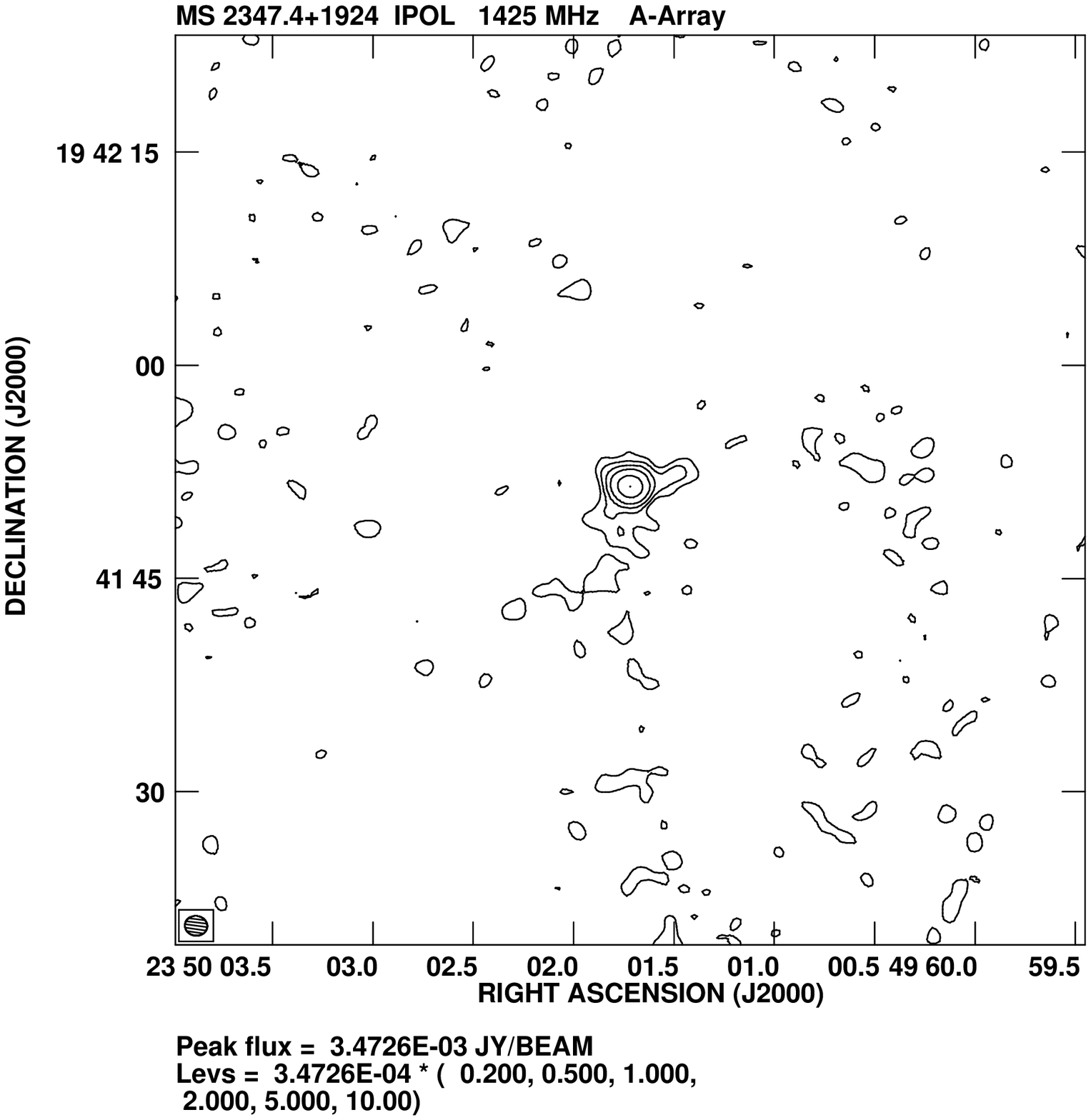}{3.5in}{0.}{45.}{45.}{-150}{-50}
\caption{Deep VLA 20cm A-array radio maps of MS 2336.5+0517 
and MS 2347.4+1924 and a ``snapshot" of MS 1207.9+3945. 
The B-array map of MS 1050.7+4946 is shown in RSP99.  The base level of each map is
set at the 2$\sigma$ noise level.  The beam is shown in the lower left corner. 
The radio properties are given in Table~\ref{tbl-4.5}. 
\label{fig-4.3}}
\end{figure}

\begin{figure}
\plotone{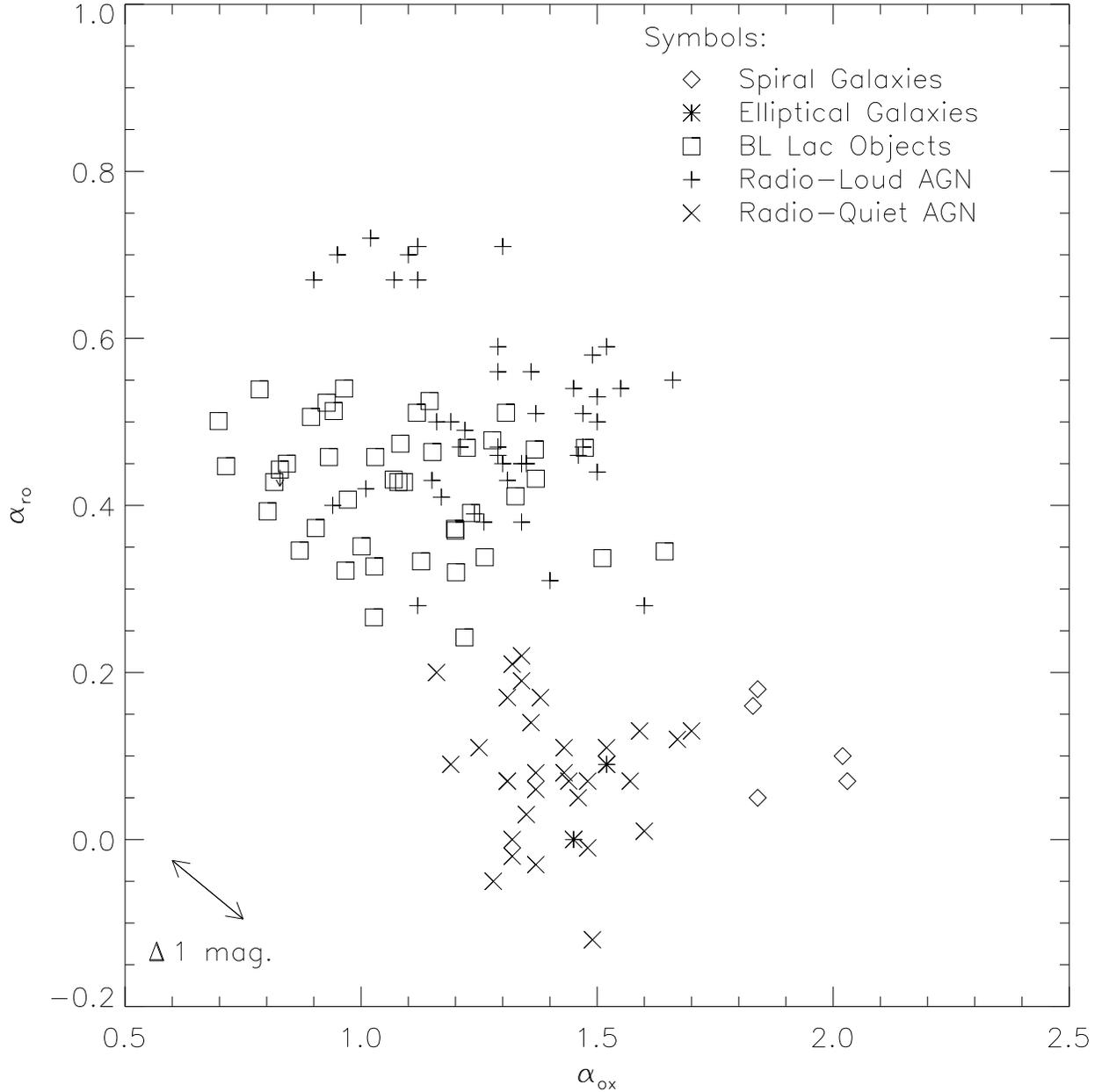}
\caption{Overall 
spectral energy distributions for extragalactic EMSS classes.  The axes are
the radio-to-optical and X-ray-to-optical spectral indices ($\alpha_{ro}$ and
$\alpha_{ox}$ respectively) as defined in Stocke et al. (1991).  Radio-loud objects
are at the top of the plot and X-ray-loud objects are to the left.  The
($\alpha_{ox},\alpha_{ro}$) values for the BL Lac objects have been corrected to account for
the presence of the host elliptical galaxy (see \S 4.2 for a discussion).  Data for
non BL-Lac classes were obtained from Stocke et al. (1991).
The diagonal arrow in the lower left shows the effect of 1 magnitude of optical
variability, which is typical of EMSS XBLs (Jannuzi et al. 1993).
\label{fig-4.2}}
\end{figure}

\begin{figure}
\plotone{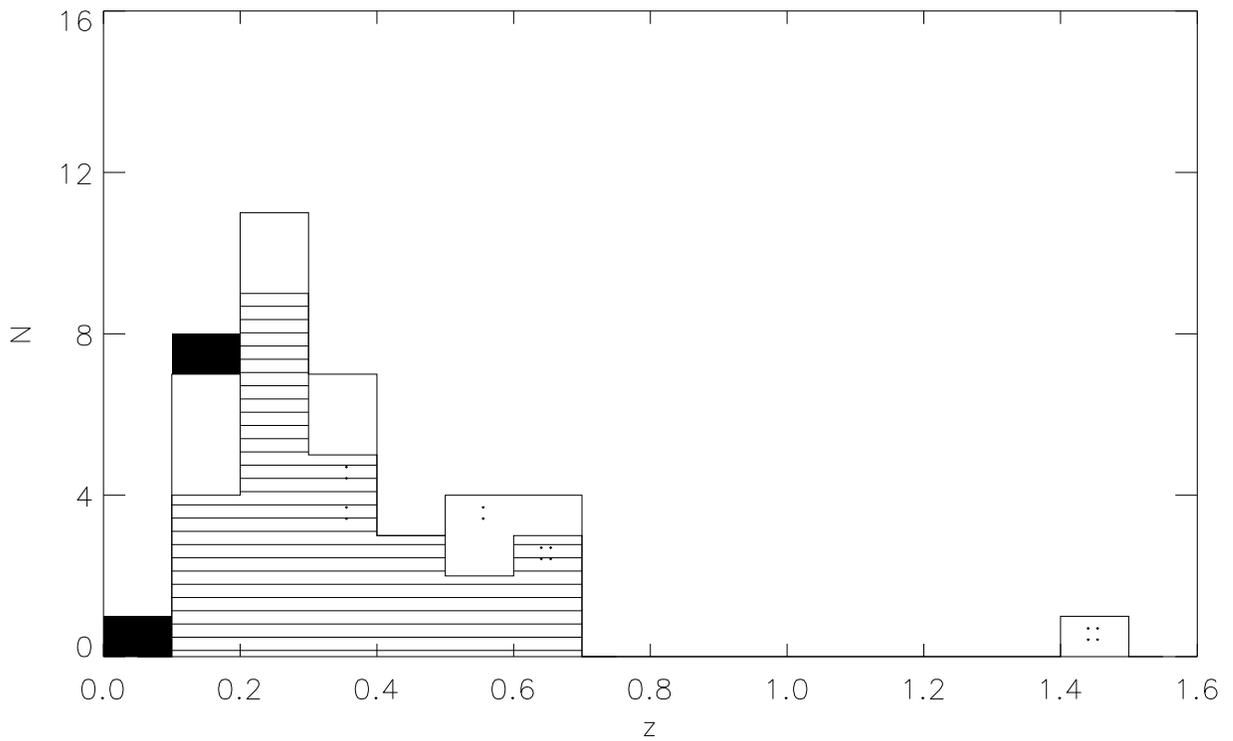}
\caption{Distribution of redshifts for the EMSS BL Lac sample.  Tentative redshifts
are marked with a colon; the highly uncertain redshifts for MS 1256.3+0151 and MS
2336.5+0517 are marked with two colons.  The hatched area represents the M91
sample, which is a subsample of the D40.  The solid areas represent
two of the three objects which are not part of either sample (the third does not
have a measured redshift).  Note that no XBLs with $z<0.1$ are seen in the M91 or
D40 samples.  With the exception of MS 1256.3+0151, which is identified with a
very uncertain redshift of $z=1.4$, no XBLs are seen at $z> 0.7$.
\label{fig-4.7}}
\end{figure}
\clearpage

\begin{figure}
\plottwo{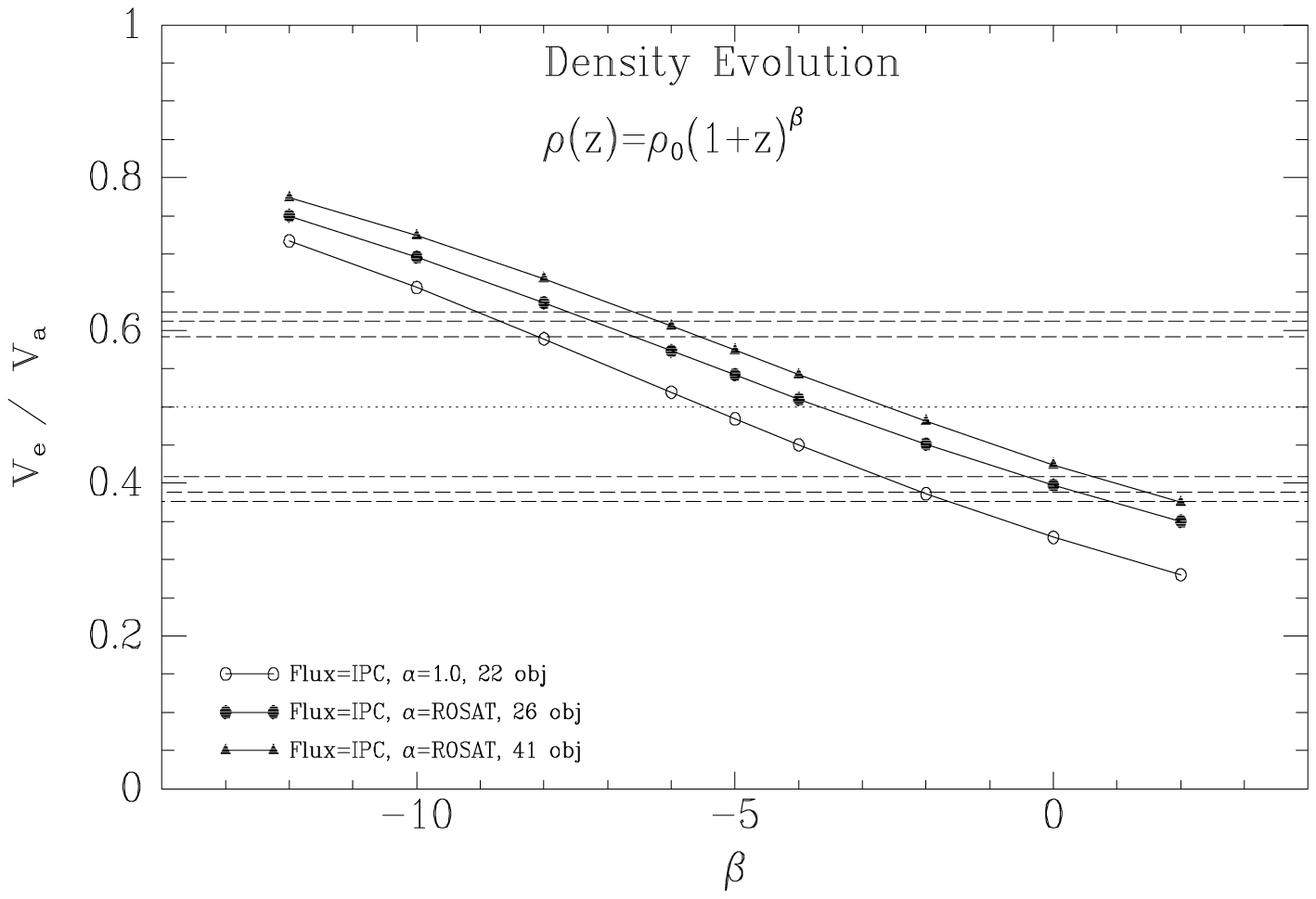}{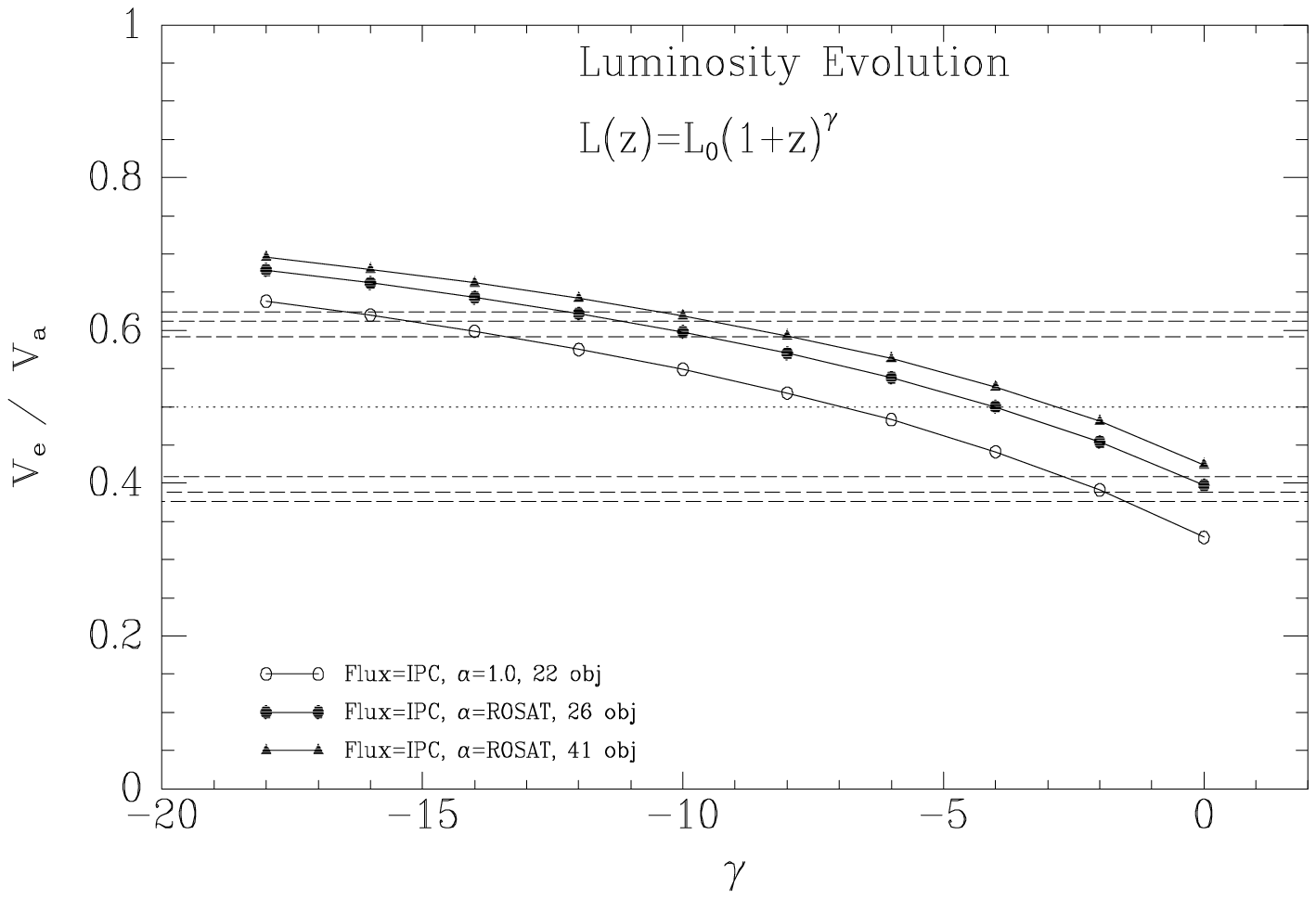}
\caption{The variation of $V_e/V_a$ with density evolution parameter $\beta$ (left)
and with luminosity evolution parameter $\gamma$ (right).  Open circles represent
values first derived for the M91 sample of 22 objects in M91; and solid circles
represent the updated complete M91 sample of 26 objects.  Triangles represent the
new D40 sample of 41 objects.  Solid lines fit the data.  The short dashed line is
for \vvmax\ $= 0.5$.
\label{fig-4.4}}
\end{figure}

\begin{figure}
\plottwo{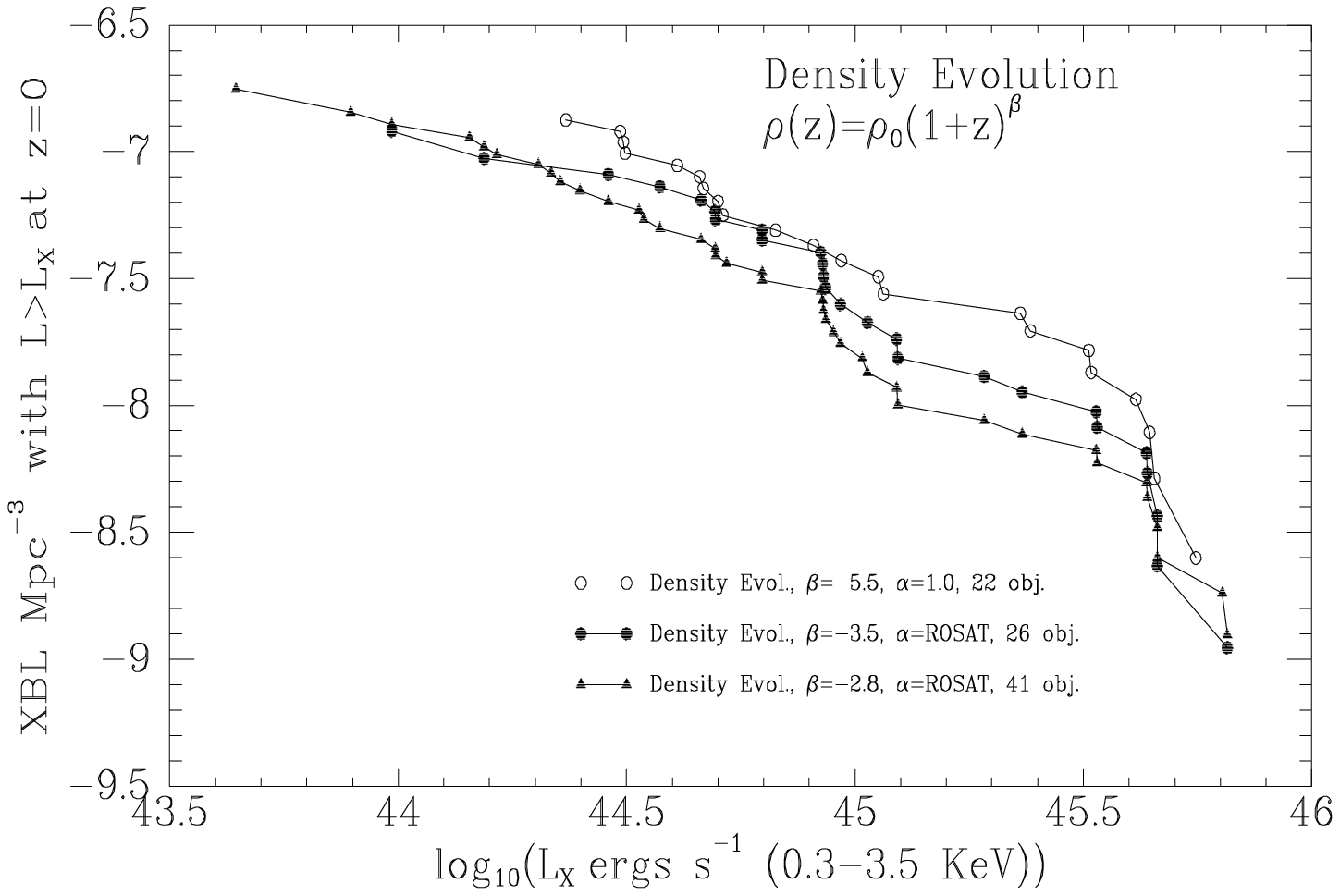}{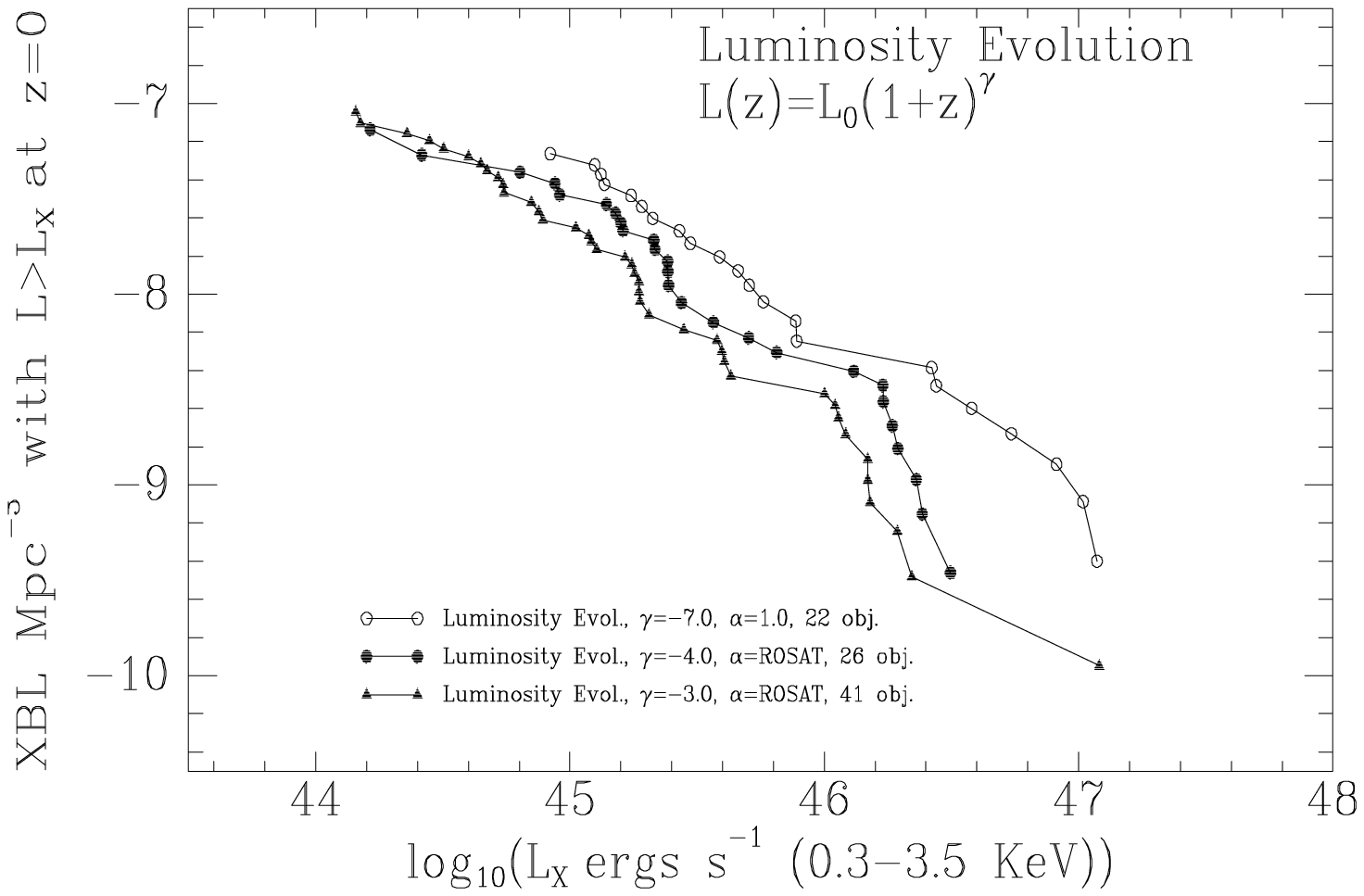}
\caption{The EMSS BL Lac integral X-ray luminosity function for density evolution
(left) and luminosity evolution (right).  Each data point represents an object in
the sample.  Open circles represent values first derived for the M91 sample of 22
objects; and solid circles represent the updated complete M91 sample of 26
objects.  Triangles represent the new D40 sample of 41 objects.
\label{fig-4.5}}
\end{figure}
\clearpage

\begin{figure}
\plottwo{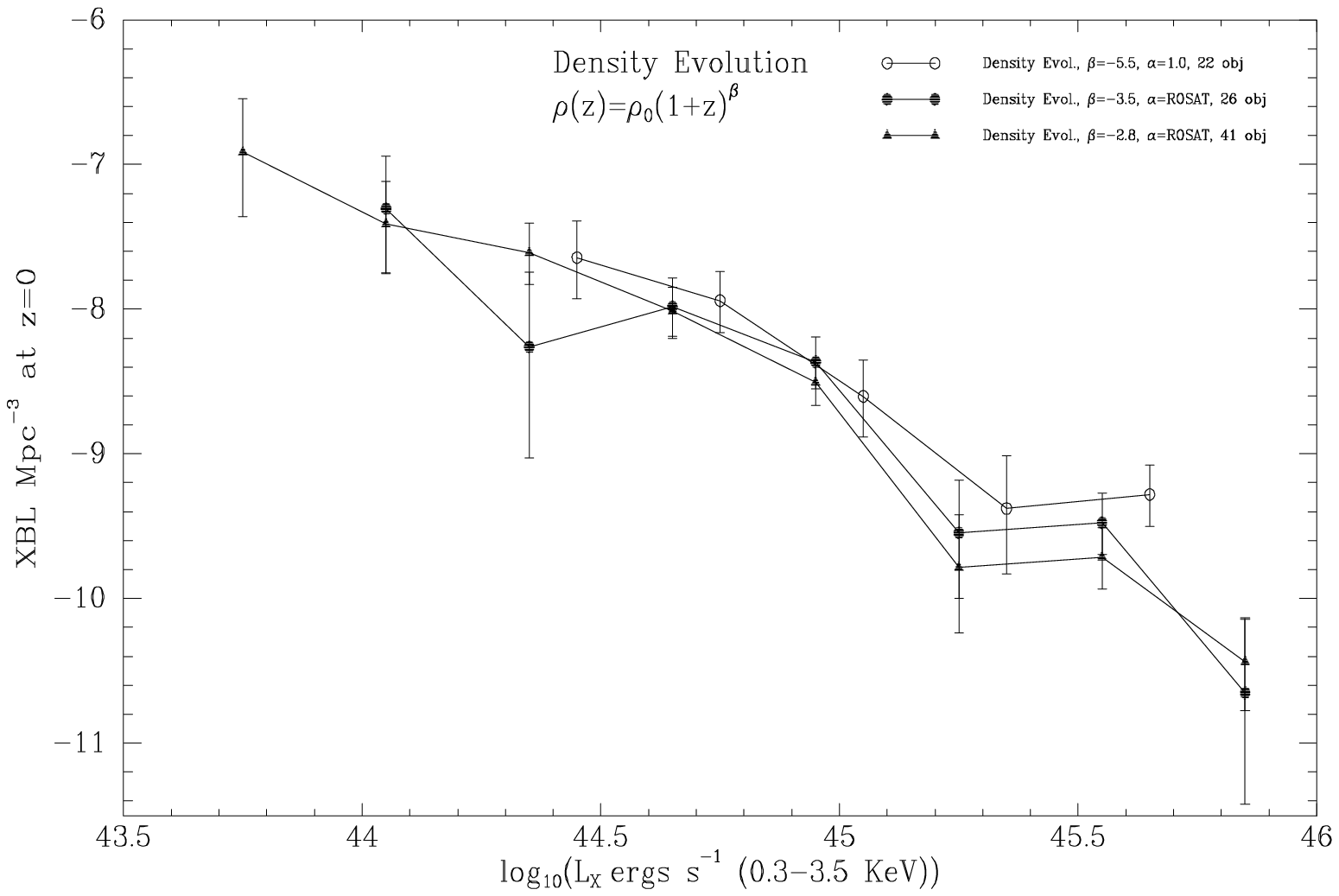}{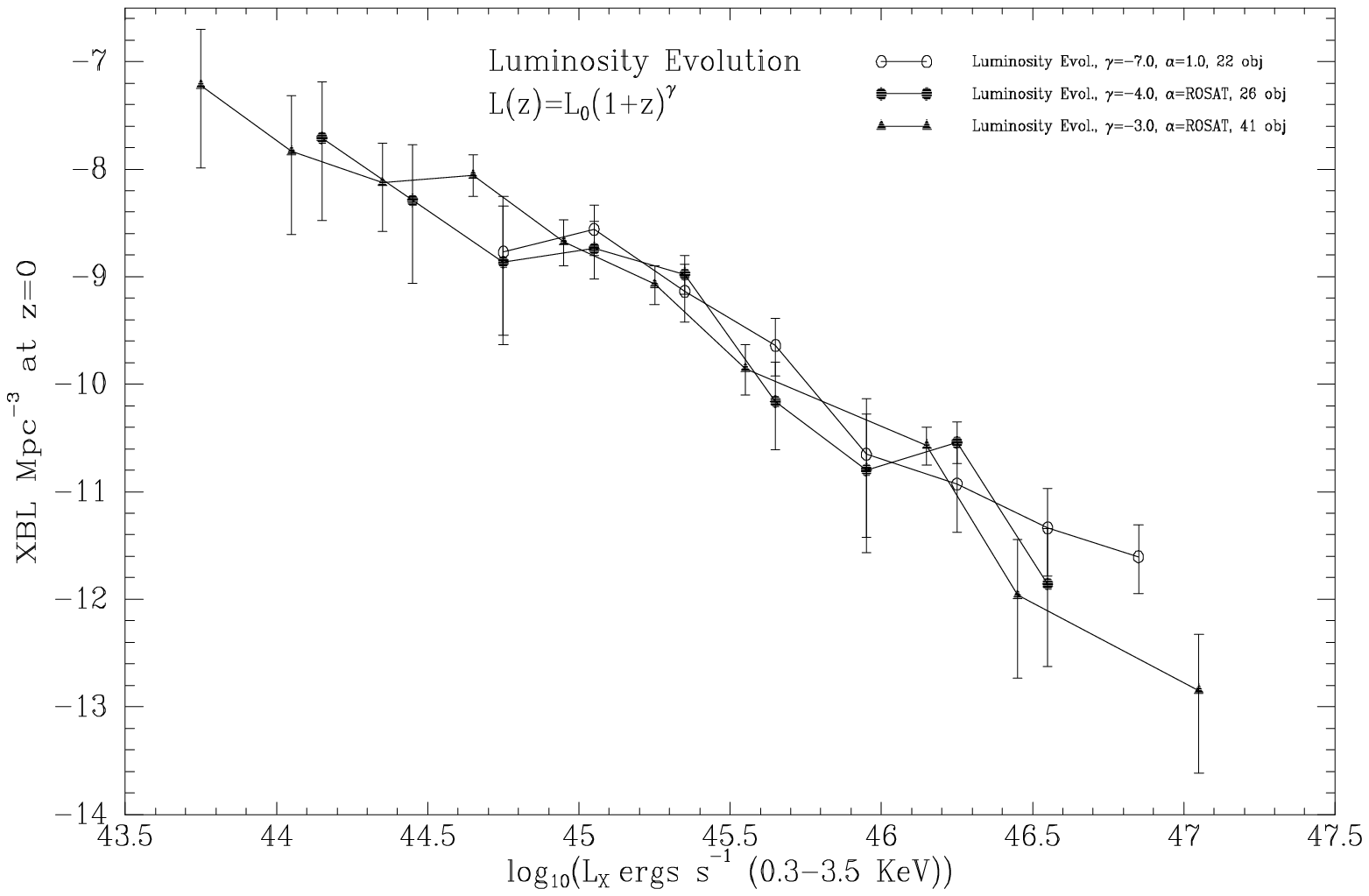}
\caption{The EMSS BL Lac differential X-ray luminosity function for density
evolution (left) and luminosity evolution (right).  Each bin is $10^{44}$ erg
s$^{-1}$ in size; the error bars reflect Poisson statistics based upon the number
of objects within that bin.  Open circles represent values first derived for the M91
sample of 22 objects in M91; and solid circles represent the updated complete
M91 sample of 26 objects.  Triangles represent the new D40 sample.
\label{fig-4.6}}
\end{figure}
\clearpage

\clearpage
\begin{figure}
\plotone{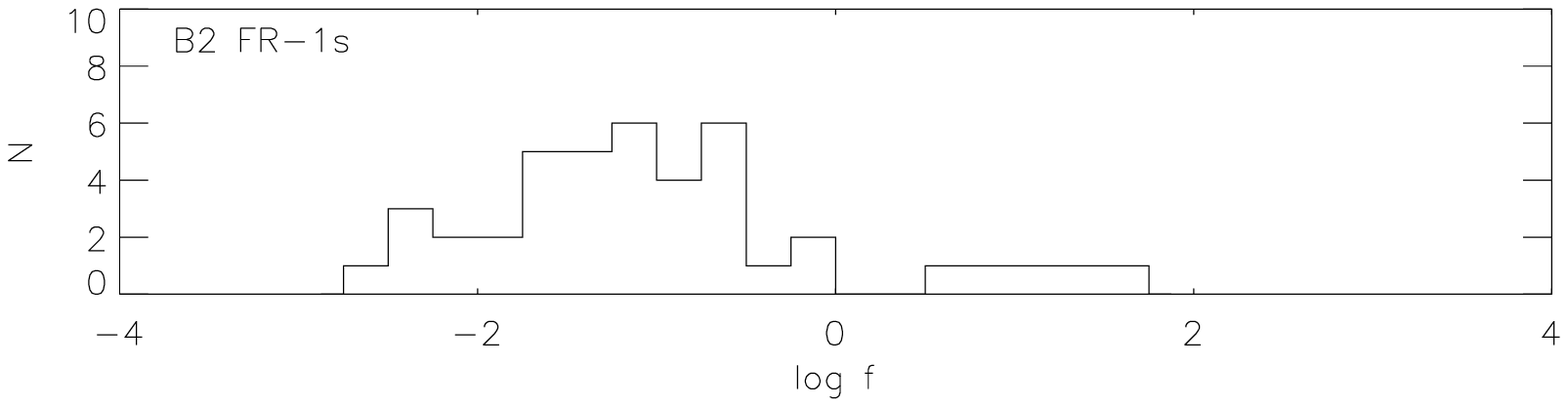} \\
\plotone{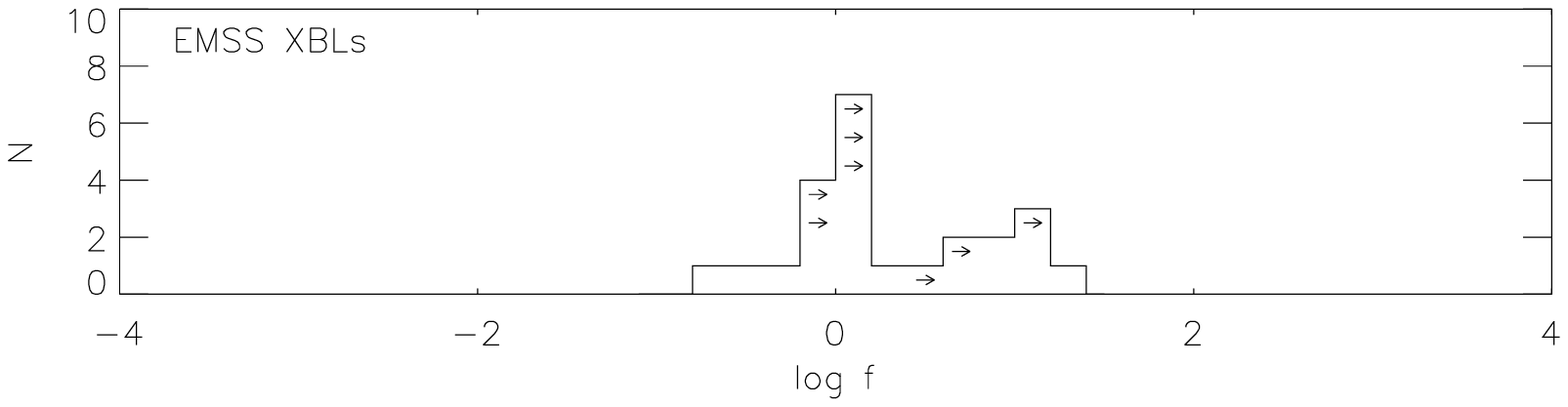} \\
\plotone{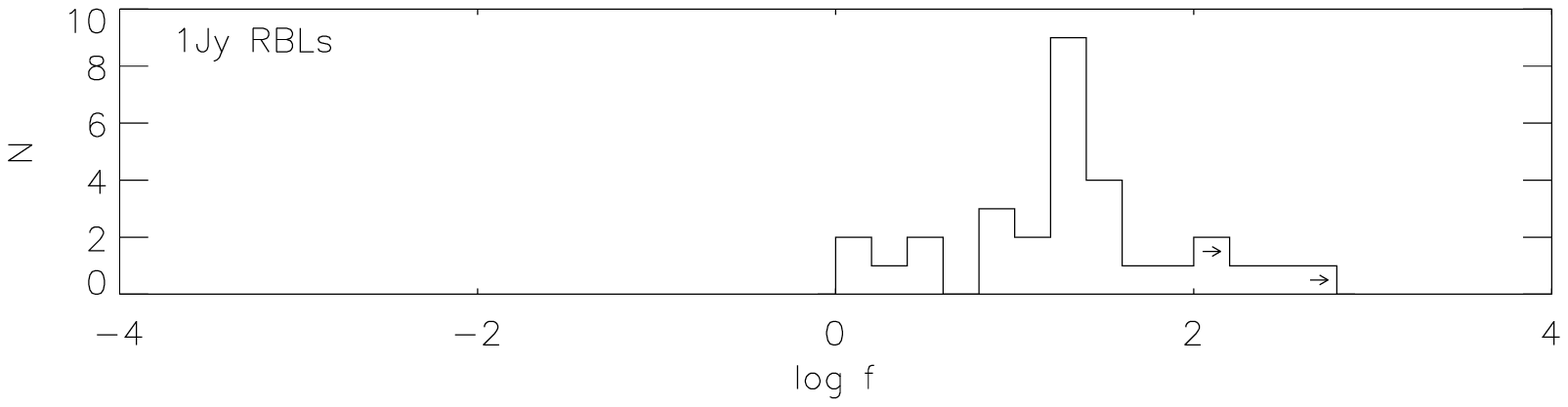}
\figcaption{The distribution of core-to-extended flux ratios ($f$) for the B2 FR--1
sample (Ulrich 1989), the M91 XBL sample and the entire 1Jy RBL sample (RS00).  Objects
marked with an arrow were unresolved; their values indicate a lower limit based upon the upper
limits on the corrected extended radio power as discussed in \S 3.2.
\label{fig-4.9}}
\end{figure}
\clearpage

\begin{figure}
\plotone{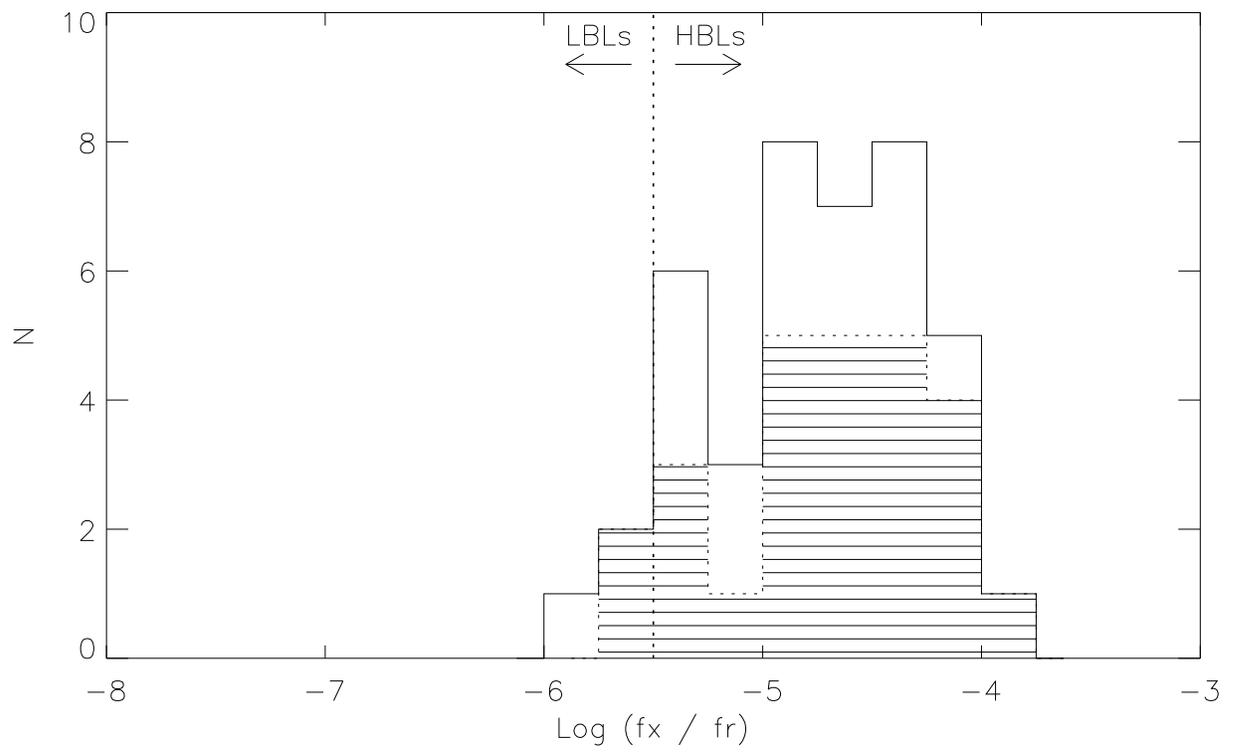}
\figcaption{The distribution of logarithmic X-ray (1 keV) to radio (5 GHz)
flux ratio values log(\fxfr) for the D40 sample.  The hatched area represents the M91
sample, which is a subsample of the D40. The putative dividing line between the HBL and LBL
classes at log(\fxfr) $= -5.5$ (e.g., Wurtz et al. 1996) is also shown.
\label{fig-4.11}}
\end{figure}



\makeatletter
\def\jnl@aj{AJ}
\ifx\revtex@jnl\jnl@aj\let\tablebreak=\nl\fi
\makeatother


\begin{deluxetable}{lcllrrrrrr}
\scriptsize
\tablecaption{Basic Properties of the EMSS BL Lac sample \label{tbl-4.1}}
\tablewidth{0pt}
\tablehead{
\colhead{Source Name} & \colhead{Sample\tablenotemark{a}} & \colhead{$z$} & \colhead{$\alpha_x$} & \colhead{$f_x$ (0.3-3.5 keV)} &
\colhead{$f_x$ (2 keV)} & \colhead{$V$} & \colhead{$f_r$ (5 GHz)} & \colhead{$a_{ox}$} & \colhead{$a_{ro}$} \\
\colhead{} & \colhead{} & \colhead{} & \colhead{} & \colhead{(x $10^{-13}$ cgs)} & \colhead{(x $10^{-2}$ $\mu$Jy)} & \colhead{} &
\colhead{(mJy)} & \colhead{} & \colhead{}
}
\startdata

MS 0122.1+0903	&	M91 & 0.338 & -0.62	&	8.23	&	8.43	&	19.98	&	1.4	&	0.80	&	0.39	\\
MS 0158.5+0019	&	M91 & 0.298 & -1.46	&	80.68	&	46.74	&	17.96	&	11.3	&	0.90	&	0.37	\\
MS 0205.7+3509	&	M91 & 0.318: & -1.71	&	6.48	&	2.82	&	19.24	&	3.6	&	1.26	&	0.34	\\
MS 0257.9+3429	&	M91 & 0.246 & -1.67	&	16.47	&	7.59	&	18.53	&	10.0	&	1.07	&	0.43	\\
MS 0317.0+1834	&	M91 & 0.190 & -1.32	&	54.06	&	34.28	&	18.12	&	17.0	&	0.71	&	0.45	\\
MS 0331.3--3629	&	D40 & 0.308 & -1.64	&	5.22	&	2.57	&	18	&	8.7	&	1.15	&	0.46	\\
MS 0350.0--3712	&	D40 & 0.165 & -1.52	&	23.21	&	12.82	&	17	&	16.8	&	1.20	&	0.37	\\
MS 0419.3+1943	&	M91 & 0.516 & -0.72	&	26.94	&	24.69	&	20.26	&	8.0	&	0.70	&	0.50	\\
MS 0607.9+7108	&	M91 & 0.267 & -1.21	&	15.13	&	10.56	&	19.61	&	18.2	&	0.94	&	0.51	\\
MS 0622.5--5256	&	\nodata & \nodata & -1.94	&	4.74	&	1.67	&	19.5	&	24.1	&	1.31	&	0.51	\\
MS 0737.9+7441	&	M91 & 0.314 & -1.34	&	104.80	&	67.20	&	16.89	&	24.0	&	0.97	&	0.32	\\
MS 0922.9+7459	&	M91 & 0.638 & -0.78	&	12.34	&	11.77	&	19.74	&	3.3	&	0.82 &	0.43	\\
MS 0950.9+4929	&	M91 & \nodata & -1.76	&	13.59	&	5.95	&	19.30	&	3.3	&	1.00	&	0.35	\\
MS 0958.9+2102	&	D40 & 0.344 & -1.50: &	2.03	&	1.13	&	19.84	&	1.5	&	1.08	&	0.43	\\
MS 1019.0+5139	&	M91 & 0.141 & -0.52	&	16.74	&	17.93	&	18.09	&	2.4	&	0.87	&	0.35	\\
MS 1050.7+4946	&	M91 & 0.140 & -1.62	&	9.25	&	4.65	&	16.86	&	53.8	&	1.37	&	0.47	\\
MS 1133.7+1618	&	D40 & 0.574: & -1.50:	&	3.52	&	1.97	&	20.04	&	9.0	&	1.12	&	0.51	\\
MS 1154.1+4255	&	D40 & 0.172 & -1.68:	&	2.64	&	1.25	&	17.67	&	9.8	&	1.37	&	0.43	\\
MS 1205.7--2921	&	D40 & 0.249 & -1.48:	&	3.78	&	2.13	&	17.5:	&	4.4	&	1.23	&	0.39	\\
MS 1207.9+3945	&	M91 & 0.616 & -1.13	&	15.45	&	11.68	&	19.12	&	5.8	&	0.97	&	0.41	\\
MS 1209.0+3917	&	D40 & 0.616 & -1.25	&	2.85	&	1.98	&	20	&	4.6	&	1.08	&	0.47	\\
MS 1219.9+7542 & D40 & 0.240 & -0.70 &  2.70 & 2.50 & 18.70 &  1.9 & 1.14 & 0.36  \\
MS 1221.8+2452	&	M91 & 0.218 & -1.47	&	10.46	&	6.04	&	17.65	&	26.4	&	1.33	&	0.41	\\
MS 1229.2+6430	&	M91 & 0.164 & -0.99	&	36.61	&	30.60	&	16.89	&	42.0	&	1.09	&	0.43	\\
MS 1235.4+6315	&	M91 & 0.297 & -1.91	&	13.62	&	5.05	&	18.59	&	7.0	&	1.20	&	0.37	\\
MS 1256.3+0151	&	D40 & 1.4:: & -1.63:	&	2.06	&	1.02	&	20	&	8.0	&	1.28	&	0.48	\\
MS 1258.4+6401	&	D40 & \nodata & -1.53:	&	3.45	&	1.96	&	20.22	&	12.0	&	1.15	&	0.53	\\
MS 1312.1--4221	&	\nodata & 0.105 & -1.07:	&	190.82	&	148.46	&	16.6	&	18.5	&	1.03	&	0.27	\\
MS 1332.6--2935	&	D40 & 0.513 & -1.03	&	35.21	&	28.49	&	19.1	&	11.7	&	0.84	&	0.45	\\
MS 1333.3+1725 &	D40 & 0.465 & -1.21: &	3.42	&	2.45	&	20 &	0.9	&	0.83	&	0.44	\\
MS 1402.3+0416	&	M91 & 0.344: & -1.85	&	9.35	&	3.70	&	17.08	&	20.8	&	1.51	&	0.34	\\
MS 1407.9+5954	&	M91 & 0.496 & -1.74	&	15.26	&	6.79	&	19.67	&	16.5	&	0.96	&	0.54	\\
MS 1443.5+6348	&	M91 & 0.298 & -1.10	&	16.49	&	12.78	&	19.65	&	11.6	&	0.79	&	0.54	\\
MS 1458.8+2249	&	M91 & 0.235 & -2.31	&	9.57	&	2.19	&	16.79	&	29.8	&	1.64	&	0.35	\\
MS 1534.2+0148	&	M91 & 0.311 & -0.89	&	23.03	&	20.38	&	18.70	&	34.0	&	0.93	&	0.52	\\
MS 1552.1+2020	& M91 & 0.273 &	-0.79	&	48.73	&	45.80	&	17.70	&	37.5	&	0.93	&	0.46	\\
MS 1704.9+6046	&	D40 & 0.277 & -1.32	&	4.90	&	3.21	&	19.12	&	1.8	&	1.13	&	0.33	\\
MS 1757.7+7034	&	M91 & 0.406 & -2.38	&	26.58	&	16.39	&	18.27	&	7.2	&	1.06	&	0.33	\\
MS 2143.4+0704	&	M91 & 0.235 & -1.91	&	25.83	&	9.63	&	18.04	&	50.0	&	1.03	&	0.47	\\
MS 2306.1--2236	&	D40 & 0.137 & -1.35:	&	22.45	&	14.23	&	16	&	4.4	&	1.22	&	0.24	\\
MS 2316.3--4222	&	\nodata & 0.045 & -1.30	&	34.03	&	22.70	&	14.5	&	540.0	&	1.47	&	0.47	\\
MS 2336.5+0517	&	M91 & 0.74:: & -1.33:	&	5.45	&	3.48	&	20.3	&	4.9	&	1.03	&	0.46	\\
MS 2342.7--1531	&	D40 & \nodata & -2.17	&	3.59	&	0.99	&	19.22	&	2.3	&	1.20	&	0.32	\\
MS 2347.4+1924	&	M91 & 0.515 & -1.59	&	5.35	&	2.73	&	20.78	&	3.2	&	0.89	&	0.51	\\

\tablenotetext{a}{Note: All BL Lacs in the M91 sample also belong to the D40 sample.}
\enddata
\end{deluxetable}



\makeatletter
\def\jnl@aj{AJ}
\ifx\revtex@jnl\jnl@aj\let\tablebreak=\nl\fi
\makeatother


\begin{deluxetable}{llll}
\tablecaption{Luminosities (W Hz$^{-1}$) of the EMSS BL Lac sample
\label{tbl-4.2}}
\tablewidth{0pt}
\tablehead{
\colhead{Source Name} & \colhead{$L_x$} & \colhead{$L_o$} & \colhead{$L_r$} \\
\colhead{} & \colhead{(2 keV)} & \colhead{(5500\AA)} & \colhead{(5 GHz)}
}
\startdata

0122.1+0903	&	19.55	&	22.25	&	23.79	\\
0158.5+0019	&	20.10	&	22.95	&	24.59	\\
0205.7+3509	&	18.90:	&	22.50:	&	24.15:	\\
0257.9+3429	&	19.15	&	22.57	&	24.38	\\
0317.0+1834	&	19.63	&	22.51	&	24.39	\\
0331.3--3629	&	18.84	&	22.96	&	24.50	\\
0350.0--3712	&	19.08	&	22.84	&	24.27	\\
0419.3+1943	&	20.34	&	22.49	&	24.90	\\
0607.9+7108	&	19.40	&	22.20	&	24.71	\\
0622.5--5256	&	\nodata	&	\nodata	&	\nodata	\\
0737.9+7441	&	20.51	&	23.42	&	24.96	\\
0922.9+7459	&	20.18	&	22.87	&	24.69	\\
0950.9+4929	&	\nodata	&	\nodata	&	\nodata	\\
0958.9+2102	&	18.58	&	22.32	&	23.83	\\
1019.0+5139	&	19.16	&	22.27	&	23.29	\\
1050.7+4946	&	18.51	&	22.76	&	24.63	\\
1133.7+1618	&	19.18:	&	22.67:	&	25.03:	\\
1154.1+4255	&	18.09	&	22.61	&	24.07	\\
1205.7--2921	&	18.33	&	22.67	&	23.72	\\
1207.9+3945	&	20.07	&	23.09	&	24.90	\\
1209.0+3917	&	19.28	&	22.74	&	24.80	\\
1219.9+7542 & 18.74 & 22.48 & 23.64 \\
1221.8+2452	&	18.97	&	22.82	&	24.70	\\
1229.2+6430	&	19.49	&	22.88	&	24.66	\\
1235.4+6315	&	19.08	&	22.70	&	24.38	\\
1256.3+0151	&	19.40::	&	23.42::	&	25.72::	\\
1258.4+6401	&	\nodata	&	\nodata	&	\nodata	\\
1312.1--4221	&	19.81	&	22.62	&	23.93	\\
1332.6--2935	&	20.35	&	22.95	&	25.06	\\
1333.3+1725	&	19.18	&	22.51	&	23.86	\\
1402.3+0416	&	19.05:	&	23.42:	&	24.97:	\\
1407.9+5954	&	19.57	&	22.69	&	25.18	\\
1443.5+6348	&	19.58	&	22.28	&	24.60	\\
1458.8+2249	&	18.51	&	23.22	&	24.81	\\
1534.2+0148	&	19.84	&	22.69	&	25.10	\\
1552.1+2020	&	19.93	&	22.81	&	24.87	\\
1704.9+6046	&	18.90	&	22.43	&	23.73	\\
1757.7+7034	&	19.82	&	23.09	&	24.65	\\
2143.4+0704	&	19.19	&	22.72	&	25.04	\\
2306.1--2236	&	18.99	&	23.08	&	23.53	\\
2316.3--4222	&	18.28	&	22.73	&	24.67	\\
2336.5+0517	&	19.63::	&	22.77::	&	24.98::	\\
2342.7--1531	&	\nodata	&	\nodata	&	\nodata	\\
2347.4+1924	&	19.23	&	22.28	&	24.50	\\

\enddata
\end{deluxetable}



\makeatletter
\def\jnl@aj{AJ}
\ifx\revtex@jnl\jnl@aj\let\tablebreak=\nl\fi
\makeatother


\begin{deluxetable}{llccccccc}
\tablecaption{Optical Spectroscopic Data for EMSS BL Lac Sample \label{tbl-3}}
\tablewidth{0pt}
\tablehead{
\colhead{} & \colhead{} & \colhead{} & \colhead{} & \colhead{} & \multicolumn{4}{c}{log $L$ ($3\sigma$) Limits} \\
\cline{6-9} \\
\colhead{Source Name}  & \colhead{$z$} & \colhead{$R$} & \colhead{$B_{4000}$} & \colhead{Max $W_{\lambda}$
(\AA)} & \colhead{Mg II} & \colhead{[O II]} & \colhead{[O III]} & \colhead{H$\alpha$}
}
\startdata

0122.1+0903	&	0.338	&	7.2	& $	0.22	\pm	0.18	$ &	3.5 & 41.6	&	41.3	&	40.7	&	41.0	\\
0158.5+0019	&	0.298	&	5.4	& $	0.09	\pm	0.02	$ &	1.7 & 41.6	&	41.1	&	41.0	&	41.1	\\
0205.7+3509	&	0.318:	&	\nodata	& $	0.00	\pm	0.03	$ &	1.6 & 41.3	&	41.1	&	40.9	&	41.2	\\
0257.9+3429	&	0.246	&	7.3	& $	0.25	\pm	0.06	$ &	1.8 & 41.3	&	41.0	&	40.6	&	40.7	\\
0317.0+1834	&	0.190	&	\nodata	& $	0.21	\pm	0.06	$ &	1.0 & \nodata	&	40.3	&	40.0	&	\nodata	\\
0331.3--3629	&	0.308	&	\nodata	& $	0.37	\pm	0.10	$ &	2.3 & \nodata	&	41.2	&	41.3	&	\nodata	\\
0350.0--3712	&	0.165	&	\nodata	& $	0.18	\pm	0.12	$ &	3.0 & \nodata	&	40.9	&	40.9	&	\nodata	\\
0419.3+1943	&	0.516	&	4..0	& $	0.11	\pm	0.07	$ &	2.8 & 41.5	&	41.0	&	40.8	&	\nodata	\\
0607.9+7108	&	0.267	&	5.4	& $	0.09	\pm	0.04	$ &	0.9 & 41.5	&	40.9	&	40.6	&	40.9	\\
0622.5--5256	&	\nodata	&	\nodata	& \nodata &	\nodata	&	\nodata	&	\nodata	&	\nodata	&	\nodata	\\
0737.9+7441	&	0.314	&	4.3	& $	0.00	\pm	0.01	$ & 1.2 & 	41.5	&	41.3	&	41.2	&	41.4	\\
0922.9+7459	&	0.638	&	3.8	& $	0.20	\pm	0.07	$ &	5.5 & 41.6	&	41.5	&	41.6	&	\nodata	\\
0950.9+4929	&	\nodata	&	\nodata	& \nodata & 4.1 &	\nodata	&	\nodata	&	\nodata	&	\nodata	\\
0958.9+2102	&	0.344	&	12.1	& $	0.31	\pm	0.12	$ &	2.8 & 41.5	&	41.0	&	40.5	&	40.4	\\
1019.0+5139	&	0.141	&	\nodata	& $	0.23	\pm	0.05	$ &	1.2 & \nodata	&	40.7	&	40.1	&	{\bf 40.9}	\\
1050.7+4946	&	0.140	&	\nodata	& $	0.32	\pm	0.07	$ &	1.5 & \nodata	&	41.7	&	40.9	&	{\bf 41.3:}	\\
1133.7+1618	&	0.574:	&	4.0	& $	0.08	\pm	0.06	$ &	2.4 & 41.8	&	41.9	&	\nodata	&	\nodata	\\
1154.1+4255	&	0.172	&	\nodata	& $	0.33	\pm	0.10	$ &	3.1 & \nodata	&	41.8	&	41.1	&	40.9	\\
1205.7--2921	&	0.249	&	\nodata	& $	0.39	\pm	0.18	$ &	6.0 & \nodata	&	40.8	&	40.9	&	\nodata	\\
1207.9+3945	&	0.616	&	7.0	& $	0.07	\pm	0.04	$ &	1.8 & 41.0	&	40.9	&	40.8	&	\nodata	\\
1209.0+3917	&	0.616	&	\nodata	& $	0.15	\pm	0.13	$ &	4.0 & \nodata	&	41.1	&	40.8	&	\nodata	\\
1219.9+7542 & 0.240 & \nodata & $ 0.39 \pm 0.16 $ & 4.0 & \nodata & {\bf 40.8} & \nodata & \nodata \\
1221.8+2452	&	0.218	&	5.7	& $	0.02	\pm	0.02	$ &	\nodata	&	1.4 & 40.8	&	41.0	&	\nodata	\\
1229.2+6430	&	0.164	&	\nodata	& $	0.18	\pm	0.09	$ &	6.0 & 41.7	&	41.3	&	42.0	&	\nodata	\\
1235.4+6315	&	0.297	&	\nodata	& $	0.05	\pm	0.07	$ &	4.0 & 41.3	&	{\bf 40.9}	&	\nodata	&	\nodata	\\
1256.3+0151	&	1.4::	&	\nodata	& 	\nodata	 &	4.7 & {\bf 43.0::}	&	\nodata	&	\nodata	&	\nodata	\\
1258.4+6401	&	\nodata	&	\nodata	& \nodata &	4.5 & \nodata	&	\nodata	&	\nodata	&	\nodata	\\
1312.1--4221	&	0.105	&	7.2	& $	0.01	\pm	0.03	$ &	3.6 & \nodata	&	\nodata	&	41.2	&	41.1	\\
1332.6--2935	&	0.513	&	4.0	& $	0.09	\pm	0.10	$ &	4.9 & \nodata	&	42.0	&	41.9	&	42.6	\\
1333.3+1725	&	0.465	&	\nodata	& $	0.38	\pm	0.11	$ &	6.7 & \nodata	&	41.5	&	41.8	&	\nodata	\\
1402.3+0416	&	0.344:	&	4.5	& $ 0.00	\pm	0.00	$ &	0.8 & 41.8	&	41.7	&	41.6	&	41.8	\\
1407.9+5954	&	0.496	&	8.6	& $	0.10	\pm	0.05	$ &	1.8 & 41.3	&	41.1	&	40.9	&	\nodata	\\
1443.5+6348	&	0.298	&	7.0	& $	0.22	\pm	0.06	$ &	1.5 & 41.1	&	40.6	&	40.3	&	40.3	\\
1458.8+2249	&	0.235	&	4.5	& $	0.00	\pm	0.01	$ &	2.4 & \nodata	&	42.0	&	41.7	&		\\
1534.2+0148	&	0.311	&	9.2	& $	0.11	\pm	0.06	$ &	0.9 & 41.2	&	40.8	&	40.3	&	40.6	\\
1552.1+2020	&	0.273	&	\nodata	& $	0.13	\pm	0.09	$ &	3.0 & 41.9	&	41.9	&	41.8	&	\nodata	\\
1704.9+6046	&	0.277	&	6.8	& $	0.19	\pm	0.17	$ &	2.8 & 42.0	&	41.2	&	40.7	&	41.0	\\
1757.7+7034	&	0.406	&	8.5	& $	0.01	\pm	0.00	$ &	1.4 & 41.2	&	41.0	&	40.8	&	41.2	\\
2143.4+0704	&	0.235	&	6.6	& $	0.00	\pm	0.02	$ &	1.1 & 41.6	&	41.1	&	40.7	&	40.9	\\
2306.1--2236	&	0.137	&	5.5	& $	0.34	\pm	0.03	$ &	6.5 & \nodata	&	41.3	&	41.0	&	41.2	\\
2316.3--4222	&	0.045	&	\nodata	& $	0.22	\pm	0.11	$ &	1.5 & \nodata	&	{\bf 40.7}	&	{\bf 40.4}	&	\nodata	\\
2336.5+0517	& 0.74::	&	\nodata	& $	0.00	\pm	0.05	$ &	3.6 & 42.0	&	42.1	&	\nodata	&	\nodata	\\
2342.7--1531	&	\nodata	&	\nodata	& \nodata & 3.6 &	\nodata	&	\nodata	&	\nodata	&	\nodata	\\
2347.4+1924	&	0.515	&	\nodata	& $	0.18	\pm	0.16	$ & 4.0 &	42.0	&	41.3	&	41.3	&	\nodata	\\

\enddata
\label{tbl-4.3}
\end{deluxetable}



\makeatletter
\def\jnl@aj{AJ}
\ifx\revtex@jnl\jnl@aj\let\tablebreak=\nl\fi
\makeatother


\begin{deluxetable}{lrrrrrrr}
\tablecaption{20cm Radio Properties of EMSS BL Lacs \label{tbl-4.5}}
\tablewidth{0pt}
\tablehead{
\colhead{Object}	& \colhead{($\alpha, \delta$)} & \colhead{$S_{core}$} & \colhead{$S_{ext}$} & \colhead{log $P_{core}$}
& \colhead{log $P_{ext}$} & \colhead{$f$} 	\\				
\colhead{}		& \colhead{(J2000)} & \colhead{(mJy)} & \colhead{(mJy)} & \colhead{(W Hz$^{-1}$)} & \colhead{(W Hz$^{-1}$)}
& \colhead{} 		
}
\startdata

0122.1+0903	&	01:24:44.54, +09:18:49.4	&	$2.03\pm0.1$	&	$< 1.5$	&	23.91 &	$< 23.85$	&	$> 1.4$		\\				
	&		&	(1.67 x 1.44)	&		&		&	$<24.11$	&	$> 0.7$			\\				
0205.7+3509	&	02:08:38.17, +35:23:12.8	&	$5.04\pm0.1$	&	$< 2.0$	&	24.26	&	$<23.91$	&	$> 2.5$			\\				
	&		&	(1.62 x 1.29)	&		&		&	$<24.16$	&	$> 1.4$			\\				
0419.3+1943	&	04:22:18.34, +35:23:12.8	&	$9.61\pm0.1$	&	$ < 1.2$	&	24.91	&	$<24.09$	&	$> 8.3$			\\				
	&		&	(1.61 x 1.33)	&		&		&	$<24.56$	&	$> 2.8$		\\				
0922.9+7459	&	09:28:22.64, +74:45:30.8	&	$4.99\pm0.1$	&	$< 0.8$	&	24.80	&	$<24.10$	&	$> 6.5$			\\				
	&		&	(1.94 x 1.54)	&		&		&	$<24.80$	&	$> 1.3$			\\				
1019.0+5139	&	10:22:12.61, +51:24:00.3	&	$3.13\pm0.1$	&	$< 0.6$	&	23.39	&	$<22.69$	&	$> 5.3$			\\				
	&		&	(4.55 x 4.19)	&		&		&	$<22.74$	&	$> 4.8$			\\		
1050.7+4946	&	10:53:44.13, +49:29:55.8	&	$52.84\pm0.1$	&	$13.2\pm0.4$	&	24.61	&	24.04	&	4.0	 \\		
	&		&	(5.97 x 5.60)	&		&		&	24.84	&	2.5		\\	
1207.9+3945 & 12:10:26.66, +39:29:09.0 & $17.91\pm0.4$ &  $4.5\pm1.9$ & 25.33 &   24.83 & 4.0  \\
	&  &  (4.66 x 4.34)      &  &  & 25.53 & 0.8  \\
1235.4+6315	&	12:37:39.04, +62:58:42.4	&	$13.84\pm0.1$	&	$< 0.7$	&	24.64	&	$<23.39$	&	$> 20.5$			\\		
	&		&	(4.72 x 4.23)	&		&		&	$<23.60$	&	$> 12.6$			\\		
1757.7+7034	&	17:56:37.48, +70:33:24.4	&	$10.82\pm0.2$	&	$< 3.4$	&	24.79	&	$< 24.35$	&	$> 3.2$			\\		
	&		&	(1.40 x 1.31)	&		&		&	$<24.67$	&	$> 1.5$			\\		
2336.5+0517	&	23:39:07.40, +05:34:26.3	&	$4.79\pm0.1$	&	$4.2\pm0.2$	&	24.90	&	24.96	&	1.2		\\		
	&		&	(1.66 x 1.60)	&		&		&	25.71	&	0.1		\\		
2347.4+1924	&	23:50:01.72, +19:41:51.5	&	$3.86\pm0.1$	&	$1.7\pm0.2$	&	24.52	&	24.26	&	2.2		\\		
	&		&	(1.58 x 1.39)	&		&		&	23.63	&	0.3		\\

\enddata
\end{deluxetable}



\makeatletter
\def\jnl@aj{AJ}
\ifx\revtex@jnl\jnl@aj\let\tablebreak=\nl\fi
\makeatother


\begin{deluxetable}{lcccl}
\tablecaption{Unidentified EMSS Sources at $\delta \geq -40$\arcdeg
\label{tbl-4.4}}
\tablewidth{0pt}
\tablehead{
\colhead{Source Name} & \colhead{$f_x$ (0.3-3.5 keV)} & \colhead{$f_r$ (5 GHz)} &
\colhead{Suggested} & \colhead{Comments} \\
\colhead{} & \colhead{(x $10^{-13}$ cgs)} & \colhead{(mJy)} & \colhead{Identification} &
\colhead{}}
\startdata

0134.4+2043  &   1.12    &      3.1     &    ?     &    IPC \& PSPC \\
             & & & &                                      positions consistent \\
                    & & & &                                with a radio source \\
 \\
0354.2--3658  &   3.09    &     $<0.6$     &  Cluster  &   Possible 21st mag. \\
              & & & &                                      cD galaxy \\
\\
0501.0--2237  &   3.26    &     17.9    &   AGN    &     Radio source well \\
              & & & &                                      outside both IPC \& \\
              & & & &                                      PSPC error circles \\
\\         
1237.9--2927   &  3.48    &     $<0.8$   &    Star   &     Possible flare on \\
              & & & &                                      mid-K star \\
\\
1317.0--2111  &   5.23    &      1.8   &     ?      &    ROSAT HRI non- \\
             & & & &                                       detection (RSP99) \\
\\
1411.0--0310  &   1.97    &     $<1.0$   &     ?    &      BSO in IPC \& PSPC \\
              & & & &                                      error circles  \\
\\
2144.2+0358  &   1.93    &     $<0.5$   &     ?    &      ROSAT HRI \& PSPC \\
              & & & &                                        confirm source; \\
              & & & &                                        blank field \\
\\
2225.7--2100  &   1.09    &     13.2   &    BL Lac?   &     IPC \& PSPC positions \\
              & & & &                                      consistent with a \\
              & & & &                                      radio source; poor   \\
              & & & &                                      spectrum shows AGN: \\

\enddata
\end{deluxetable}



\makeatletter
\def\jnl@aj{AJ}
\ifx\revtex@jnl\jnl@aj\let\tablebreak=\nl\fi
\makeatother


\def\vvmax{$\langle V/V_{max} \rangle$}
\def\veva{$\langle V_e/V_a \rangle$}
\def\avgz{$\langle z \rangle$}

\begin{deluxetable}{lrrr}
\tablecaption{\veva\ for EMSS BL Lac Subsamples \label{tbl-4.6}}
\tablewidth{0pt}
\tablehead{
\colhead{Subsample} & \colhead{$N$} & \colhead{\veva} & \colhead{\%\tablenotemark{a}}
}
\startdata

D40           & 41 & $0.427\pm0.045$ & 4\% \\
M91           & 26 & $0.399\pm0.057$ & 7\% \\
$f_x \geq 10$\tablenotemark{b}   & 22 & $0.464\pm0.062$ & 68\% \\
$f_x \geq 30$\tablenotemark{b}   &  6 & $0.547\pm0.118$ & 72\% \\
\tablenotetext{a}{Probability of no evolution result.}
\tablenotetext{b}{$f_x$ is in units of $10^{-13}$ ergs cm$^{-2}$ s$^{-1}$
($0.3-3.5$ keV)}
\enddata
\end{deluxetable}



\makeatletter
\def\jnl@aj{AJ}
\ifx\revtex@jnl\jnl@aj\let\tablebreak=\nl\fi
\makeatother


\def\vvmax{$\langle V/V_{max} \rangle$}
\def\veva{$\langle V_e/V_a \rangle$}
\def\avgz{$\langle z \rangle$}

\begin{deluxetable}{lrrr}
\tablecaption{\veva\ for EMSS BL Lac Subsamples \label{tbl-4.7}}
\tablewidth{0pt}
\tablehead{
\colhead{Subsample} & \colhead{$N$} & \colhead{\veva} & \colhead{\%\tablenotemark{a}}
}
\startdata

$-6.0 <$ \fxfr\ $ \leq -5.0$          & 13 & $0.550\pm0.080$ & 54\% \\
$-5.0 <$ \fxfr\ $ \leq -4.5$          & 14 & $0.470\pm0.077$ & 53\% \\
$-4.5 <$ \fxfr\ $ \leq -3.9$          & 14 & $0.271\pm0.077$ & 0.4\% \\

\tablenotetext{a}{Probability of no evolution result.}
\enddata
\end{deluxetable}

\end{document}